\documentclass[fleqn,usenatbib,useAMS]{mnras}
\pdfoutput=1

\usepackage{graphicx}	
\usepackage{amsmath}	
\usepackage{amssymb}	
\usepackage{pdflscape}	

\usepackage{caption}

\usepackage[T1]{fontenc}
\usepackage{ae,aecompl}
\usepackage{txfonts}

\usepackage{paralist}
\usepackage{afterpage}
\usepackage[normalem]{ulem}
\usepackage{subcaption}

\def\UW{$^{1}$}
\def\UMelbourne{$^{2}$}
\def\CAASTRO{$^{3}$}
\def\UWeSci{$^{4}$}
\def\ASU{$^{5}$}
\def\Brown{$^{6}$}
\def\SKASA{$^{7}$}
\def\RU{$^{8}$}
\def\ANU{$^{9}$}
\def\Haystack{$^{10}$}
\def\MIT{$^{11}$}
\def\Curtin{$^{12}$}
\def\Dunlap{$^{13}$}
\def\USydney{$^{14}$}
\def\CfA{$^{15}$}
\def\Victoria{$^{16}$}
\def\UWisc{$^{17}$}
\def\UMichigan{$^{18}$}
\def\CASS{$^{19}$}
\def\Tata{$^{20}$}
\def\ASTRON{$^{21}$}
\def\RRI{$^{22}$}
\def\NRAO{$^{23}$}
\def\UWA{$^{24}$}
\def\Berkeley{$^{25}$}
\def\SKAUK{$^{26}$}

\def\Cand{$7466$~}              

\def\Rejects{$45$~}
\def\EFlagged{$66$~}

\def\Outliers{$205$~}
\def\MFlagged{$900$~}

\def\vTenOut{$90$~}
\def\vFiveIn{$192$~}

\def\NewSrcs{$25$~}
\def\Deleted{$24$~}
\def\Total{$7394$~}

\begin{document}

\title[MWA EoR Foreground Survey] {A High Reliability Survey of Discrete Epoch of Reionization Foreground Sources in the MWA EoR0 Field}

\author[P.~Carroll, J.~Line, et al.]
{P.~A.~Carroll\UW,
J.~Line\UMelbourne$^,$\CAASTRO,
M.~F.~Morales\UW$^\dagger$,
N.~Barry\UW,
A.~P.~Beardsley\ASU$^,$\UW,
B.~J.~Hazelton\UW$^,$\UWeSci,
\newauthor
D.~C.~Jacobs\ASU,
J.~C.~Pober\Brown$^,$\UW,
I.~S.~Sullivan\UW,
R.~L.~Webster\UMelbourne$^,$\CAASTRO,
G.~Bernardi\SKASA$^,$\RU,
\newauthor
J.~D.~Bowman\ASU,
F.~Briggs\ANU$^,$\CAASTRO,
R.~J.~Cappallo\Haystack, 
B.~E.~Corey\Haystack, 
A.~de~Oliveira-Costa\MIT,
\newauthor
J.~S.~Dillon\Berkeley$^,$\MIT
D.~Emrich\Curtin,
A.~Ewall-Wice\MIT,
L.~Feng\MIT,
B.~M.~Gaensler\Dunlap$^,$\USydney$^,$\CAASTRO,
\newauthor
R.~Goeke\MIT,
L.~J.~Greenhill\CfA,
J.~N.~Hewitt\MIT,
N.~Hurley-Walker\Curtin,
M.~Johnston-Hollitt\Victoria,
\newauthor
D.~L.~Kaplan\UWisc, 
J.~C.~Kasper\UMichigan$^,$\CfA, 
HS.~Kim\UMelbourne$^,$\CAASTRO,
E.~Kratzenberg\Haystack, 
E.~Lenc\USydney$^,$\CAASTRO,
A.~Loeb\CfA,
\newauthor
C.~J.~Lonsdale\Haystack, 
M.~J.~Lynch\Curtin, 
B.~McKinley\UMelbourne$^,$\CAASTRO,
S.~R.~McWhirter\Haystack,
D.~A.~Mitchell\CASS$^,$\CAASTRO, 
\newauthor
E.~Morgan\MIT, 
A.~R.~Neben\MIT,
D.~Oberoi\Tata,
A.~R.~Offringa\ASTRON$^,$\CAASTRO, 
S.~M.~Ord\Curtin$^,$\CAASTRO,
\newauthor
S.~Paul\RRI,
B.~Pindor\UMelbourne$^,$\CAASTRO,
T.~Prabu\RRI,
P.~Procopio\UMelbourne$^,$\CAASTRO,
J.~Riding\UMelbourne$^,$\CAASTRO,
A.~E.~E.~Rogers\Haystack, 
\newauthor
A.~Roshi\NRAO, 
N.~Udaya~Shankar\RRI, 
S.~K.~Sethi\RRI,
K.~S.~Srivani\RRI, 
R.~Subrahmanyan\RRI$^,$\CAASTRO, 
\newauthor
M.~Tegmark\MIT,
Nithyanandan~Thyagarajan\ASU,
S.~J.~Tingay\Curtin$^,$\CAASTRO, 
C.~M.~Trott\Curtin$^,$\CAASTRO,
\newauthor
M.~Waterson\SKAUK$^,$\UWA,
R.~B.~Wayth\Curtin$^,$\CAASTRO, 
A.~R.~Whitney\Haystack, 
A.~Williams\Curtin, 
C.~L.~Williams\MIT,
\newauthor
C.~Wu\UWA,
J.~S.~B.~Wyithe\UMelbourne$^,$\CAASTRO
\\
$^{1}$University of Washington, Seattle, USA\\
$^{2}$The University of Melbourne, Melbourne, Australia\\
$^{3}$ARC Centre of Excellence for All-sky Astrophysics (CAASTRO)\\
$^{4}$University of Washington, eScience Institute, Seattle, USA\\
$^{5}$Arizona State University, Tempe, USA\\
$^{6}$Brown University, Department of Physics, Providence, USA\\
$^{7}$Square Kilometre Array South Africa (SKA SA), Pinelands, South Africa\\
$^{8}$Department of Physics and Electronics, Rhodes University, Grahamstown, South Africa\\
$^{9}$The Australian National University, Canberra, Australia\\
$^{10}$MIT Haystack Observatory, Westford, USA\\
$^{11}$MIT Kavli Institute for Astrophysics and Space Research, Cambridge, USA\\
$^{12}$International Centre for Radio Astronomy Research, Curtin University, Perth, Australia\\
$^{13}$Dunlap Institute for Astronomy and Astrophysics, University of Toronto, Toronto, Canada\\
$^{14}$University of Sydney, Sydney, Australia\\
$^{15}$Harvard-Smithsonian Center for Astrophysics, Cambridge, USA\\
$^{16}$Victoria University of Wellington, School of Chemical \& Physical Sciences, Wellington, New Zealand\\
$^{17}$University of Wisconsin--Milwaukee, Milwaukee, USA\\
$^{18}$University of Michigan, Department of Atmospheric, Oceanic and Space Sciences, Ann Arbor, USA\\
$^{19}$CSIRO Astronomy and Space Science (CASS), Epping, Australia\\
$^{20}$National Centre for Radio Astrophysics, Tata Institute for Fundamental Research, Pune, India\\
$^{21}$Netherlands Institute for Radio Astronomy (ASTRON), Dwingeloo, Netherlands\\
$^{22}$Raman Research Institute, Bangalore, India\\
$^{23}$National Radio Astronomy Observatory, Charlottesville and Greenbank, USA\\
$^{24}$International Centre for Radio Astronomy Research, University of Western Australia, Crawley, Australia\\
$^{25}$University of California, Berkeley, California, USA\\
$^{26}$SKA Organization Headquarters, Jodrell Bank, United Kingdom\\
}

\label{firstpage}
\pagerange{\pageref{firstpage}--\pageref{lastpage}}\pubyear{2016}
\maketitle

\clearpage

\begin{abstract}
Detection of the Epoch of Reionization HI signal requires a precise understanding of the intervening galaxies and AGN, both for instrumental calibration and foreground removal. We present a catalogue of \Total extragalactic sources at 182~MHz detected in the RA=0 field of the Murchison Widefield Array Epoch of Reionization observation programme. Motivated by unprecedented requirements for precision and reliability we develop new methods for source finding and selection. We apply machine learning methods to self-consistently classify the relative reliability of 9490 source candidates. A subset of \Cand are selected based on reliability class and signal-to-noise ratio criteria. These are statistically cross-matched to four other radio surveys using both position and flux density information. We find 7369 sources to have confident matches, including 90 partially resolved sources that split into a total of 192 sub-components. An additional \NewSrcs unmatched sources are included as new radio detections. The catalogue sources have a median spectral index of $-$0.85. Spectral flattening is seen toward lower frequencies with a median of $-$0.71 predicted at 182~MHz. The astrometric error is $7$" compared to a 2.3' beam FWHM. The resulting catalogue covers $\sim$1400~deg$^2$ and is complete to approximately 80~mJy within half beam power. This provides the most reliable discrete source sky model available to date in the MWA EoR0 field for precision foreground subtraction.
\end{abstract}


\section{Introduction}
\label{sec:Intro}

Observations of the Epoch of Reionization (EoR) 21~cm neutral Hydrogen (HI) signal are one of the key science goals of the Murchison Widefield Array \citep[MWA;][]{Lonsdale2009,Tingay2013,Bowman2013}. Astrophysical foreground sources are estimated to be 4--5 orders of magnitude brighter than this signal, presenting a major obstacle and motivating a careful dedicated survey to be used for both calibration and foreground power removal within the EoR analysis. 
 
During the commissioning phase of the MWA and early development of the EoR analysis pipeline, the Molonglo Reference Catalogue (MRC; \citealt{Large1991}) was used to model the foregrounds. The MRC is complete to 1~Jy at 408~MHz or about 2~Jy at 182~MHz with an assumed average spectral index of $-$0.8. This is not only much shallower than desired, but large errors are introduced by this naive flux density extrapolation. 

The MRC catalogue was later replaced by the MWA Commissioning Survey \citep{Hurley-Walker2014}, giving us a deeper and frequency-specific sky model that greatly improved calibration and foreground power subtraction. Unfortunately the MWACS does not cover the northernmost $5^{\circ}$ of the EoR fields, and as an early product of a partially built instrument it contains large flux density and astrometric uncertainties. While the GaLactic and Extragalactic All-sky MWA survey \citep[GLEAM;][Hurley-Walker {\it in prep.}]{Wayth2015} was still underway, an extragalactic survey tuned to the requirements of EoR foreground subtraction in the MWA EoR0 (RA=0, Dec=-27) field was initiated.

Processing MWA data is different in a number of ways from traditional interferometric radio data processing. Traditional radio analysis recipes are well tuned for arrays with narrow fields of view, stable beams, sparse instantaneous $uv$ coverage, and limited source confusion. MWA data break these assumptions. The MWA primary beam is $\sim$1400~deg$^2$ and changes with time as the field drifts through, the instantaneous $uv$ coverage is excellent, and the continuum confusion limit is reached very quickly. Background sky regions reach a signal-to-noise ratio (SNR) $\sim$~10 in a 112~sec integration. 

For these reasons most MWA analyses have deconvolved sources on times scales of a few minutes or less to minimize time-dependent beam effects and leverage the snapshot $uv$ coverage. Many different deconvolution algorithms may be chosen, though typically a pixel based algorithm such as CLEAN is used (i.e. source components are located at pixel centres and may take either positive or negative values). 

Going from radio deconvolution products to a source catalogue is often performed by fitting sources in a restored image. For the MWA this has meant restoring the individual snapshot observations, mapping to celestial coordinates to remove widefield distortions, and then co-adding or mosaicing (e.g. \citealt{Wayth2015}). Combining snapshots through co-adding increases the SNR of real sources and drives down the amplitude of false side lobe sources as the array beam (PSF) rotates. While this reduces contamination, it removes information that is valuable for determining reliability, that is whether or not a source is in fact true and real. True sources should be detected consistently across all observations, while noise and side lobes will vary in time. This information is lost if source finding is not performed prior to image stacking. Reliability is a primary concern for the EoR foreground model as it will be used for both calibration and subtraction \citep{Barry2016}.

Various source extractors have been developed that isolate flux density peaks in an image, fit an assumed PSF or morphological shape, and measure the integrated flux density and background rms (e.g. \citealt{Hancock2012}). This approach does well but is not without its limitations. \citet{Hopkins2015} demonstrate clear variations in the performance of eleven different source finders in terms of completeness, reliability, and measurement accuracy. Blended or extended sources were particularly troublesome despite the fact that all sources in the test images were artificially positioned at pixel-centers.   
In this work we take a novel approach to source finding that does not require the use of restored images or an assumed source shape. Individual snapshots are deconvolved using the Fast Holographic Deconvolution \citep[][Sullivan et al., {\it in prep.}]{Sullivan2012} software package. FHD deconvolution is similar to CLEAN in that sources are iteratively removed using point-like source components with a CLEAN gain. However, FHD differs in using a full direction-dependent PSF, centroiding each source component (not fixed at pixel centers), and using only positive components. For a bright unresolved point source, FHD will create a set of positive source components, each with a floating-point precision position, tightly scattered about the actual source position ($\ll$ PSF FWHM). Diffuse sources will be modeled by a cloud of source components approximating the extended flux density distribution.

The FHD source components trace the sub-resolution flux density distribution of sources remarkably well. We can therefore identify and extract sources simply by spatially clustering the components regardless of shape. This does not require a restored image, nor the assumption of a PSF or morphological model. Machine-learning classification methods are then used to self-consistently assess source reliability. This process for source finding, measurement, and classification has been termed KATALOGSS (KDD Astrometry, Trueness, and Apparent Luminosity of Galaxies in Snapshot Surveys; hereafter abbreviated to KGS). We use the KGS results in combination with a typical signal-to-noise detection threshold and cross-matching to maximize the overall completeness and reliability of the final catalogue. 

This paper presents these methods and the resulting source survey of the MWA EoR0 field. The observations and pre-processing are described in \S\ref{sec:Data}; the process of source finding and association across snapshots in \S\ref{sec:SourceFinding}; and the reliability classification is detailed in \S\ref{sec:Classification}. In \S\ref{sec_XMatch} we introduce the Positional Update and Matching Algorithm (PUMA) used to cross-match the catalogue to other radio surveys and fit for the power-law spectral index. The final catalogue is described in \S\ref{sec:catalogue} along with measures of the astrometric accuracy, spectral index distribution, and completeness. We further identify \NewSrcs new sources previously undetected at radio frequencies and discuss potential associations in \S\ref{sec:newsources}.

\section{Data \& Processing}
\label{sec:Data}

The MWA EoR0 field is centred at $RA=0$~hr and $Dec=-27^{\circ}$, and was chosen because it has no bright complex sources in the primary field of view. The FWHM of the antenna beam is approximately 20$^\circ$, but sources in the edges of the beam and first few side lobes are clearly visible and should be subtracted \citep{Pober2016,Thyagarajan2015a,Thyagarajan2015b}. For this catalogue we concentrate on identifying sources in the primary beam but go out to the 5\% power point (nearly first beam null, $\sim 1400$~deg~$^2$). The data for this catalogue include seventy-five~2~min snapshot observations (112~sec consecutive integrations with 8 second gaps) from the night of August~23, 2013. The observations were made at 182~MHz with 31~MHz bandwidth and cover 2.5~hours in total. 

The MWA antenna can only point in discrete locations in elevation and azimuth \citep{Tingay2013}, leading to a `drift-and-shift' observation pattern where the sky drifts through the antenna beam for approximately half an hour before the antenna is re-pointed to follow the field. These 75 snapshots cover 5 antenna pointings centred on field transit. The visibility data were averaged in time and frequency to 2 seconds and 40~kHz, and flagged for RFI with Cotter \citep{Offringa2010}. Due to the remoteness of the Murchison Radio Observatory, less than 3\% of the data were flagged for RFI.  

Fast Holographic Deconvolution (FHD; \citealt{Sullivan2012}) was used for calibration, imaging, and deconvolution
FHD is an efficient implementation of A-projection \citep{Morales2009,Bhatnagar2008,Myers2003}, and enables direction dependent beam corrections by using the antenna holographic beam pattern in gridding. During each deconvolution iteration, the brightest pixels are identified and the sources are fit with a Gaussian-approximated synthesized beam shape to determine the flux density. Unlike standard CLEAN, the position is centroided to a floating point location and all components are strictly positive-valued. A gain of 0.2 is applied, and the source components are forward modelled through the direction dependent instrument model to update the residual sky image. The deconvolution loop is stopped if the rms of the residual image increases or if the source fitting fails and no valid components can be extracted. Otherwise, deconvolution halts after a maximum of 500 iterations or a maximum of $3\times10^5$ components are subtracted.

In four of the 75 snapshots, deconvolution stopped early due to lack of convergence on a single bright and extended source (starburst galaxy, NGC 253). These four snapshots were excluded from the remaining analysis. Among the remaining 71 snapshots, the deconvolution threshold ranged from 45--76~mJy with an average of $57\pm6$~mJy. The background noise level of the snapshot images is stable over time, averaging $9-11$~mJy/pixel at beam-center.

\section{Source Finding}
\label{sec:SourceFinding}

The FHD source components trace the sub-resolution flux density distribution of sources remarkably well. We can therefore identify and extract sources simply by identifying clusters of components. This does not require a restored image, nor the assumption of a PSF or morphological model, thus avoiding potential sources of error inherent to producing a restored image and using an image-based source finder (see \S\ref{sec:Intro}).

A driving factor in distinguishing true sources from contamination is the consistency of detection. True sources should appear in most snapshots, while false sources (noise peaks and side lobe sources) vary in time and should appear in few. This is a valuable metric for determining source reliability and is lost when images are co-added. Source finding is therefore performed individually for each 112~sec snapshot observation to preserve this information.

DBSCAN \citep{Ester1996}, a density-based hierarchical clustering algorithm, is used to identify spatially isolated clusters of source components produced by FHD. DBSCAN works by first identifying local maxima in the density distribution of components and then building clusters hierarchically outward from these cores. The input parameters for DBSCAN are the neighborhood radius within which a point is considered to be a part of the same cluster, and the minimum number of points required within that radius for a new core to be formed. For this application the minimum number of components was set to one so as not to exclude sources with only a single component extracted, and the neighborhood radius was set to the half-width-half-max (HWHM) of the PSF. 

Approximately 5000 clusters were identified in each snapshot, for a total of $3.58\times10^5$ across all snapshots. Each cluster is considered a single detection. For each detection, we calculated the position centroid, standard deviation, and rms width of the radial scatter of the component positions. The flux density, $S$, was estimated according to Equation~\ref{eqn:flux-corr}, where $S_i$ and $g_i$ are the flux density and clean gain factor for $i^{th}$ component and $N_{\rm comp}$ is the total number of components. This corrects for the fraction the flux density that was not deconvolved for a point source. 

\begin{equation}
\label{eqn:flux-corr}
S =   \frac{\sum_{i=1}^{N_{\rm comp}}{S_i} }{ 1 - \prod_{i=1}^{N_{\rm comp}}(1- g_i)}
\end{equation}

A second round of spatial clustering was then performed to match source detections across snapshots and create a single set of source candidates with detection rates. In choosing a neighborhood radius, we found that using the PSF HWHM was maximized the number of sources detected in all snapshots, while a radius of one-quarter the beam width maximized the number of sources detected uniquely in all snapshots (i.e. not blended with another source in the same snapshot). 

To understand this, consider the case of a radio galaxy with two lobes separated by an angular distance close to the PSF width. Due to variations in the beam response and positional uncertainty between snapshots, the lobes may be clustered into a single sources in some snapshots. On cross-association, a larger clustering radius will merge these into a single source detected in all snapshots but with multiple detections in those where the lobes were individually resolved. A smaller radius will find three sources, each detected in only a fraction of the snapshots, and the total flux density will be double counted. 

We therefore use the PSF HWHM to maximize the number of sources detected in all snapshots for reliability determination. Although this means that some close pairs will be blended, we later use the quarter-width radius to split those indicated as having multiple components through cross-matching. (\S\ref{subsec_visinspec}).

A total of 9490 clusters, or spatially isolated source candidates, were identified with detections in at least two snapshots. A roughly equivalent number were detected in only a single snapshot and discarded. For each cluster, if multiple detections were found within any snapshot, the flux densities were summed and the position centroided into a single equivalent detection for that snapshot. 

A $3\sigma$ clip was applied to the flux density distribution of all detections in a cluster, where $\sigma$ is the standard deviation of the distribution, to exclude spatially coincident noise or side lobe sources. The mean position and amplitude gain of the primary beam (hereafter referred to simply as beam response or beam power) were then calculated. 

Since sources drift through the beam over time, and the rms noise scales inversely with the beam response, we weight the mean and standard deviation of flux density by the inverse of the estimated variance in the residual image at the source position. A local estimate of the pixel rms in a region around the source position is difficult given the low resolution of the images (1 pixel = 2.24' at beam center). We instead use the estimated thermal component of the variance by fitting pixel rms (Jy/beam) as a function of beam response for each snapshot and using the rms value predicted by source's position in the beam.

We use both the standard deviation $\sigma_{S}$ and standard deviation of the mean $\sigma_{\overline{S}}$ as measures of uncertainty because $\sigma_{S}$ is poorly constrained for small $N_{\rm det}$ while $\sigma_{\overline{S}}$ is based on an indirect estimate of the thermal rms in each snapshot. These differing limitations give complementary measures of uncertainty that, along with other observables, can be used to learn how true sources, noise fluctuations, and side lobes sources behave. Rather than apply a cut based on a strict SNR detection threshold, we classify each source in terms of it's overall reliability to inform our selection. 

\section{Reliability Classification}
\label{sec:Classification}

In determining the reliability of a source, the signal-to-noise ratio, or SNR, is a valuable statistic. For example, a $5\sigma$ confidence level is a common choice for robust statistical significance of a detection, however this operates on the assumption of a Gaussian noise distribution that can be sufficiently measured or predicted. Non-thermal sources of noise can lead to false confidence in spurious detections. Particularly problematic in radio images are the side lobes of bright sources that result from imperfect calibration and deconvolution. These side lobes can be significantly brighter than the background rms noise. They may also be spatially coincident between consecutive snapshots due to the short integrations (small $uv$ rotation) and low resolution. This false confidence and seemingly non-spurious behavior make it challenging to identify contamination in an automated way. 

We expect a non-negligible number of false sources contaminating the sample and consider a $5\sigma$ selection to be insufficient on its own to exclude the brightest and most problematic contaminants. Cross-matching to overlapping surveys can help to identify contamination, but this is limited by the reliability of the comparison survey and the incidence of false matches. Requiring all sources to be previously detected in another survey can also result in a loss of completeness.    

Here we summarize a classification scheme developed to assess source ``trueness" and measurement reliability in a self-consistent way. The resulting classifications are used in conjunction with cross-matching to inform the final catalogue selection. Details of the classification steps and intermediate results can be found in Appendix~\ref{apx:machine-learning} for the interested reader.

Machine learning based classification algorithms require a set of input features that describe the population. Feature engineering is the process of scaling, combining, or otherwise manipulating fundamental parameters to increase the predictive potential of the model. The process of feature engineering and selection is somewhat of an art, driven by domain insight to the question at hand as much as the data available. 

Features input to complex models are typically developed and honed through many iterations of trial and error. We ultimately define 9 features based on observable measurements that result in a well-modeled distribution:

\begin{enumerate}
\item {\bf Log Flux Density} The log of the weighted mean flux density of all source detections, $\log_{10}(S/\rm Jy)$. 
\item {\bf Log Signal to Noise} The log of the ratio of the mean flux density to the standard deviation, $\log_{10}(S/\sigma_{S})$. 
\item {\bf Log Signal to Noise of the mean} The log of the ratio of the mean flux density to the standard deviation of the mean, $\log_{10}(S/\sigma_{\overline{S}})$.
\item {\bf Number of Detections} The number of snapshots in which a source was detected, $N_{\rm det}$.
\item {\bf Expected Number of Detections} The estimated cumulative probability that the mean source flux density lies above the deconvolution limit in each snapshot, $N_{\rm exp}$.
\item {\bf Reliability Metric} We define a normalized reliability metric that takes a value between 0 and 1, as $r_{\rm det} = N_{\rm det}/\sqrt{71\,N_{\rm exp}}$.This is designed to down-weight the relative reliability of sources that drift out of the field  (i.e. 71 of 71 expected is more reliable than 2 of 2 expected).

\item {\bf Local Density.} The number density of sources within a $1^{\circ}$ radius of the source candidate $\rho_N$ $(\pi$ deg$^2)^{-1}$. 
\item {\bf Distance to Brightest Neighbour.} Distance to the brightest source within a $1^{\circ}$ radius of the source candidate, $d_{\rm bright}$ (\rm deg). 
\item {\bf Flux Density Ratio to Brightest Neighbour.} The flux density ratio between the source candidate and the brightest neighbour within a $1^{\circ}$ radius, $S/S_{\rm bright}$.
\end{enumerate}

The final three features offer additional information that help to differentiate likely side lobe sources which typically occupy regions of high number density in close proximity to a much brighter source. Principal component analysis was then used to reduce the parameter space from nine dimensions to three. Three components explain 83\% of the total feature variance and allow for simple visualization and model fitting. The 17\% information loss is later regained. 

A model consisting of ten independent three-dimension Gaussian components was fit to the data in the reduced parameter space. Ten components were found to model the distribution well without over-fitting. Each source candidate was labeled by the Gaussian component it was most probably associated with. The Gaussian labels were then used to train a more robust ensemble classifier. This allowed the classification boundaries to adjust according to the true feature distribution rather than the forced Gaussian approximation. 

The classification labels were ordered according to the median $r_{\rm det}$. Lower numbered classes therefore tend to be more reliable ($R0-R2$) while mid-range classes tend to be fainter or further from field-center ($R3-R6$). The majority of false positives appear to be captured in classes $R8$ and $R9$ along with the faintest sources that are detected too few times to be considered reliable. A more detailed interpretation is given in Appendix~\ref{apx:machine-learning}. 

\begin{table}
\centering
	\begin{tabular}{l l l l l l l l}
	\hline
      $R_{\rm class}$ & $N$      & $r_{\rm det}$      & $N_{\rm det}$ & $N_{\rm exp}$ & $S(\rm Jy)$ & $S/\sigma_S$ & $S/\sigma_{\overline{S}}$ \\
    \hline
    \hline
    0	&	2353	&	0.99	&	70	&	71	&	0.35	&	10.74	&	  57.71 \\
    1	&	  416	&	0.99	&	70	&	71	&	0.93	&	19.45	&	131.68 \\
    2	&	  172	&	0.97	&	69	&	71	&	0.27	&	  7.86	&	  39.11 \\
    3	&	  794	&	0.80	&	56	&	71	&	0.20	&	  6.68	&	  26.44 \\
    4	&	1162	&	0.60	&	38	&	60	&	0.20	&	  6.54	&	  18.35 \\
    5	&	  204	&	0.38	&	21	&	38	&	0.31	&	  6.61	&	  11.21 \\
    6	&	1435	&	0.32	&	18	&	50	&	0.12	&	  6.29	&	  10.83 \\
    7	&	   21   &	0.11	&	  3	&	  6	&	0.21	&	  4.49	&	    3.80 \\
    8	&	2630	&	0.09	&	  4	&	33	&	0.14	&	  6.88	&	    4.63 \\
    9	&	  303	&	0.05	&	  2	&	32	&	0.11	&	36.68	&	    3.29 \\
    \hline
    \end{tabular}
\caption{The resulting ten relative reliability classifications. Columns from left to right are the reliability class, number of sources in that class, and the median values of detection rate, number of detections, expected number of detections, flux density, SNR, and SNR of the mean. Lower classes are more reliable while higher classes tend to be fainter or far from field center. Classes $R8$ and $R9$ appear to capture sporadic and side lobe contaminants, but also faint sources near the detection threshold. The reliability classification is used to inform the final catalogue selection.}
\label{tab:classes}
\end{table}

We make an initial cut on the set of 9490 source candidates to include only those detected with high confidence ($S > 5\sigma_S$ and $S>5\sigma_{\overline{S}}$) or reliability ($R_{\rm class} < 7$; all of which meet the $S>5\sigma_{\overline{S}}$ criteria). This reduces the sample to \Cand source candidates. Looser criteria may be explored in the future in terms of the effectiveness of foreground power subtraction. 

\section{Radio Cross Matching}
\label{sec_XMatch}

To check the reliability of the sources, a cross match was performed with the following catalogues: the 74~MHz Very Large Array Low Frequency Sky Survey redux~\citep[VLSSr,][]{Lane2014}; the 408~MHz Molonglo Reference Catalogue~\citep[MRC,][]{Large1991}; the 843~MHz Sydney University Molonglo Sky Survey~\citep[SUMSS,][]{Mauch2003}; and the 1.4~GHz NRAO VLA Sky Survey~\citep[NVSS,][]{Condon1998}. A summary of these catalogues is given in Table~\ref{other_cats}. 

The EoR0 field is centred at RA~0hr and Dec~$-27^{\circ}$ and the KGS sources reach a depth of $\sim$60~mJy at 182~MHz in the centre of the field. This coverage does not match well to any one of the comparison surveys. The VLSSr and NVSS catalogues cover the northern half of our field ($\gtrsim-30^{\circ}$) while SUMSS covers the southern half ($\lesssim-30^{\circ}$). The MRC catalogue covers the full sky area but is only complete to 1~Jy at 408~MHz or $\sim2$~Jy at 182~MHz.

\begin{table*}
\large
\centering
	\begin{tabular}{l c r c c l }
	\hline
	Survey & $\nu$~(MHz) & $N_{\rm sources}$ & Dec & PSF~FWHM & $S_{\rm complete}$ \\
	\hline \hline
	VLSSr & 74 & 92,696 & $\delta>-30^{\circ}$ & 75" &   $\sim$1~Jy/beam   \\
	MRC & 408 & 12,141 & $\delta_{1950}>-85^{\circ}$  & $\sim$3' &$\sim$1~Jy/beam           \\
	SUMSS & 843 & 211,050 & $\delta<-30^{\circ}$ & 45" &  18~mJy/beam   \\
	NVSS & 1400 & 1,773,484 & $\delta>-40^{\circ}$ & 45" & 2.5~mJy/beam   \\
	\hline
	\end{tabular}
\caption{A summary of the catalogues matched with PUMA to the KGS outputs including the reported or estimated completeness limit.}
\label{other_cats}
\end{table*}

Cross matching was performed using the Positional Update and Matching Algorithm (Line et al., {\it in prep.}). PUMA uses a combination of positional and spectral information to statistically test whether sources from multiple surveys in close proximity to one-another are true matches. We briefly describe the matching steps here, but for a full explanation of each step please refer to Appendix~\ref{app_bayesmatch} and the code documentation\footnote{PUMA code is open source. The code and documentation can be found here - \url{https://github.com/JLBLine/PUMA}}. Note this process is automated through \S\ref{subsec_visinspec} where we manually investigate outliers.

\subsection{PUMA}

Initially, PUMA attempts to match sources purely by position. In the first stage, a positional cross match is performed using STILTS \citep{Taylor2006}. All sources within a radius of 2.3' from the base KGS source are selected. The choice of this radius is somewhat arbitrary (equal to the PSF FWHM) and intentionally liberal. For each cross match result, the probability $P$ that all catalogues are describing the same source is calculated following \citet{Budavari2008}, taking account of the positional errors. At this point, if a KGS source is matched to only one source from any catalogue and $P>0.95$ it is accepted without further investigation. This is labelled as an isolated match.

The quoted uncertainties on position, particularly for blended and complex sources, are not necessarily accurate or directly comparable between surveys. If $0.8<P<0.95$, or all cross-matched sources lie within the resolution of the MWA (i.e. half of the original STILTS search radius), we investigate the spectral energy distribution (SED) by fitting a power law spectral model, $S\propto\nu^{\alpha}$, using weighted least squares. If the fit is good, the source is accepted as an isolated match. Note that if a match is only found in one other catalogue, this fit always passes as there can be no residuals. Steps are taken later in \S\ref{subsec_visinspec} to account for any issues that could arise here.

Multiple matches to a single catalogue may occur due to confusion at the lower resolution or coincidental false source contamination. In the case where multiple sources from a comparison catalogue are matched to a single KGS source, PUMA first attempts to remove any false matches by fitting the spectral model to each possible combination of sources. If one match combination has smaller residuals than all others, as well as having $P>0.95$, it is accepted as the dominant match. 

If no dominant match is found, it is possible that a source is resolved into multiple components in the higher resolution catalogues. This is the common case for radio galaxies and star-forming galaxies with structure that is unresolved by the MWA. To test this, the spectral model is fit to the cumulative flux density of the matches at each frequency. If the fit is good, the source is accepted as a multiple match. 

Table~\ref{tab_PUMAmatch} details the number of each type of PUMA match decision. The flux density and uncertainty of matched sources are included in the KGS catalogue along with the measured broad band spectral index (SI). The spectral index distribution is discussed in \S\ref{sec_SIdist}. The position of matches to the NVSS or SUMSS catalogues are also included for reference (a flux density weighted mean position is reported for multiple matches) and used to assess astrometric precision in \S\ref{sec:astrometry}.

\begin{table}
\centering
	\large
	\begin{tabular}{l r l r l r }
	\hline
		Match result & Count&(\%) & Modified & (\%) \\
	\hline
	\hline
	isolated &  6119 & (82.8)  & 75 & (1.2) \\ 
	dominant &   310 & (4.2)  & 11 & (3.5) \\
	multiple &   940 & (12.7)  & 153 & (16.3) \\
	none     &    25 & (0.34)  &  &    \\
	Total    &  7394 & (100)   & 239 & (3.2)  \\
	\hline
	\end{tabular}
	\caption{The total number and percent of catalogue sources in each PUMA decision category, and the number and percent of each category for which the match was flagged and manually modified in \S\ref{subsec_visinspec}. The majority of sources (87\%) are matched to a single counterpart in other catalogues (isolated or dominant) with a 98.6\% automatic success rate. When confusion occurred (mulitple matches), PUMA chose the proper match combination in 84\% of cases. Most modifications were required for complex and extended sources.}
	\label{tab_PUMAmatch}
\end{table}

\subsection{Visual Inspection}
\label{subsec_visinspec}

To check the robustness of the PUMA decisions we visually inspected any potentially suspect matches or atypical sources. These include \EFlagged sources automatically flagged by PUMA when a confident match decision could not be made and \Rejects sources with a STILTS match that was automatically rejected by PUMA. For the sake of reliability, we also double check PUMA accepted matches that we manually flagged as outliers. These include all \MFlagged sources accepted by PUMA as a multiple match and \Outliers defined as having either
\begin{inparaenum}[\itshape a\upshape)]
\item ~spectral index in the 1\% tails of the distribution, $\alpha<-1.46$ or $\alpha>-0.17$;
\item ~positional offset from NVSS or SUMSS $>3\sqrt{\sigma_{\rm RA}^2+\sigma_{\rm Dec}^2}$
\end{inparaenum}.

\begin{figure*}
\centering
\subcaptionbox{Example of visualized position (top) and SED (bottom) information for a complicated match before (left) and after (right) manual modification. Ellipses indicate the reported major/minor axis and position angle.\label{fig:example-multiple:a}}
	{
	\includegraphics[width=\textwidth]{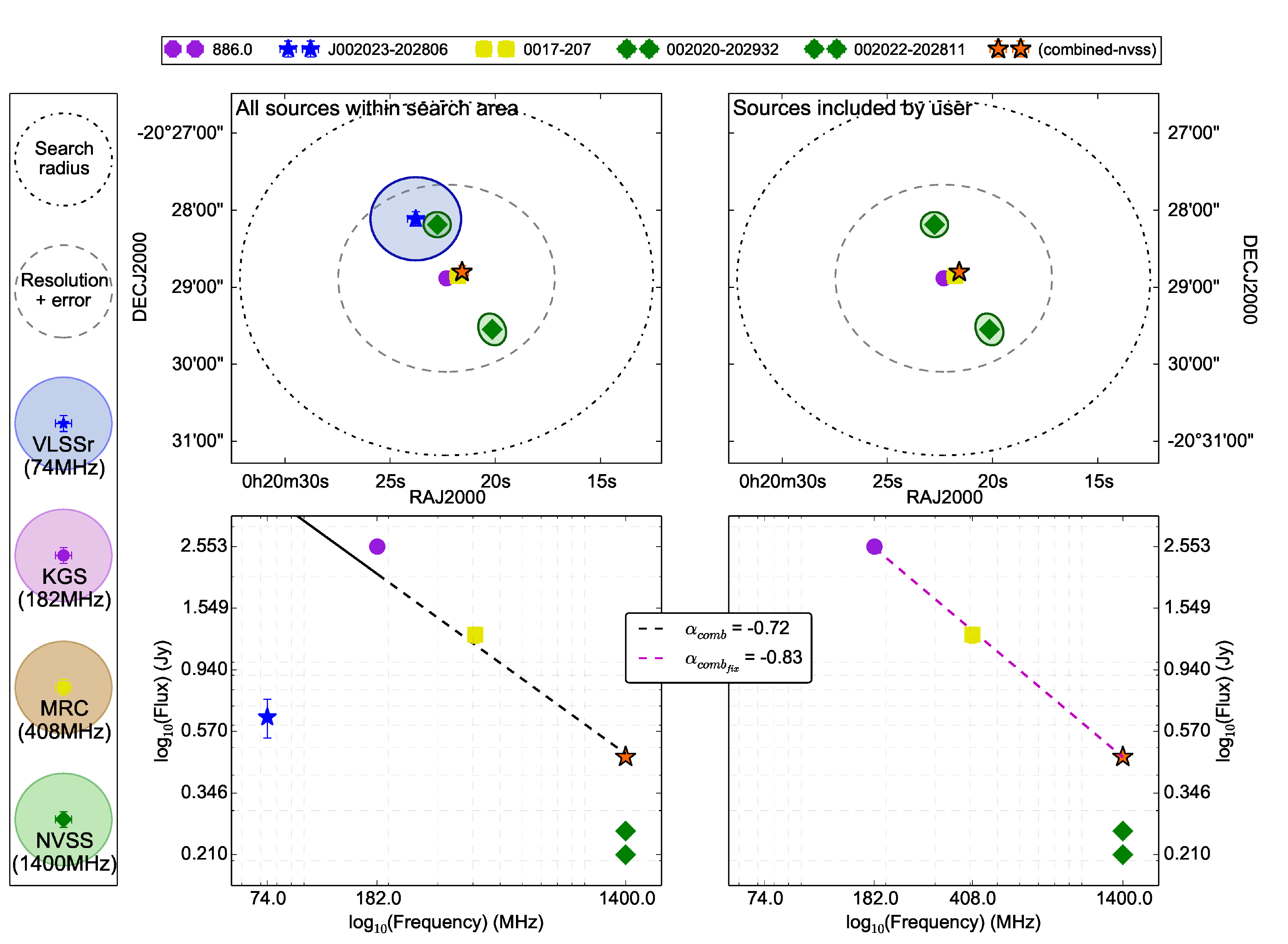}
	
	}
\subcaptionbox{Example postage stamp images inspected for complicated matches. The white dash/dotted circles correspond to the search radius and resolution+error as indicated in \label{fig:example-multiple:b}}
	{
	\includegraphics[width=\textwidth]{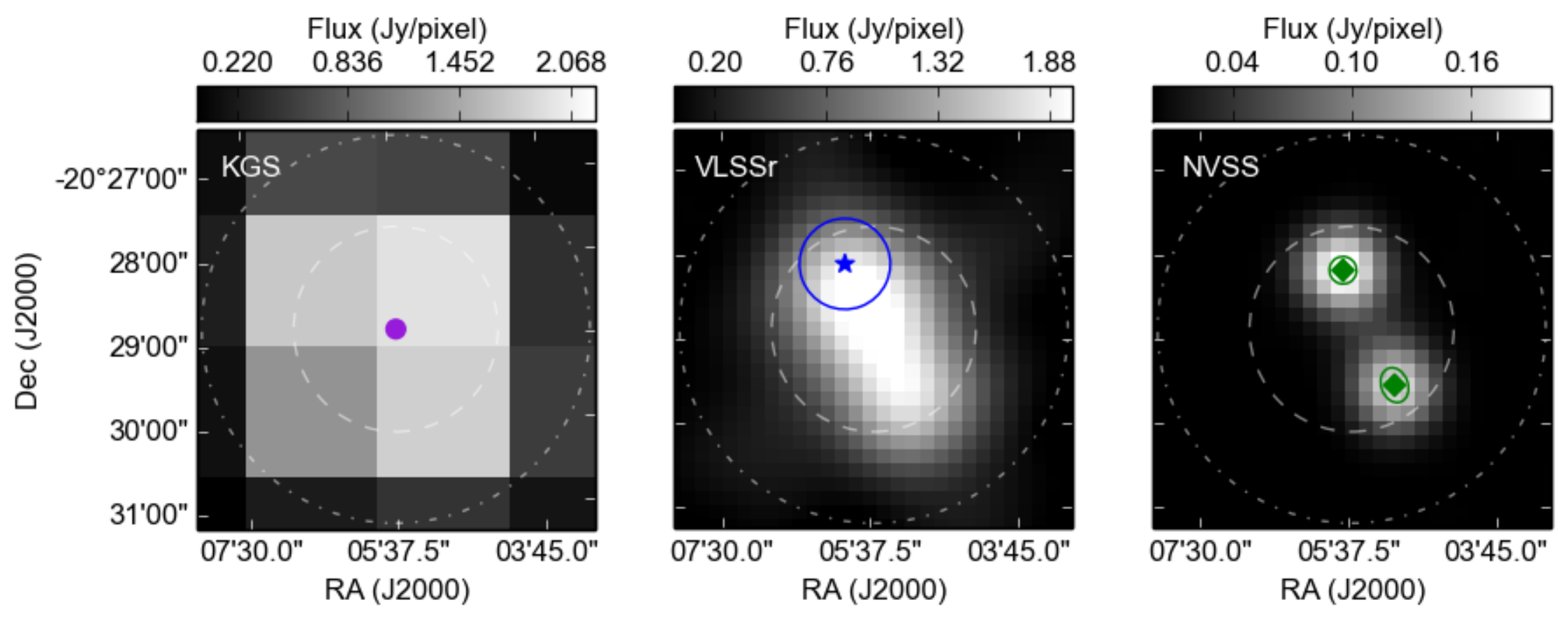}
	}
\caption{To investigate flagged matches, catalogue information on source position, shape, and flux density were plotted (\subref{fig:example-multiple:a}) and, for complicated sources, postage stamp images were obtained and compared (\subref{fig:example-multiple:b}). In this example of a multiple match, two sub-resolution sources appear present in both the VLSSr and NVSS images, but one is missing from the VLSSr catalogue. The centroid position of the NVSS sources agrees well with the KGS and MRC positions. The flux density of the NVSS sources are combined and the VLSSr source is excluded.}
\label{fig:example-multiple}
\end{figure*}

In addition to visualizing the catalogue information and PUMA results, we looked at postage stamps of VLSSr, SUMMS, NVSS, and MWA images. Where appropriate, the PUMA decision or match information was modified. This could include removing matched sources that appeared spurious or ignoring a catalogue in a multiple match that appeared to be missing a source visible in its image. Figure~\ref{fig:example-multiple} shows an example of a source where both the catalogue information and images were inspected and used to modify the catalogue match. 

By selecting outliers, we were able to discard \Deleted false isolated matches. These were typically sources with poor reliability, visually identified as bright side lobes, and coincidentally matched to either a true source or to apparent side lobe contamination in a comparison survey image. We erred on the side of reliability in these decisions.

Approximately 10\% of multiple matches were able to be deconstructed into two or more components. In \S\ref{sec:SourceFinding}, a radius of one-quarter the beam width maximized the number of sources detected uniquely in all snapshots. If multiple source candidates were found by using the tighter clustering radius, these were similarly cross-matched and substituted in manually if the overall result improved. A total of \vTenOut sources were replaced with \vFiveIn counterparts. The reliability class of replacement sources was independently predicted using the original classifier. 

By specifically targeting outliers, we not only discarded contamination but noted many interesting sources. For example, the unique morphology of NGC~7793 resulted in its identification as a near face-on spiral galaxy in the Sculptor group. A subsequent search on the positions of other group members revealed detections of NGC~253 and NGC~55. NGC~253, the Sculptor galaxy, is the sixth brightest source in the catalogue and its extended morphology will require a more complex treatment in the foreground model. The low frequency emission from the Sculptor group galaxies will be further investigated by Kapinska et al. ({\it in prep.}). Other noteworthy examples of morphological and spectral outliers are presented in \S\ref{sec:catalogue}.

\subsection{False Sources and New Detections}
\label{sec:unmatched}

There are 167 source candidates not matched to another catalogue within the initial STILTS search radius, and another 12 STILTS matches that were rejected by PUMA. The majority of these are classified $R7-R9$ and visual inspection supports their exclusion as noise or side lobe contaminants. Twenty-five are included in the final catalogue. Of these, 20 are reliably classified $R0-R6$. Five appear to be extended emission blended with a brighter source, but are reliably detected independently of that source. 

We chose to include five faint sources classified $R8$ after careful consideration. These are interesting and illustrative. Detected in few snapshots, the mean and standard deviation are poorly constrained. The flux density is also likely to be over-estimated due to Eddington bias (see \S\ref{sec:eddington-bias}) near the detection threshold. In combination, these effects seem to have resulted in artificially high SNR measures, allowing their inclusion in the source candidate sample. We chose to keep these simply because they appear to be real in the images and were deemed deserving of follow up. Many similar sources did not make the initial candidate selection.  In terms of a foreground model, the level of contamination we risk through their inclusion is negligible.

The new source detections are explored further in \S\ref{sec:newsources} where we consider potential associations in other wavebands.

\section{The KGS EoR0 Catalogue}
\label{sec:catalogue}

Since nearly all sources have accurate matches to other catalogues, we can explore the completeness, astrometric precision, spectral index distribution, and flux scale reliability of the catalogue.

\subsection{Flux Scale}

\begin{figure}
	\includegraphics[width=\columnwidth]{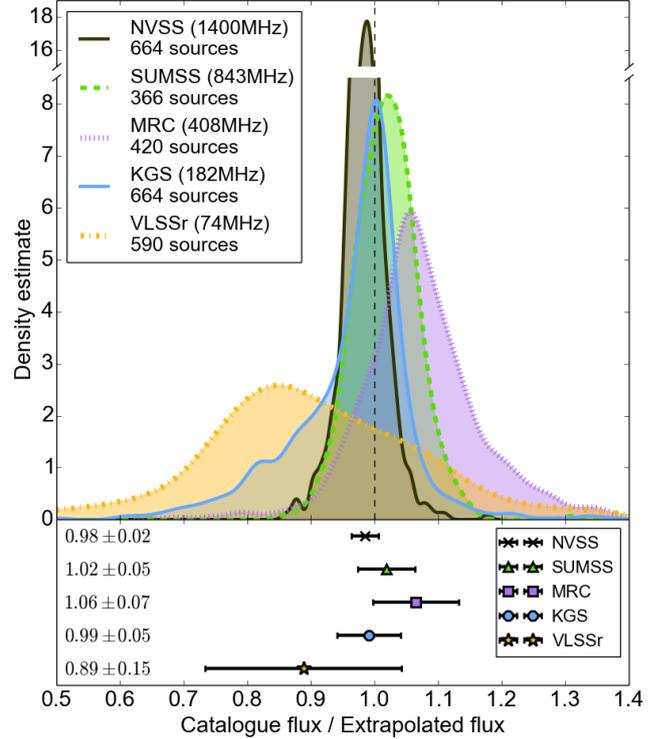}
	\caption{The ratio between observed flux density and extrapolated flux density from a fit to the SED is shown for every time a catalogue appeared in a match with at least two other catalogs for isolated sources. The upper panel shows a univariate kernal density estimation of each distribution (note broken y~axis due to the sharp peak in the NVSS ratio distribution),
while the lower panel shows the median and median absolute deviation of each distribution. The KGS spectral index agrees very well with no indication of flux bias on average.}
	\label{fig_fluxratio}
\end{figure}

To investigate how each matched catalogue contributed to the fitted spectral index, the flux density at each frequency was extrapolated using the fitted parameters. To be sure of a true matched SED, only isolated sources were used. A ratio between the reported catalogue flux density and extrapolated flux density was then calculated, as shown in Figure~\ref{fig_fluxratio}. 

On average there is no significant bias in the distributions, however it is interesting to note the width and skewness in these distributions. The VLSSr and KGS skew somewhat low and the MRC skews somewhat high. For sources detected in more than two catalogs, the spectral index fit used the quoted flux densities and uncertainties of the comparison catalogs. NVSS and SUMSS have lower uncertainties than the lower frequency catalogs and VLSSr has the largest. This is evident in the spread of the flux density ratio distributions; NVSS has a tight distribution centered at one, whereas VLSSr has a broader distribution. Clearly NVSS is being fit preferentially over the other catalogs. Although the median values are all consistent with unity, systematic effects  are clearly present. 

It is difficult to distinguish catalogue flux biases from intrinsic spectral curvature effects, but it appears that only a sub-population of sources are affected rather than there being an overall shift in the distribution. A systematic under or over estimation due to the original flux scaling or calibration could account for this. No attempt was made to match flux scales across catalogues since all catalogues except for the VLSSr are tied to or derived from the Baars scale \citep{Baars1977} based on measurements of Cassiopeia A. The VLSSr is tied to the RCB scale \cite{Roger1973}, which is considered to be more accurate at low frequencies. Due to the low weighting of the VLSSr data points and the fact that all but three VLSSr matches are also matched to the NVSS, this scale difference negligibily impacts the SED fits and overall SI distribution. Nonetheless, in the following sections we divide all VLSSr flux densities by a factor of 1.1 to place them on the Baars scale and increase the flux density uncertainty by 5\% following \citet{Lane2014}. 

Alternatively, if a significant portion of SEDs display some degree of intrinsic positive curvature, the ratio distributions would be impacted predictably. The lowest frequency flux densities (VLSSr and KGS) would typically be overestimated by a power-law fit, while the central frequencies (MRC and SUMSS) would be underestimated. This is consistent with the flux density ratio distributions observed, however a more careful treatment is needed to conclude intrinsic curvature over systematic effects. We do not find evidence for flux scale bias relative to the comparison catalogs. 

\subsection{Eddington Bias}
\label{sec:eddington-bias}

Source flux densities near the detection threshold will be systematically high due to Eddington bias. At low apparent flux densities, statistical fluctuations below the sensitivity limit go undetected resulting in an overestimation of the true mean \citep{Eddington1913}. To estimate a correction for this effect, we numerically solve for the true flux density that would most likely result in the observed average over all detections.

The probability density function of a source detection is assumed to be Gaussian, centered on the true flux density and with standard deviation equal to the background rms. The expectation value of the measurement is then the mean of the probability density function above the detection threshold. For each observation, the detection threshold is estimated as a function of position in the beam and the background rms is estimated from the residual images within a 20x20 pixel box around the source position.

The reported source flux density is the weighted mean of the snapshot detections. Using this as the initial guess for the true flux density of the source, we find the weighted mean of the expectation values for all detections as described above, and numerically solve for the true flux density that minimizes the difference between this expected mean and that observed.

We find the bias affects sources with apparent flux density ($S \cdot \rm beam$) below about 100~mJy, but few sources are changed by more than 10\%. 
To gauge the accuracy of the true flux density estimates, we looked at the 182-1400~MHz spectral index distribution for isolated sources before and after correction. Above $S \cdot \rm beam$ = 100~mJy (Figure~\ref{fig:eddington-bias}, right), there is no significant difference in the distributions before and after correction. Below this threshold, Eddington bias is evident in the shift of the spectral index distribution toward more negative values (Figure~\ref{fig:eddington-bias}, left). After correction, the median value agrees well with that of the unbiased distribution. 

Of the 2548 sources that would be corrected, the difference in flux density exceeds the standard error $\sigma_S$ for only 177 sources (50 exceed $3\sigma_S$). For this reason, Eddington bias is not a major concern. The overall median SI change is small, from $-0.850$ to $-0.843$. However, the bias is significant for individual sources within the catalog, particularly at low apparent flux density. It is important to note that the validity of the correction factor for any source is contingent on there being a large enough number of detections that the mean is sufficiently constrained. It is therefore the least reliable for the most affected sources. 

The catalog contains a column EB\_{}corr that may be multiplied by the flux density to approximate a correction for Eddingon Bias. The correction factor is 1 by default for a source if $(S \cdot \rm beam)>100$~mJy or if it is a multiple match (i.e. not a point source). We recommend adding the reported flux density uncertainty in quadrature with the absolute difference between the original and corrected flux density values. In the following sections we use the corrected flux density estimates.

\begin{figure*}
\label{fig:eddington-bias}
\centering
\includegraphics[width = \textwidth, trim={3cm 0 3cm 0},clip]{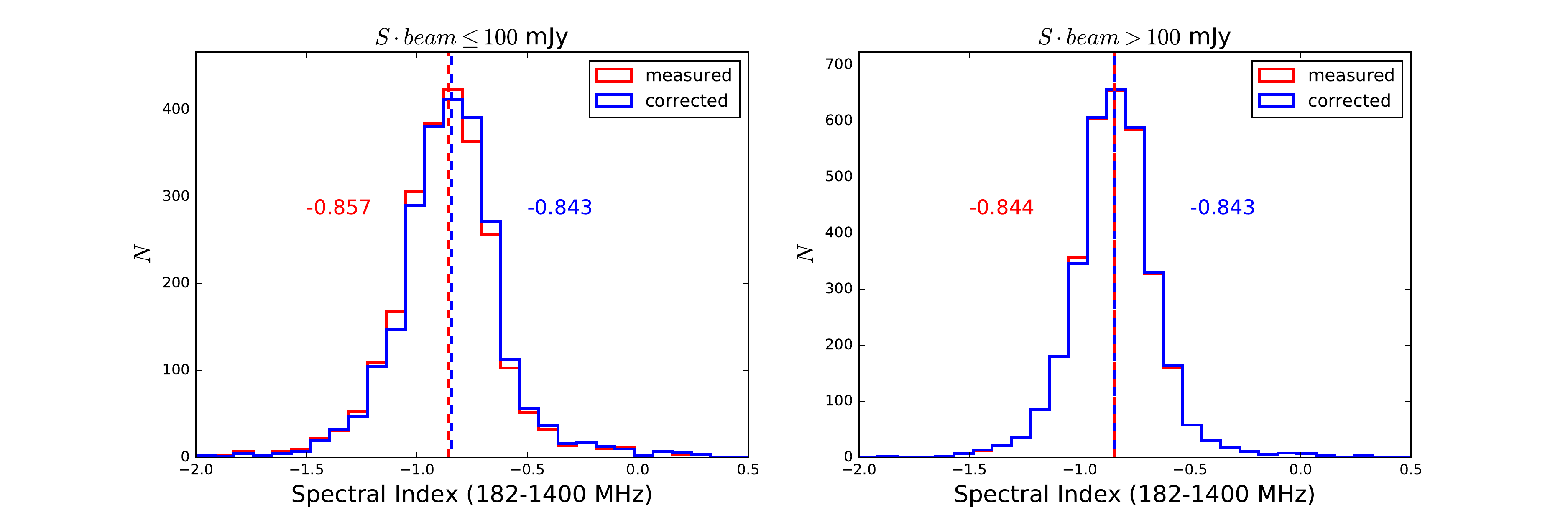}
\caption{The two-point spectral index distributions of isolated sources with matches to NVSS at 1400~MHz before (red) and after (blue) correction for Eddington bias. Sources with apparent flux density $S \cdot \rm beam<100$~mJy (left) tend to be the most affected. After correction, the median spectral index agrees well with sources above this threshold (right) for which the estimated correction is negligible.}
\label{fig:eddington-bias}
\end{figure*}

\subsection{Completeness}

\begin{figure*}
  \includegraphics[width=0.85\textwidth]{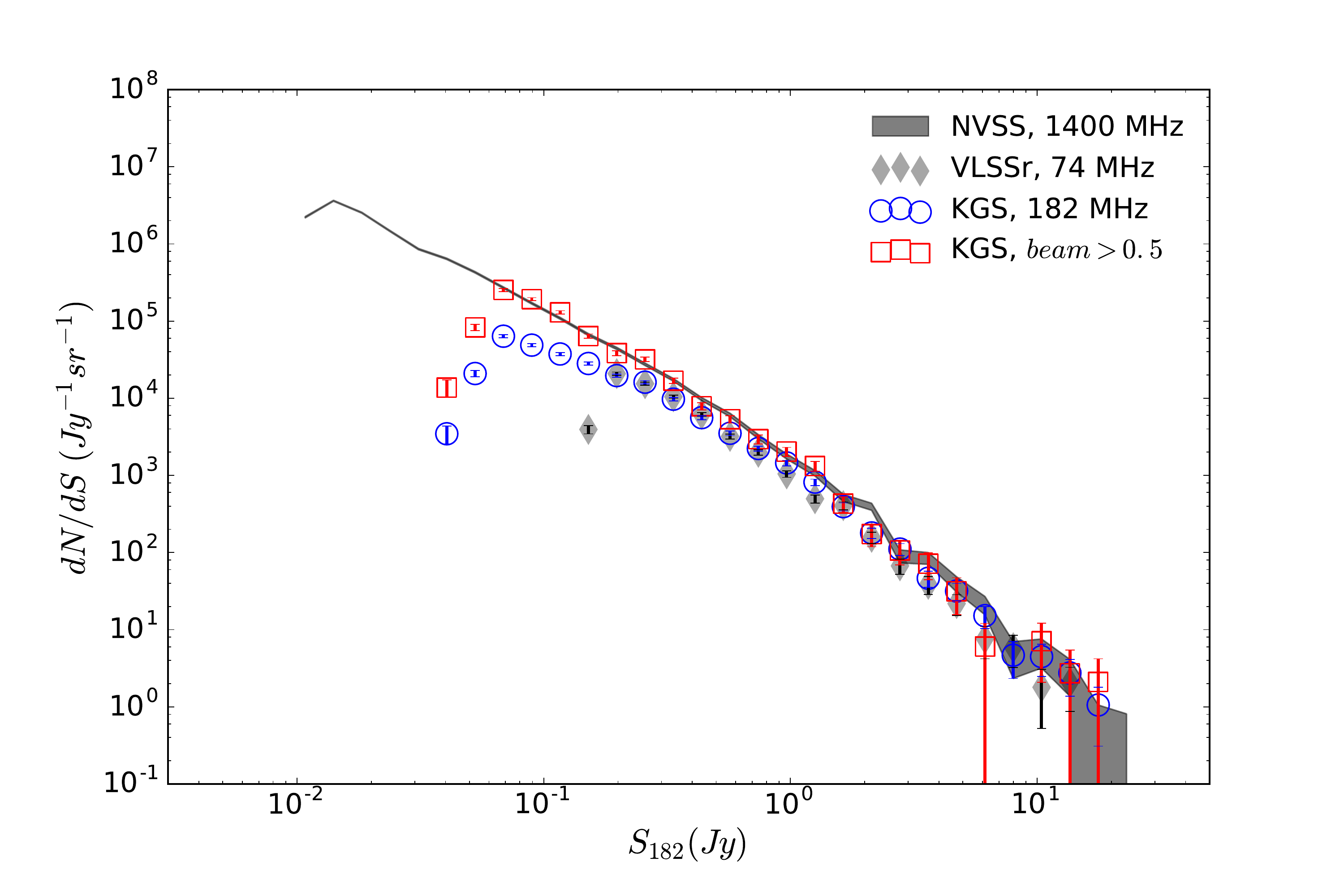}
  \caption{Differential source counts of the full catalogue (blue circles) compared to the NVSS (grey line) and VLSSr (grey diamonds). Counts are made within the overlapping footprint and the flux density values are projected to 182~MHz using the median two-point spectral index of all isolated matches. Completeness falls off as $1/\rm beam$ below ~1~Jy, however within half-beam (red squares) the catalogue is complete to approximately 80~mJy. The VLSSr shows comparable completeness to $S_{182}=200$~mJy below which KGS sources are more likely to go undetected in the at 74~MHz.}
\label{fig:completeness}
\end{figure*}

In order to assess completeness, we compare source counts to the predicted source counts of the NVSS and VLSSr surveys projected to 182~MHz (Figure~\ref{fig:completeness}). Counts are considered only within the overlapping survey areas and each catalogue flux density is projected to 182~MHz using the median two-point spectral indexes $\alpha_{74}^{182}=-0.68$ and $\alpha_{182}^{1400}=-0.85$ for all matched isolated sources. 

Because the sensitivity, and thus detection threshold, goes as $1/\rm beam$, the overall completeness falls off steadily below $\sim 1$~Jy compared to the NVSS. Within the half-power point, we find the catalogue is complete to approximately 80~mJy. Source counts appear to be comparable to the VLSSr above 200~mJy, below which KGS sources are likely to go undetected at 74~MHz within the overlapping footprint.  

\subsection{Spectral Index Distribution}
\label{sec_SIdist}

\begin{figure}
\center
\includegraphics[width=78mm]{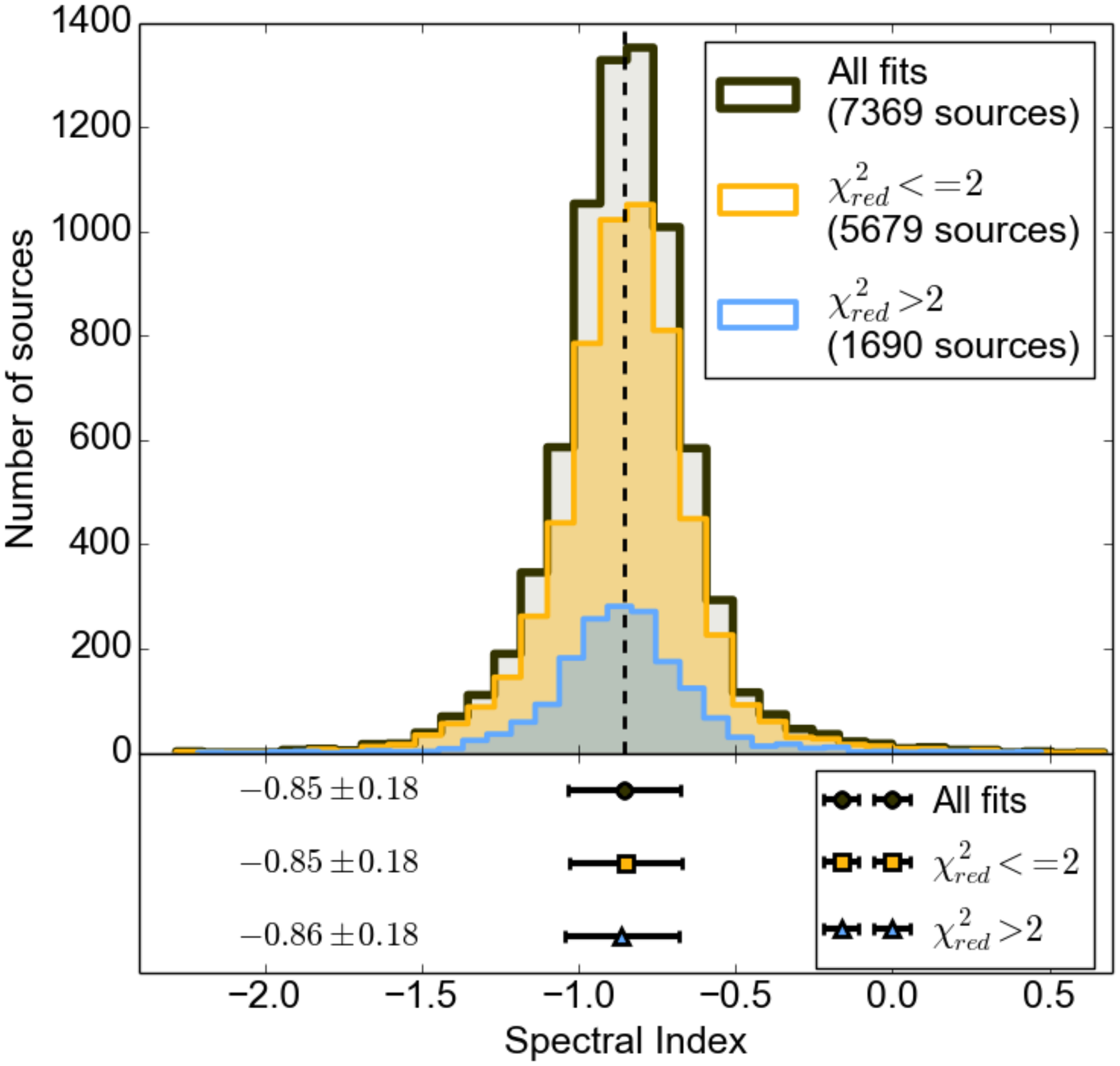}
\caption{The SI distribution derived by matching to VLSSr, MRC, SUMSS and NVSS. The full sample is shown in black, with good spectral fits ($\chi^2_{\rm red}<=2$) shown in gold and poor spectral fits ($\chi^2_{\rm red}>2$) shown in blue. The mean and standard deviation of SI distributions are shown in the lower panel.}
\label{fig_SIdist}
\end{figure}

The spectral index (SI) distribution found by the match performed in \S\ref{sec_XMatch} is shown in Figure~\ref{fig_SIdist}.  We find an overall median of $-$0.85 and no bias is seen by including or excluding sources with poor spectral fits $\chi^2_{\rm red}>2$ in this estimate. 

\begin{figure*}
\center
\includegraphics[width=\textwidth]{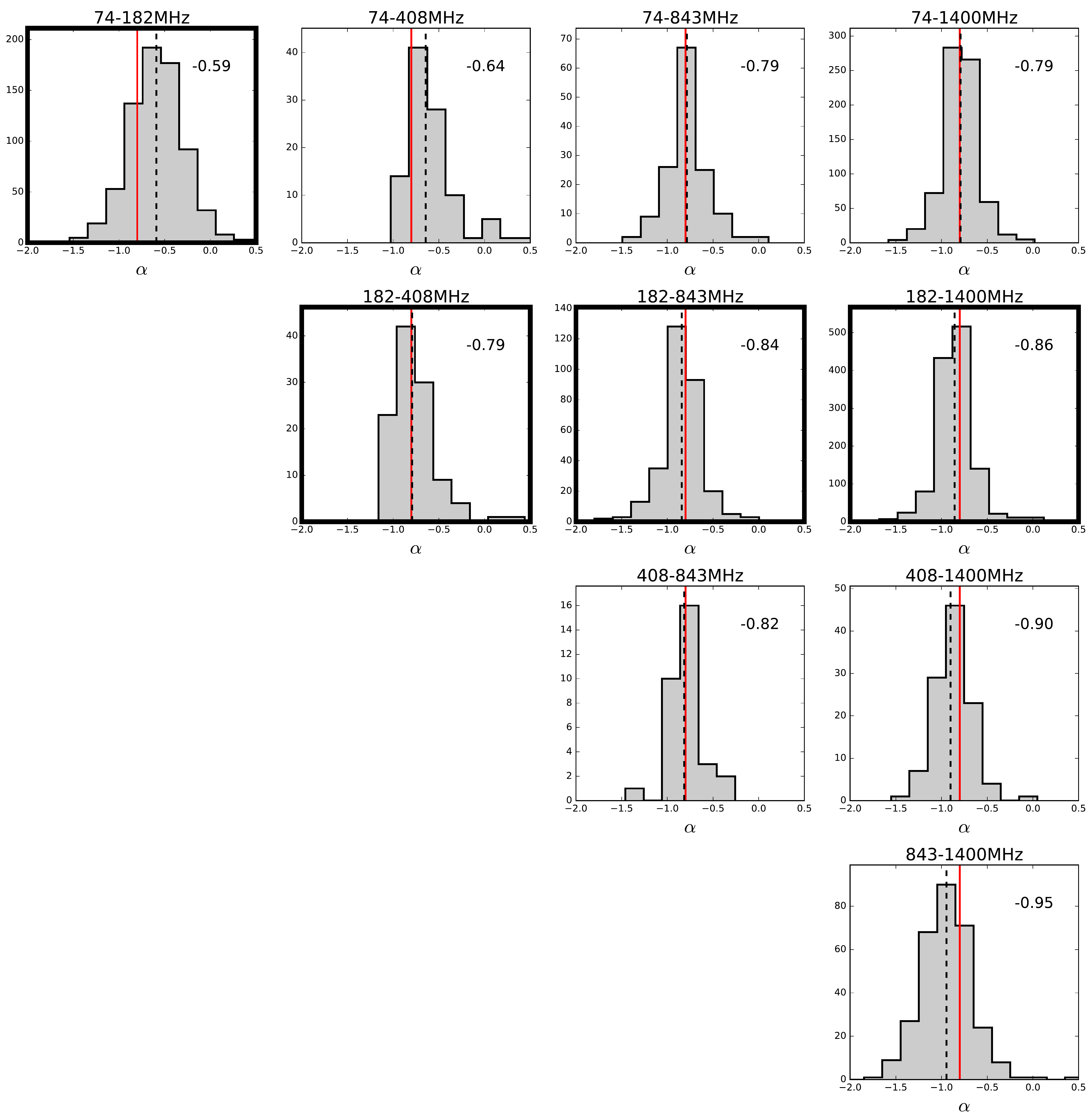}
\caption{The two-point SI distributions for all catalogue matches to isolated KGS sources with $S>200$~mJy and $\rm beam>0.5$. Bold axes indicate distributions using the KGS 182~MHz flux density. The median values are marked by dashed black lines and the red lines mark $-$0.8 for reference.  The median becomes increasingly negative toward higher frequencies.}
\label{fig:two-point-SI}
\end{figure*}

In Figure~\ref{fig:two-point-SI}, we show the SI distribution for all two-point SI measurements among matches to isolated KGS sources detected at an average beam power greater than 0.5 and flux density $S > 200$~mJy. The two-point median spectral index is seen to range considerably, from -0.59 to -0.95, with a trend toward steeper spectra at higher frequencies. The lowest frequency measurements $\alpha_{74}^{182}$ and $\alpha_{182}^{408}$ give an average of -0.70 at 182~MHz. 

By fitting a second order polynomial to the subset of 883 isolated sources detected in 3 or more catalogues, we predict a median 182~MHz spectral index of -0.71 with an interquartile range between -0.88 and -0.53 (the mean and standard deviation are $-0.71 \pm 0.32$). These results are consistent with \citet{Offringa2016}, who directly measure the sub-band (132--198~MHz) spectral index for a highly comparable set of sources in the center of the MWA EoR0 field. They find a median of -0.70 at 168~MHz  (mean of $-0.687 \pm 0.275$). For comparison, the Low Frequency Array (LOFAR) MSSS MVF survey \citep{Heald2015} finds median values of $\alpha_{30}^{158} = -0.66$ and $\alpha_{119}^{158} = -0.77$ (mean values of -0.60 and -0.70 respectively) among 628 sources with $S_{150}>200$~mJy.

\subsection{Astrometry}
\label{sec:astrometry}

Cross matching allows us to approximate the best astrometric position of a source based on higher frequency, higher resolution counterparts. Of the 7369 matched sources, all but three include a match to either the NVSS or SUMSS catalogues.

\begin{figure*}
\centering
	\subcaptionbox{\label{fig:posdist:a}}{\includegraphics[width=0.45\textwidth]{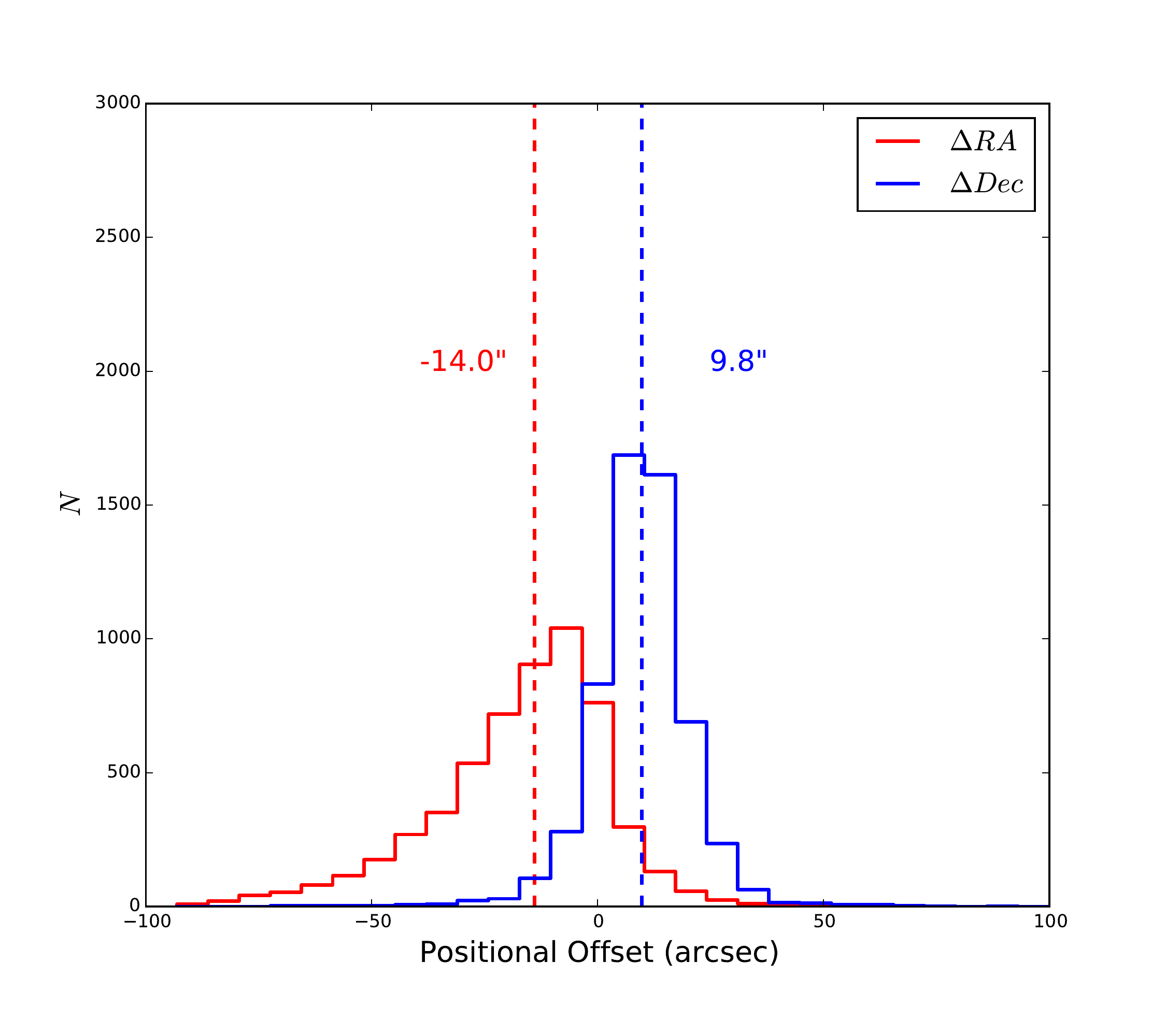}}
	\subcaptionbox{\label{fig:posdist:b}}{\includegraphics[width=0.45\textwidth]{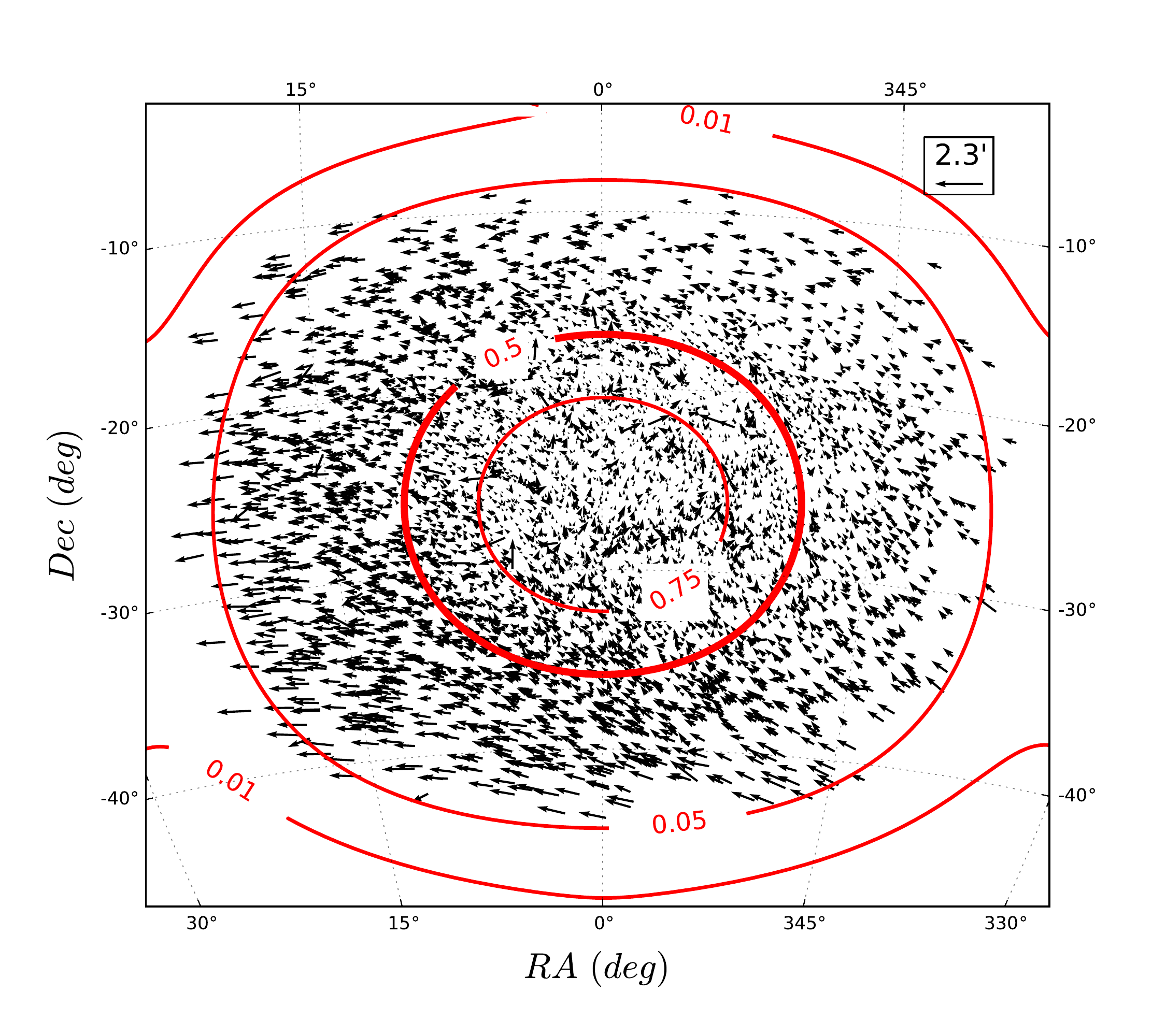}}
	\subcaptionbox{\label{fig:posdist:c}}{\includegraphics[width=0.45\textwidth]{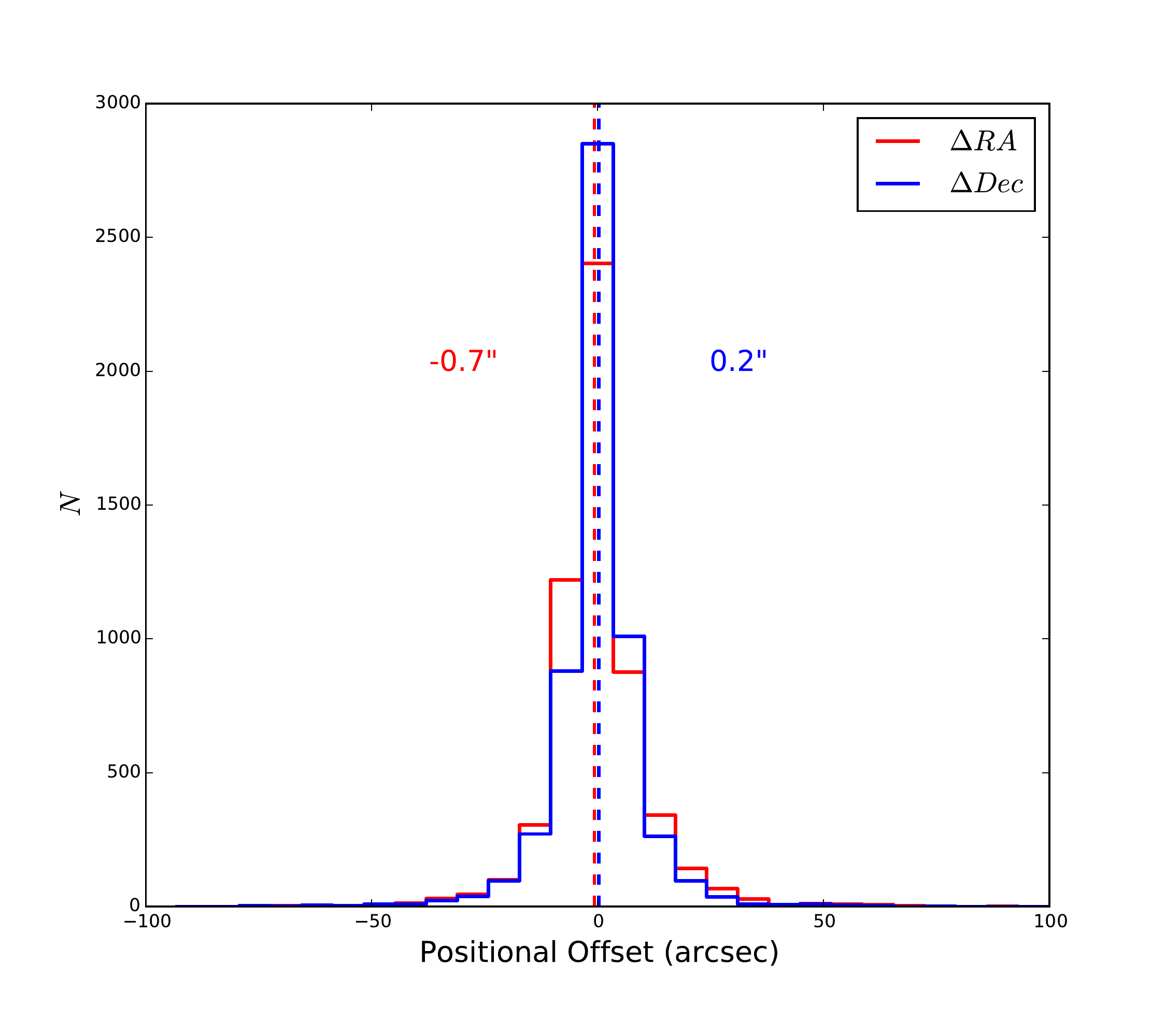}}
	\subcaptionbox{\label{fig:posdist:d}}{\includegraphics[width=0.45\textwidth]{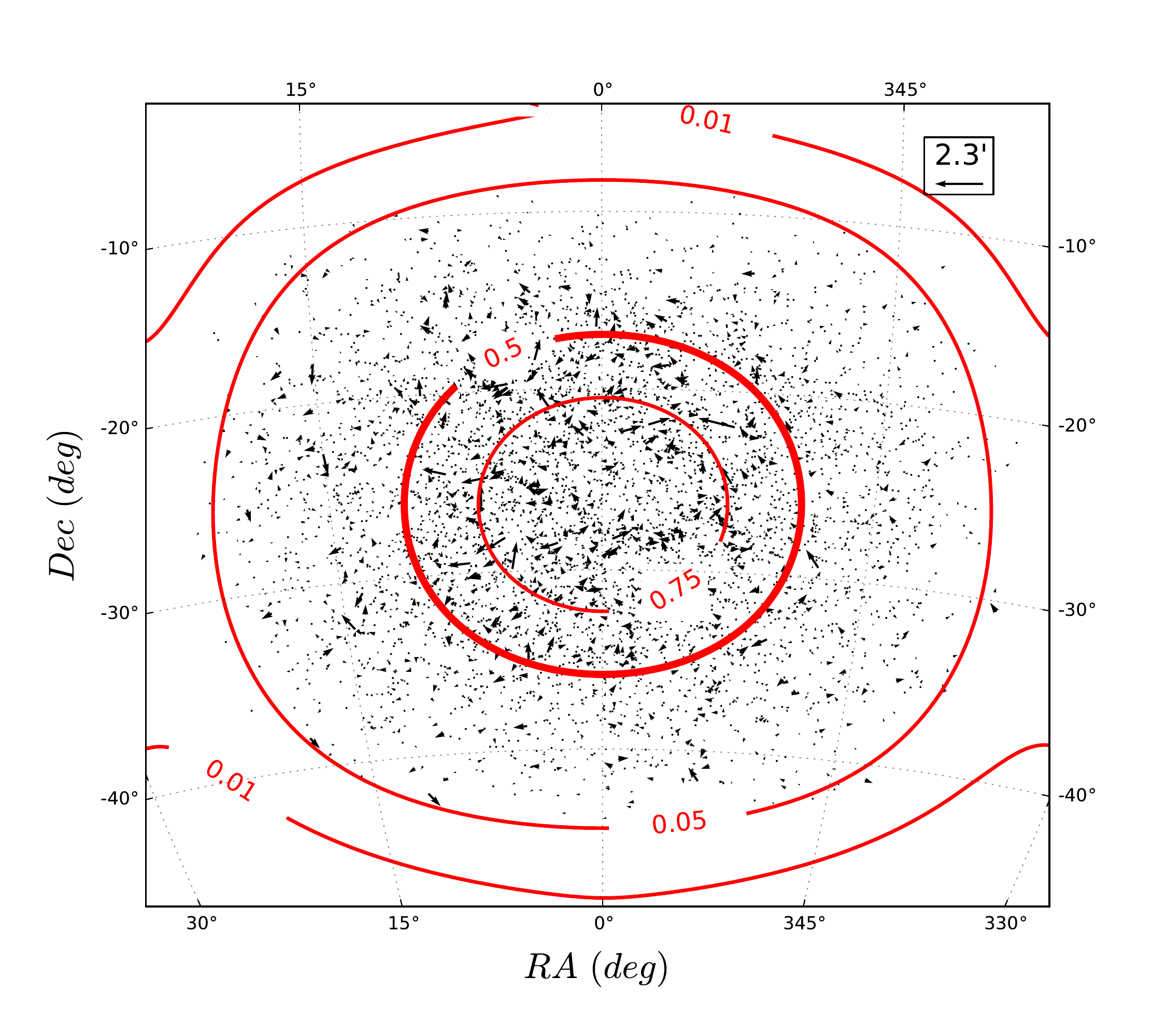}}
\caption{Left: The distribution of positional offsets compared to either NVSS or SUMSS counterparts for isolated sources before (\subref{fig:posdist:a}) and after (\subref{fig:posdist:c}) correcting for the north-eastward systematic bias. Right: Vector fields illustrating the offset magnitude and direction before (\subref{fig:posdist:b}) and after (\subref{fig:posdist:d}) correction. The average analytic MWA beam power across all snapshots is contour plotted for reference.}
\label{fig:posdist}
\end{figure*}

Figure~\ref{fig:posdist:a} shows the distribution of offset distances in RA and Dec from the NVSS or SUMSS match to isolated sources. The median offset is $\sim$10" in either dimension. While this is less than the median errors $\sigma_{\rm RA}=19"$ and $\sigma_{\rm Dec}=15"$, a north-eastward systematic bias is clearly apparent. This is illustrated by a vector field in Figure~\ref{fig:posdist:b}. 

The source of this offset can be traced to errors in the MWACS catalogue used for calibration. Considering MWACS isolated matches on position, the median offsets for MWACS within half-beam are $\Delta \rm RA = -8"$ and $\Delta \rm Dec = -10"$. KGS offsets outside of half-beam are much larger in RA, but we suspect the root of the problem is in the calibration.

The bias is found to be well modeled by a second order polynomial as a function of (RA, Dec) position. The modeled bias is used to approximate a positional correction. The distribution of offsets after correction is shown in Figure~\ref{fig:posdist:c}--\ref{fig:posdist:d}. The median offset is reduced to <1" compared to $\sim$10" in either dimension. The median absolute offsets are $|\Delta \rm RA| = 4"$ and $|\Delta \rm Dec| = 3"$. In the catalogue, we report the bias-corrected KGS position as well as the matched catalogue (NVSS or SUMSS) position for comparison. 

\subsection{The Catalogue}
\label{sec:catalogue}

\begin{figure*}
\centering
  \caption{The PUMA match results for the 10 sources listed in Table~\ref{cats_output}, selected to demonstrate a variety of possible match and source types. The left plot shows the uncorrected catalogue positions, errors, and reported shape when available. The gray dashed line indicates the approximate PSF FWHM about the KGS source position. The dot-dashed black line marks the 2.3' initial search radius. The right plots shows the SED information. The black dashed line indicates the chosen power-law fit. Other lines indicate the fit to various possible source combinations when these were considered.}
  \label{fig:interesting1}
  \includegraphics[width=0.8\textwidth]{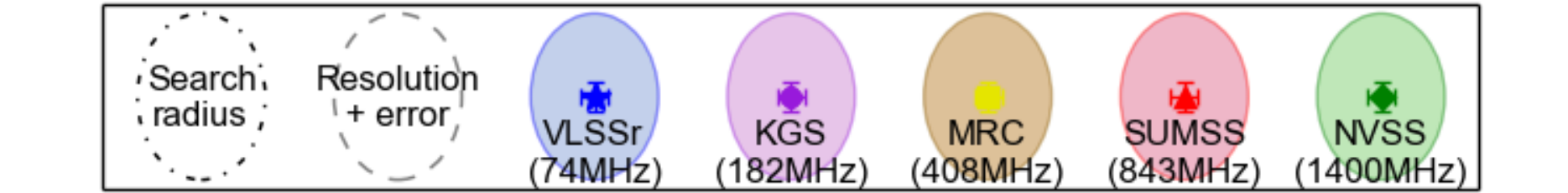}
 \end{figure*}

Table~\ref{cats_output} lists a subset of the catalogue selected to represent a diverse sample. The PUMA position and spectral cross-match information for these are illustrated in Figure~\ref{fig:interesting1}. The complete catalogue of \Total sources is included in the electronic supplement. The columns are:

\begin{enumerate}
\item {\bf Name} Source name.
\item {\bf RAJ2000} Corrected mean J2000 Right Ascension in degrees of the snapshot detections.
\item {\bf DECJ2000} Corrected mean J2000 Declination in degrees of the snapshot detections.
\item {\bf e\_{}RAJ2000} Standard deviation of the measured Right Ascension for all snapshot detections in arcseconds.
\item {\bf e\_{}DECJ2000} Standard deviation of the measured Declination for all snapshot detections in arcseconds.
\item {\bf S\_{}182} Weighted mean integrated 182~MHz flux density measured in Jy.
\item {\bf e\_{}S\_{}182} Standard deviation of the measured flux density for all snapshot detections in Jy.
\item {\bf EB\_corr} Estimated flux density correction factor for Eddington Bias.
\item {\bf R\_{}class} The reliability classification (0-9).
\item {\bf Beam} The mean relative beam response (0--1) at the source location.
\item {\bf N\_{}det} Number of snapshots the source was detected in.
\item {\bf Match\_{}Type} Type of match: isolated, dominant, multiple, combine, or none.
\item {\bf Inspected} 0 if not visually inspected; 1 if the catalogue data and images were inspected; 2 if the match was modified by the authors.
\item {\bf Match\_{}RAJ2000} J2000 Right Ascension in degrees of the catalogue match to NVSS or SUMSS.
\item {\bf Match\_{}DECJ2000} J2000 Declination in degrees of the catalogue match to NVSS or SUMSS.
\item {\bf e\_{}Match\_{}RAJ2000} Uncertainty in Right Ascension of the NVSS or SUMSS catalogue match in arcseconds.
\item {\bf e\_{}Match\_{}DECJ2000} Uncertainty in Declination of the NVSS or SUMSS catalogue match in arcseconds.
\item {\bf SI} Spectral index $\alpha$ from a power law spectral index fit $S\propto\nu^\alpha$ to all catalogue matches.
\item {\bf e\_{}SI} Error on the spectral index parameter.
\item {\bf S\_{}74} Flux density in Jy of the VLSSr catalogue match.
\item {\bf e\_{}S\_{}74} VLSSr flux density error in Jy.
\item {\bf S\_{}408} Flux density in Jy of the MRC catalougue match.
\item {\bf e\_{}S\_{}408} MRC flux density error in Jy.
\item {\bf S\_{}843} Flux density in Jy of the SUMSS catalogue match.
\item {\bf e\_{}S\_{}843} SUMSS flux density error in Jy.
\item {\bf S\_{}1400} Flux density in Jy of the NVSS catalogue match.
\item {\bf e\_{}S\_{}1400} NVSS flux density error in Jy.
\item {\bf VLSSr} VLSSr source name.
\item {\bf MRC} MRC source name.
\item {\bf SUMSS} SUMSS source name.
\item {\bf NVSS} NVSS source name.
\end{enumerate}

\clearpage

\begin{figure*}
\centering  
\subcaptionbox{{\bf KGS J001436-262208} (NVSS J001436-262216): The faintest source in the catalogue at 63~mJy with excellent positional agreement to an NVSS point source. The spectral index is steep at $\alpha$=-1.46, but this does not account for the Eddington bias correction listed in Table~\ref{cats_output}.}{\includegraphics[width=0.8\textwidth]{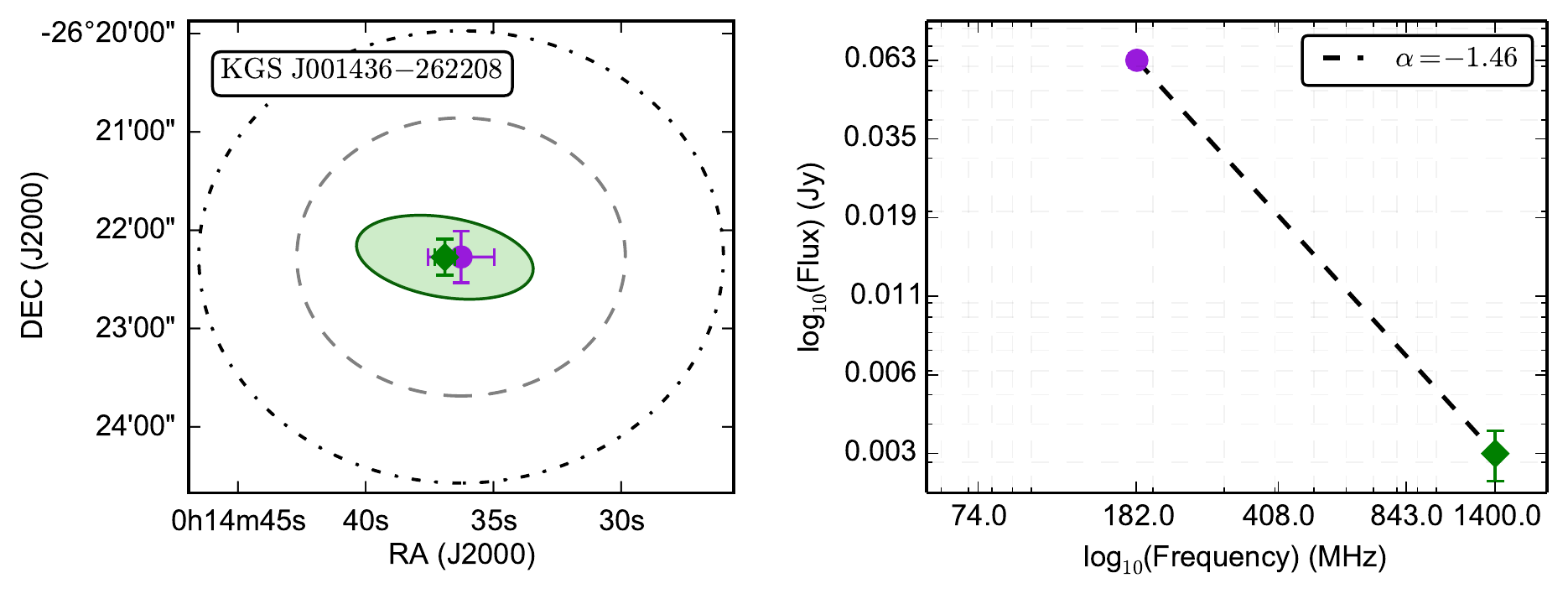}}
  \subcaptionbox{{\bf KGS J002549-260214} (PKS B0023-263): A strongly peaked spectrum source.}{\includegraphics[width=0.8\textwidth]{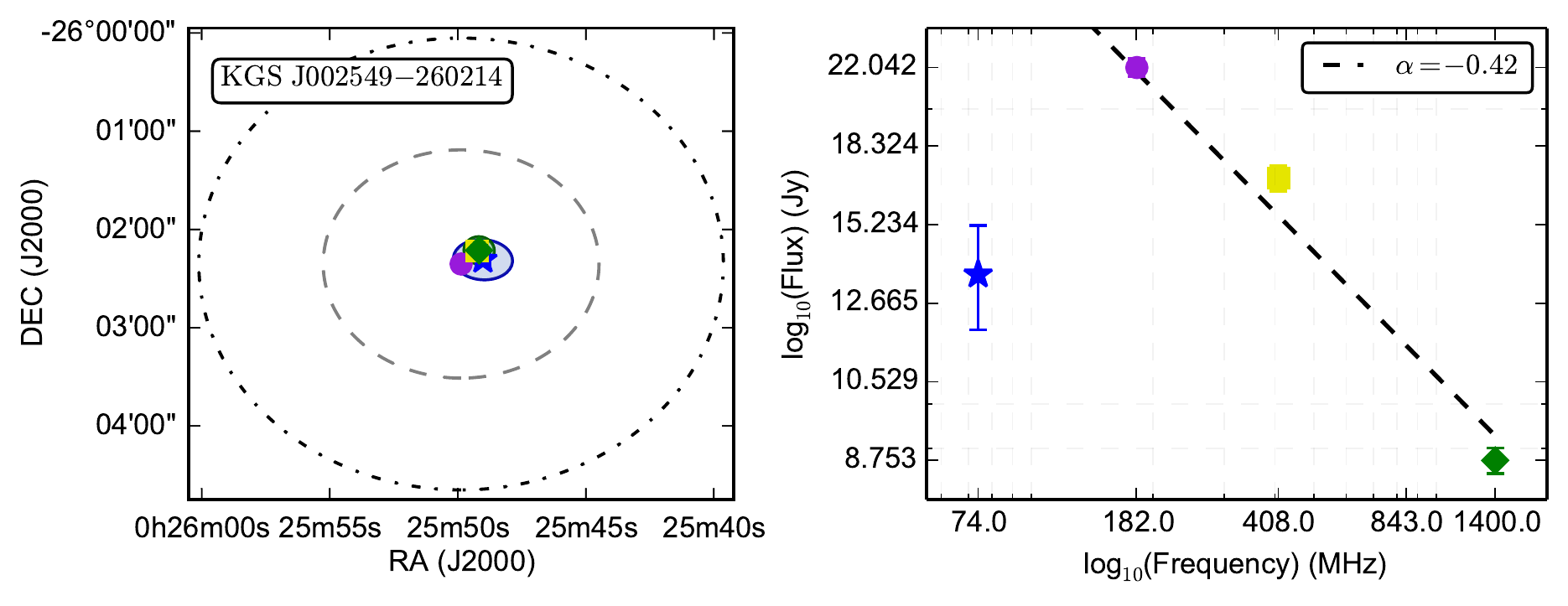}}
  \subcaptionbox{{\bf KGS J235701-344535} (PKS B2354-350): This source demonstrates excellent positional agreement with a counterpart observed in all five comparison catalogues.}{\includegraphics[width=0.8\textwidth]{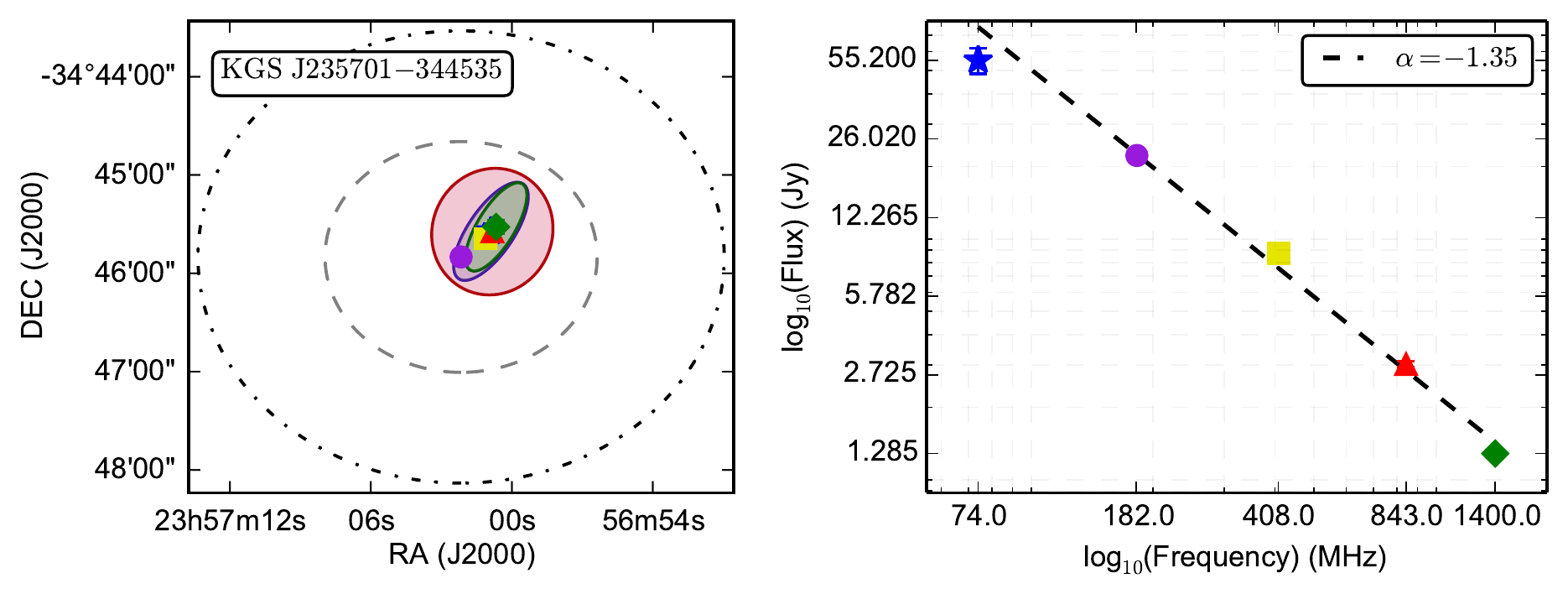}}
\end{figure*}

\begin{figure*}
\centering
  \includegraphics[width=0.8\textwidth]{figures/interesting_key.pdf}
  \subcaptionbox{{\bf KGS J234802-163113} (PKS B2345-167): A good positional match with a poorly fit spectrum. This was subsequently identified as a variable QSO \citep{Valtaoja1992}.}{\includegraphics[width=0.8\textwidth]{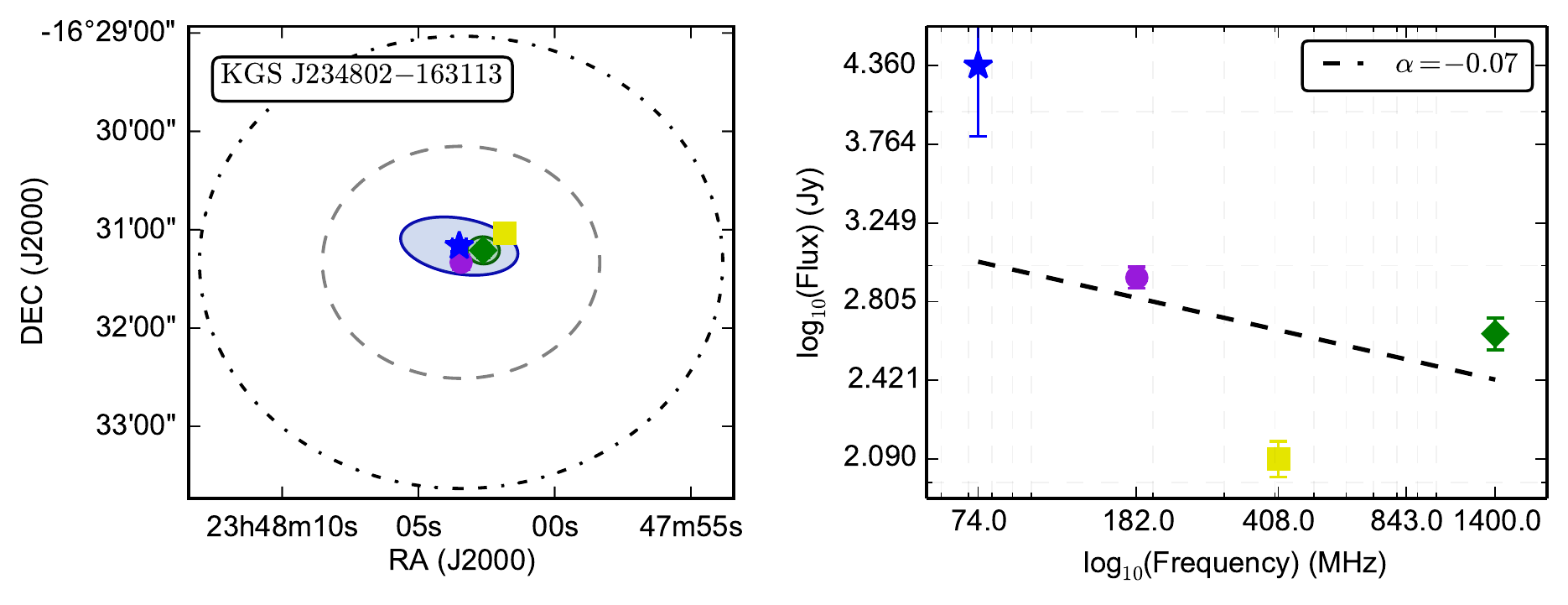}}
  \subcaptionbox{{\bf KGS J231310-315751} (PKS B2310-322): An extended source well-matched to a double at higher frequencies. The combined spectrum is well-fit by a power law. A confusing source (lower-right) within the initial search radius is ignored.}{\includegraphics[width=0.8\textwidth]{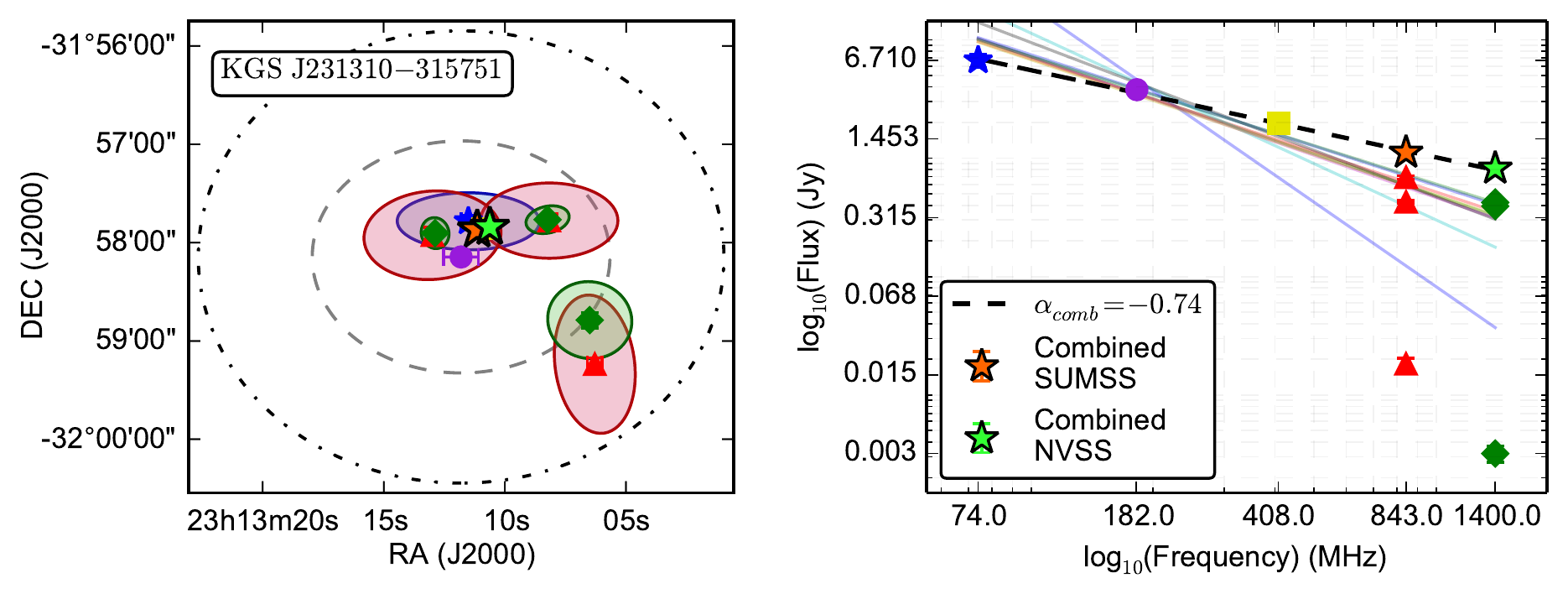}}
  \subcaptionbox{{\bf KGS J225710-362746} (PKS B2254-367): A source with a positive spectral index $\alpha\sim 0.5$.}{\includegraphics[width=0.8\textwidth]{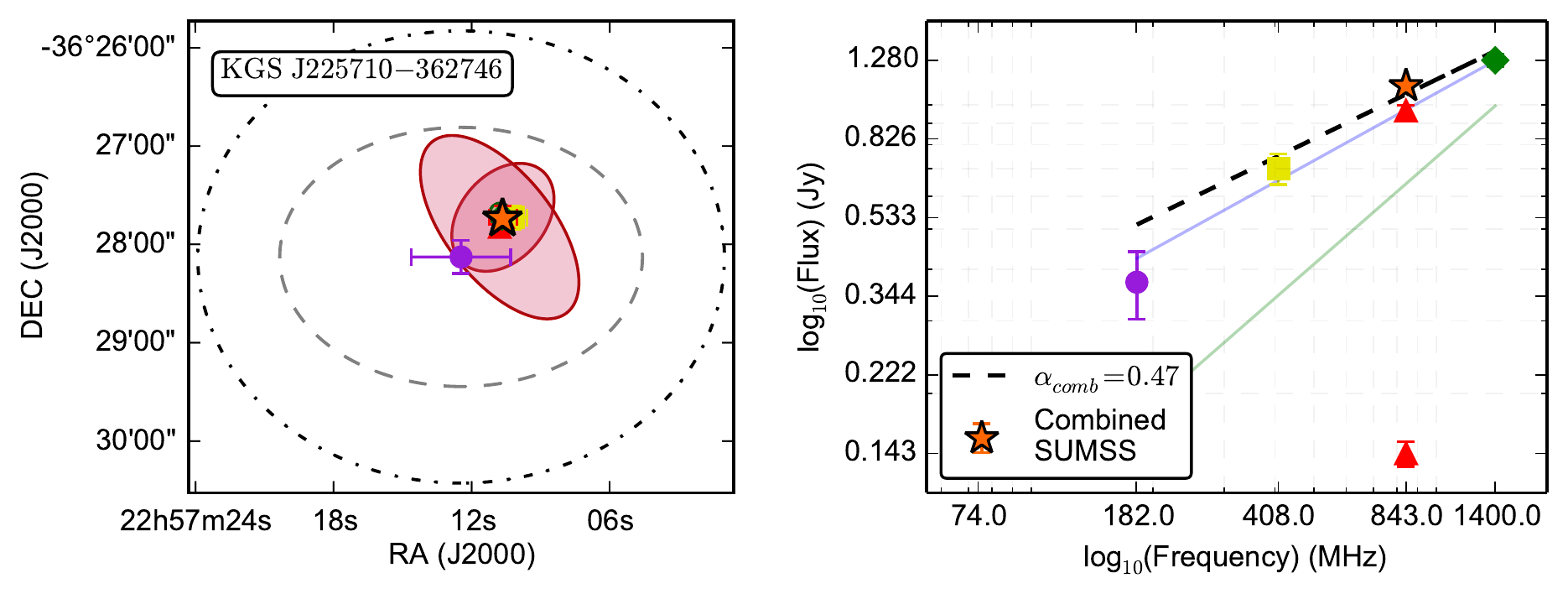}}
\end{figure*}

\begin{figure*}
\centering
  \includegraphics[width=0.8\textwidth]{figures/interesting_key.pdf}
  \subcaptionbox{{\bf KGS J011535-314621} (PKS B0113-320): An example of a multiple match detected in all catalogues. When the components are combined, there is good spectral agreement with a power law fit, confirming the match despite the positional offset.}{\includegraphics[width=0.8\textwidth]{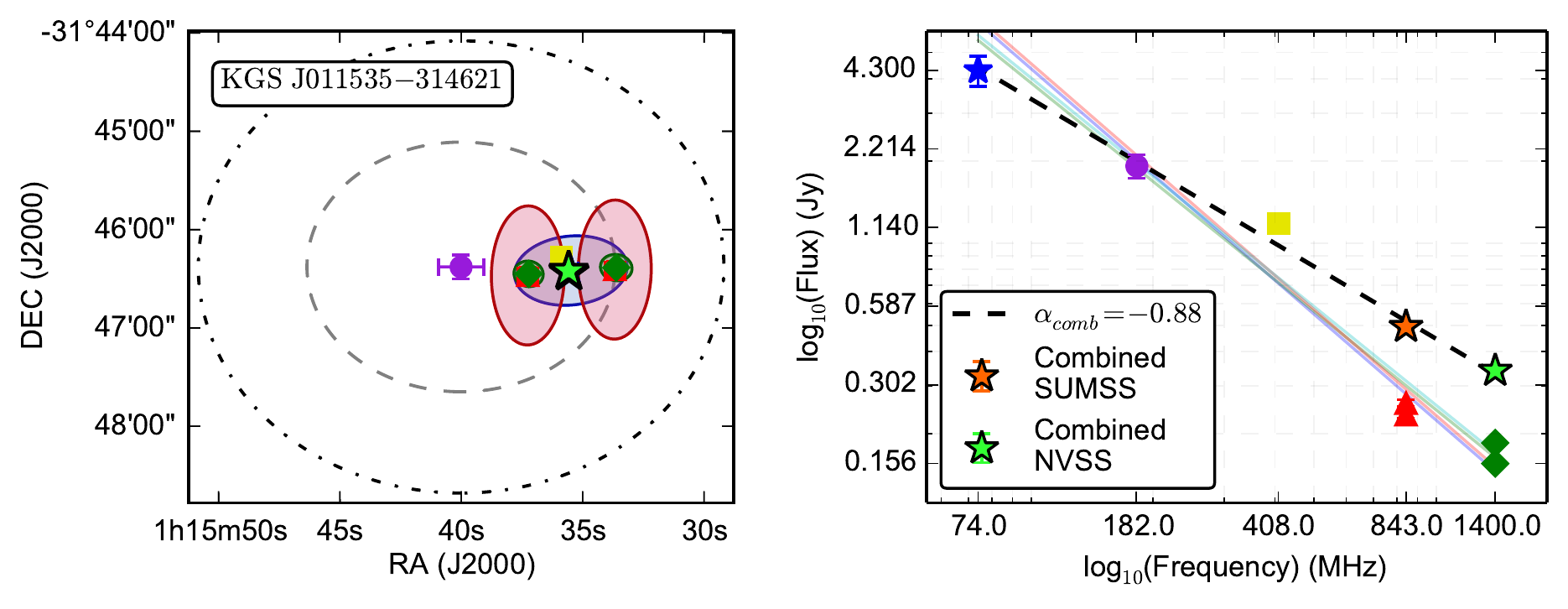}}
  \subcaptionbox{{\bf KGS J004541-420550} (SUMSS J004550-420501): This is an extremely steep spectrum source detected near the edge of the field with $\alpha$=$-$2.23. The flux density may be subject to error in the beam model.}{\includegraphics[width=0.8\textwidth]{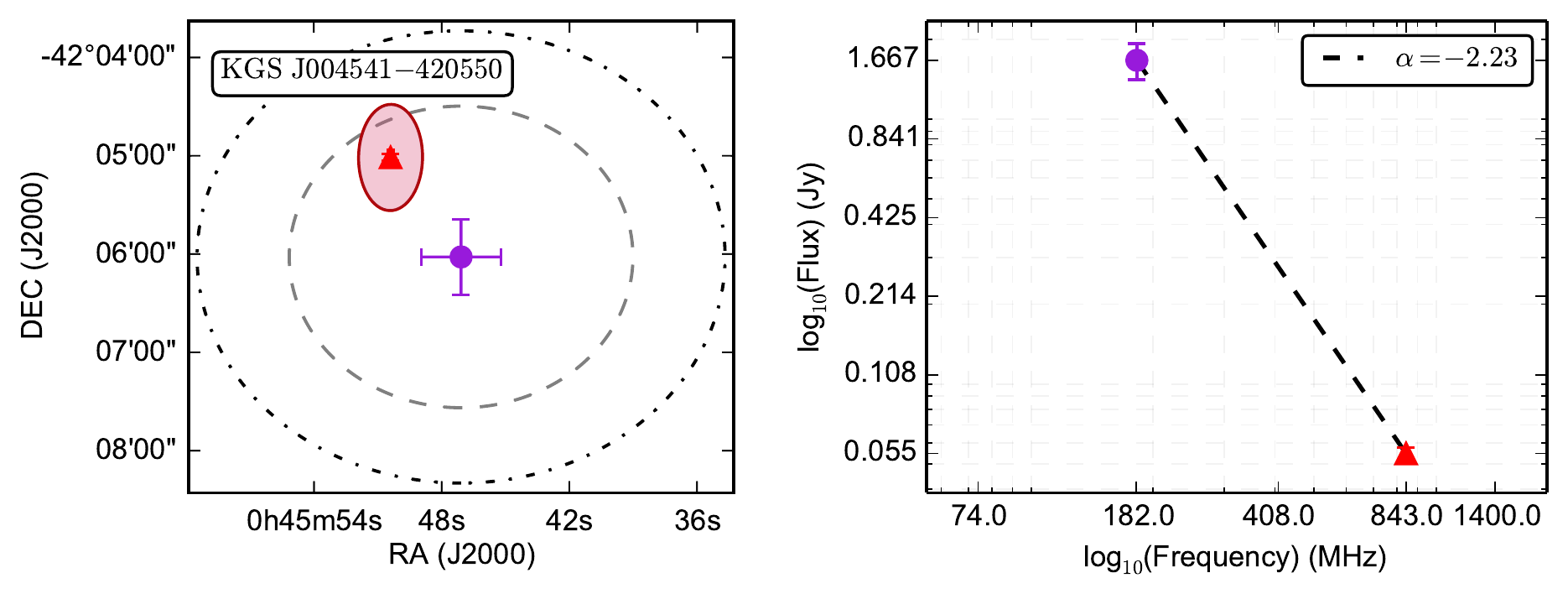}}
  \subcaptionbox{{\bf KGS J004617-420739} (PKS B0043-424): This is the brightest source in the catalogue at 28.3~Jy. It is near the edge of the field and exhibits a large positional bias.}{\includegraphics[width=0.8\textwidth]{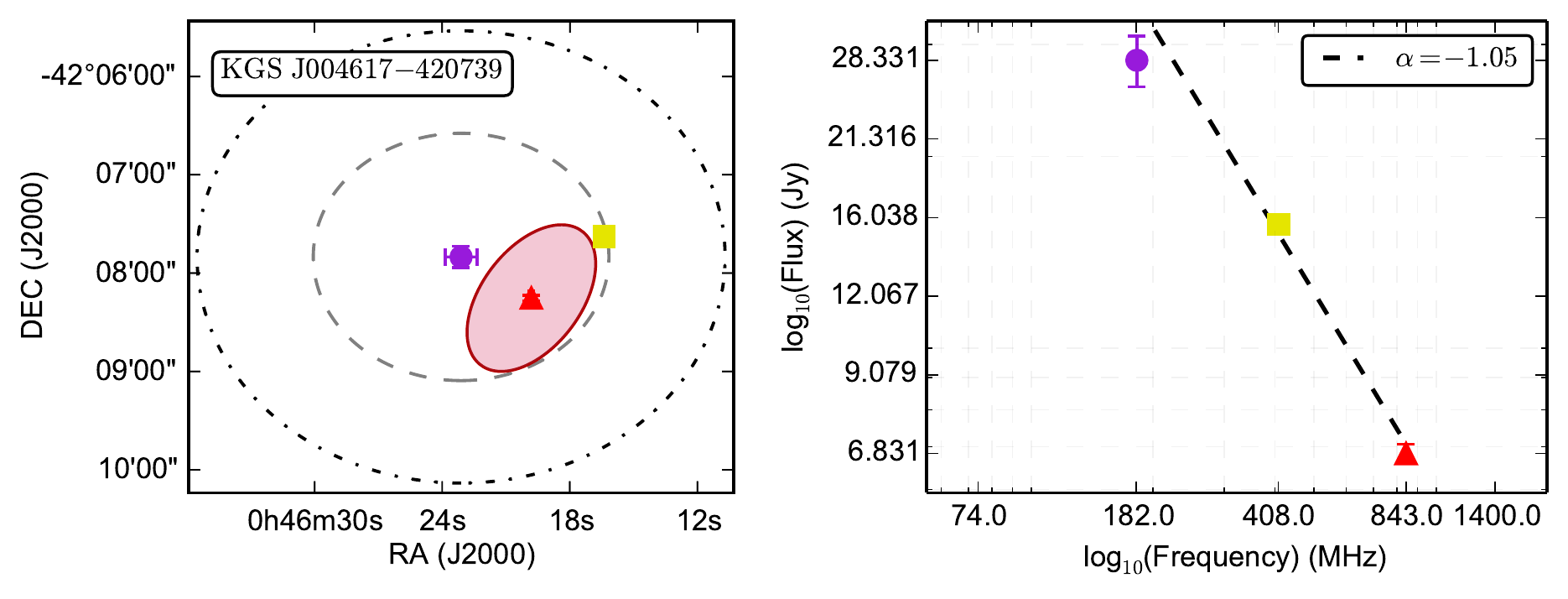}}
\end{figure*}

\begin{figure*}
\centering
  \includegraphics[width=0.8\textwidth]{figures/interesting_key.pdf}
\subcaptionbox{{\bf KGS J013028-260952} (PKS B0128-264): A good demonstration of the Bayesian positional match selection for a source with strong positional bias. The selected match has a posterior probability of 0.99 and is confirmed by the spectral fit.}{\includegraphics[width=0.8\textwidth]{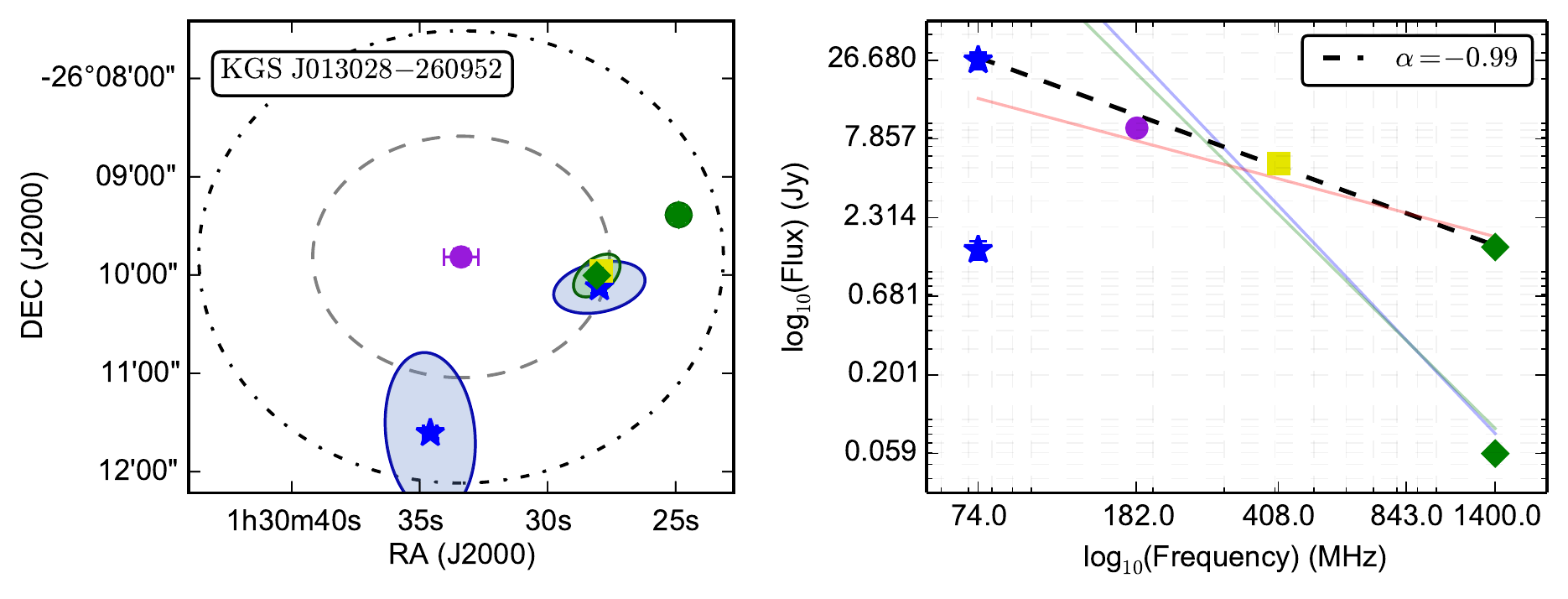}}
\end{figure*}

\begin{landscape}
\begin{table}
\large
\caption{A sample subset of the catalogue data. Ten sources were chosen to represent a diversity of characteristics. Ordering is by increasing distance from field centre (0$^h$,$-$27$^{\circ}$). The corresponding cross-match results are shown in Figure~\ref{fig:interesting1}.}
\begin{center}
	\begin{tabular}{l c c c c c c c c c} 
	\hline
	Name & KGS\_{}RAJ2000 & KGS\_{}DECJ2000 & e\_{}RAJ2000 & e\_{}DECJ2000 & S\_{}182 & e\_{}S\_{}182 & EB\_{}corr & R\_{}class & Beam \\
	\hline
	\hline
    KGS J001436-262208	& 3.65004	& $-$26.36889	& 19.4	& 15.8	& 0.063 	& 4e-3	& 0.45  &  8	& 0.94	\\
	KGS J002549-260214	& 6.45528	& $-$26.03734	& 4.0	& 0.7	& 22.04	    & 0.47	& 1.00  &  0	& 0.77	\\
	KGS J235701-344535	& 359.25422	& $-$34.75985	& 2.9	& 1.4	& 22.21 	& 0.36	& 1.00  &  0	& 0.57	\\
	KGS J234802-163113	& 357.01100	& $-$16.52027	& 4.3	& 1.8	&  2.93 	& 0.06	& 1.00  &  0	& 0.45	\\
	KGS J231310-315751	& 348.29558	& $-$31.96420	& 10.8	& 1.8	&  3.79	    & 0.22	& 1.00  &  1	& 0.38	\\
	KGS J225710-362746	& 344.29165	& $-$36.46284	& 32.4	& 10.1	&  0.37	    & 0.07	& 1.00  &  8	& 0.20	\\
	KGS J011535-314621	& 18.89879	& $-$31.77273	& 14.0	& 7.2	&  1.92	    & 0.19	& 1.00  &  4	& 0.15	\\
	KGS J004541-420550	& 11.42311	& $-$42.09745	& 28.1	& 23.0	&  1.67 	& 0.26	& 0.98  &  4	& 0.08	\\
	KGS J004617-420739	& 11.57300	& $-$42.12757	& 11.2	& 6.5	& 28.33 	& 2.59	& 1.00  &  4	& 0.08	\\
	KGS J013028-260952	& 22.61847	& $-$26.16470	& 10.1	& 4.7	&  9.30	    & 0.84	& 1.00  &  6	& 0.09	\\	
	\end{tabular}
	\linebreak
	\linebreak
	\begin{tabular}{c c c c c c c c c c c c c c c c} 
	\hline
	N\_{}det & Match\_{}Type & Inspected & Match\_{}RAJ2000 & Match\_{}DECJ2000 & e\_{}Match\_{}RAJ2000 & e\_{}Match\_{}DECJ2000 & SI & e\_{}SI \\
	\hline \hline
	5	& isolated	& 1	& 3.65380	& $-$26.37120	& 6.1	& 11.2 & $-$1.46	&       \\
	71	& isolated	& 0	& 6.45490	& $-$26.03690	& 0.4	& 0.7 & $-$0.42	& 0.10	\\
	71	& isolated	& 0	& 359.25280	& $-$34.75880	& 0.7	& 0.7 & $-$1.35	& 0.05	\\
	71	& isolated	& 1	& 357.01090	& $-$16.52020	& 0.4	& 0.7 & $-$0.07	& 0.10	\\
	71	& multiple	& 1	& 348.29430	& $-$31.96400	& 0.7	& 0.7 & $-$0.74	& 0.03	\\
	7	& multiple	& 1	& 344.29450	& $-$36.46190	& 0.7	& 0.7 &  ~0.47	& 0.550	\\
	41	& multiple	& 1	& 18.89830	& $-$31.77370	& 0.7	& 0.7 & $-$0.88	&    	\\
	31	& isolated	& 1	& 11.46000	& $-$42.08360	& 1.8	& 2.2 & $-$2.23	& 0.07	\\
	43	& isolated	& 2	& 11.58250	& $-$42.13760	& 1.4	& 1.8 & $-$1.05	& 0.14	\\
	13	& isolated	& 0	& 22.61700	& $-$26.16680	& 0.4	& 0.7 & $-$0.99	& 0.06	\\
	\end{tabular}
	\linebreak
	\linebreak	
	\begin{tabular}{c c c c c c c c c c c c} 
	\hline
	S\_{}74 & e\_{}S\_{}74 & S\_{}408 & e\_{}S\_{}408 & S\_{}843 & e\_{}S\_{}843 & S\_{}1400 & e\_{}S\_{}1400 & VLSSr & MRC & SUMSS & NVSS \\
	\hline \hline

	    	&    	&    	&    	&    	    &    	& 3.2E-3	& 6E-4  &            &          &      
	001436...  \\
	13.55	& 1.65	& 17.00	& 0.51	&    	    &    	& 8.75	    & 0.26  & J002549... & 0023-263 &            & 002549...  \\
	55.20	& 6.72	&  8.70	& 0.35	& 3.02	    & 0.09	& 1.28	    & 0.04	& J235700... & 2354-350 & J235700... & 235700...  \\
	 4.36	& 0.54	&  2.09	& 0.07	&    	    &    	& 2.64	    & 0.08	& J234803... & 2345-167 &            & 234802... \\
	 6.71	& 0.86	&  1.97	& 0.10	& 1.12	    & 0.03	& 0.82	    & 0.02	& J231311... & 2310-322 & J231308... & 231306... \\
	        &       &  0.70	& 0.06	& 1.11 	    & 0.03 	& 1.28      & 0.05	&            & 2254-367 & J225710... &
	 225710... \\
	 4.30   & 0.55  &  1.18	& 0.07	& 0.49	    & 0.01	& 0.34	    & 7E-3	& J011535... & 0113-320 & J011533... & 011533... \\
	       	&    	&    	&    	& 5.5E-2	& 3E-3	&    	    &       &            &          & J004550... & 	  \\
	    	&    	& 15.65	& 0.39	& 6.83	    & 0.237	&    	    &       &            & 0043-424 & J004613... &        \\
	26.68	& 3.26	&  5.36	& 0.17	&    	    &    	& 1.46	    & 0.05	& J013027... & 0128-264 &            & 013028...  \\
	\hline \hline
	\end{tabular}
\end{center}
\label{cats_output}
\end{table}
\end{landscape}

\subsection{Caveats}
There are two important caveats and potential sources of error we wish to emphasize to users of the catalogue.

\subsubsection{Primary Beam Model}
\label{sec:BeamModel}

The purpose of this survey was primarily to build a foreground model for the EoR analysis. As such, we've elected to include sources covering the full field, out to 5\% of the peak beam response. The accuracy of the source flux density measurements relies on the accuracy of the model of the primary beam response. In-situ measurements for beam sensitivity characterization are in progress but at the time of this analysis an analytic MWA beam shape was assumed. As sources move through the beam, trends in the light curves near the edge of the field ($\rm beam<0.2$) suggest a 10--20\% flux density uncertainty that may not be sufficiently captured by the standard error reported in the catalogue. 

\subsubsection{Extended Sources}

Among the ~13\% of the sources flagged as multiple and visually inspected, many exhibit extended morphologies that are not well represented by the sub-components indicated in the higher resolution catalogues. The MWA has many short baselines and much higher surface-brightness sensitivity than most radio telescopes. This has already led to the discovery of a number of large sources that were resolved out in previous surveys (e.g. a dying giant radio galaxy presented in \cite{Hurley-Walker2015}). Diffuse emission picked up by the MWA will make interpreting the flux densities between surveys problematic for extended sources and care should be taken in this regard. 

\subsection{New Sources}
\label{sec:newsources}

We expect point sources with power-law spectra above $-30^{\circ}$ declination to be detected in the NVSS or VLSSr surveys. Below $-30^{\circ}$, we can expect new detections of ultra-steep spectrum sources (USS;  $\alpha \lesssim -1.5$ with $S \propto \nu^{\alpha}$) that fall below the SUMSS completeness limit (18~mJy at 843~MHz) and are unobserved or undetected in the VLSSr. Transient, variable, or peaked spectrum sources may additionally lead to new detections depending on the spectral behavior. Further, the low surface-brightness sensitivity of the MWA allows for the detection of faint extended sources that may be resolved and fall below the sensitivity limit of other surveys.

\begin{figure*}
\captionsetup[subfigure]{labelformat=empty}
	\subcaptionbox{}{\includegraphics[width=0.3\textwidth]{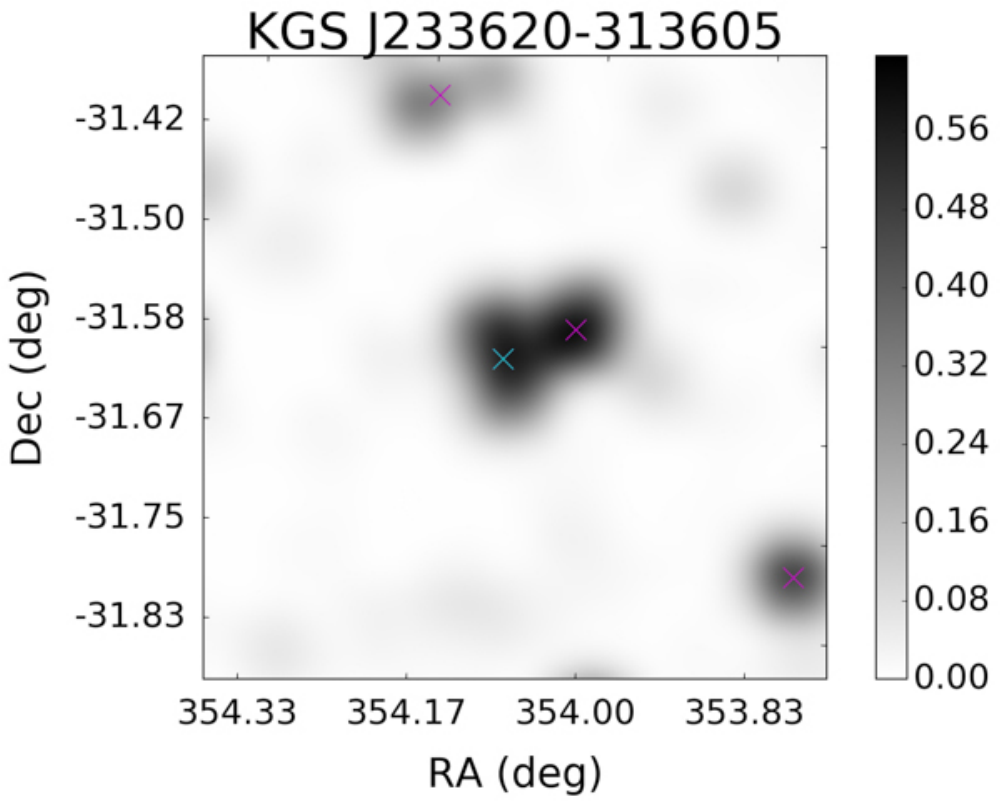}}
	\subcaptionbox{}{\includegraphics[width=0.3\textwidth]{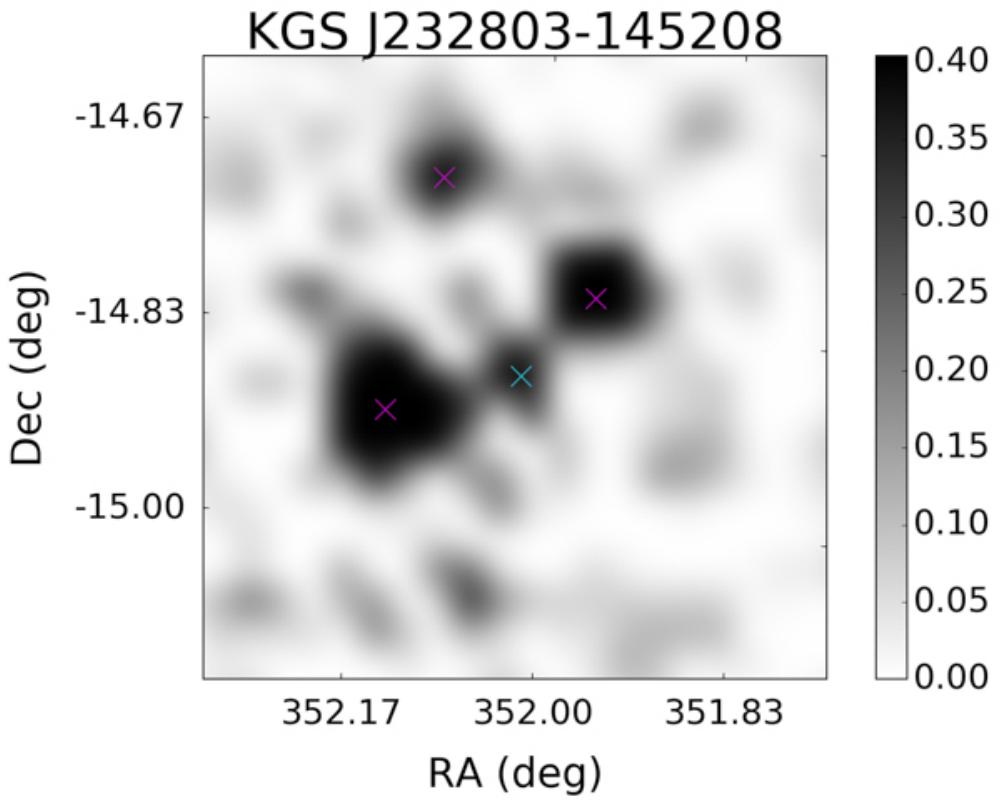}}
	\subcaptionbox{}{\includegraphics[width=0.3\textwidth]{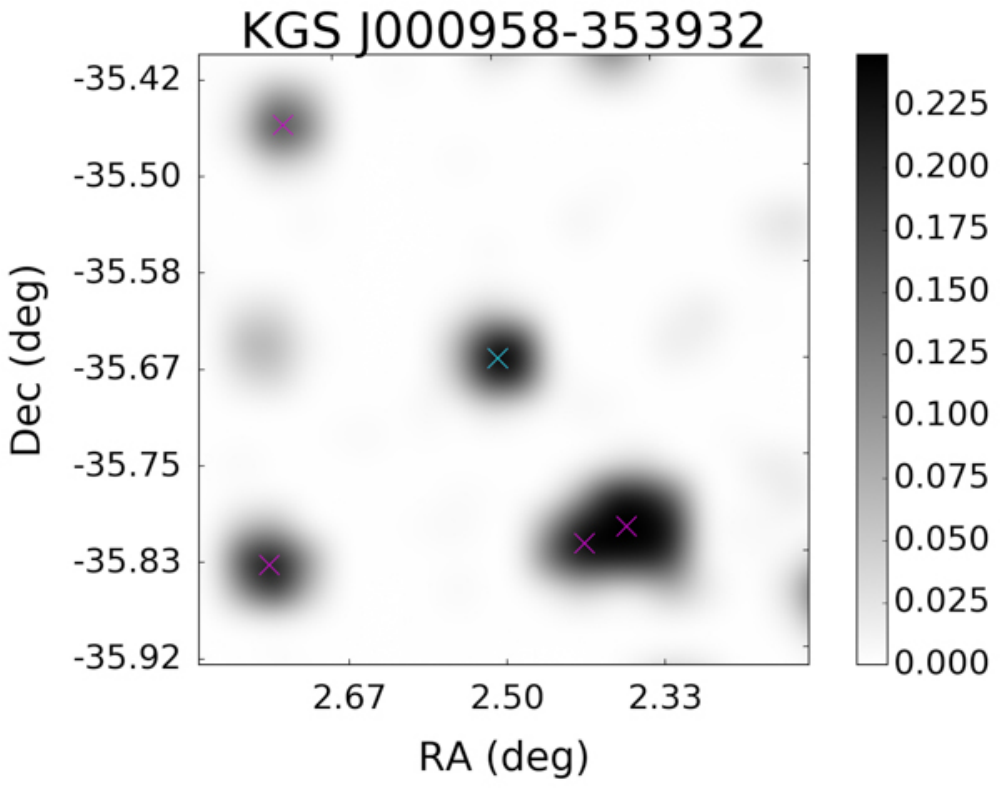}}\\
	
	\subcaptionbox{}{\includegraphics[width=0.3\textwidth]{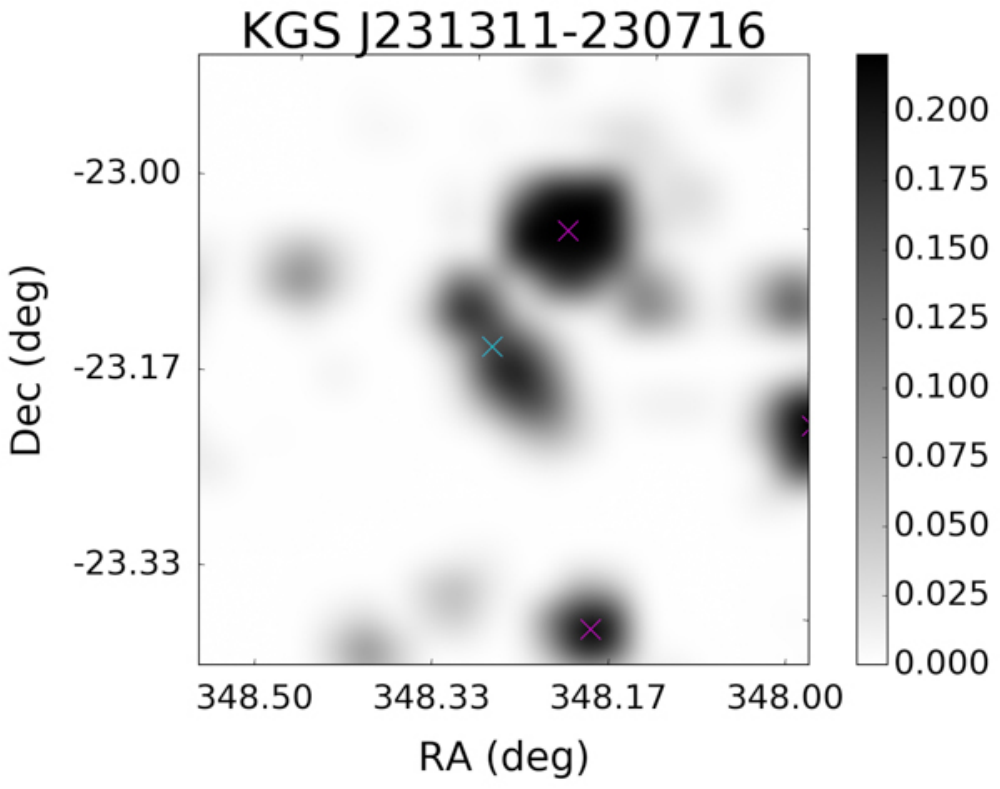}}	
	\subcaptionbox{}{\includegraphics[width=0.3\textwidth]{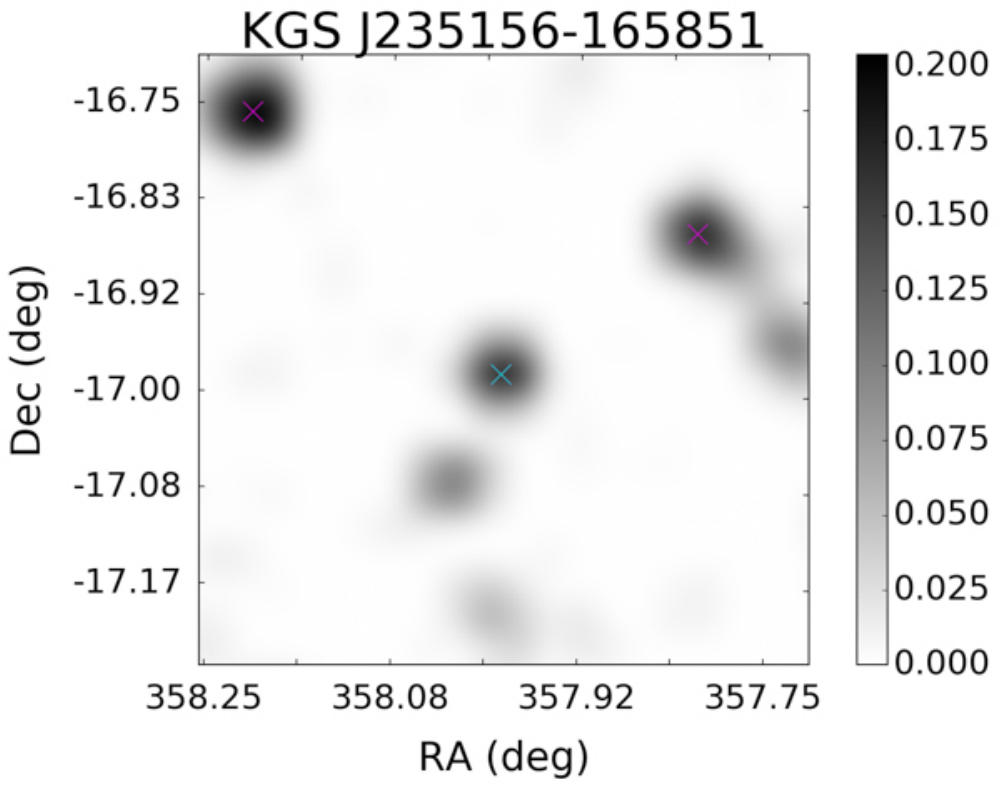}}
	\subcaptionbox{}{\includegraphics[width=0.3\textwidth]{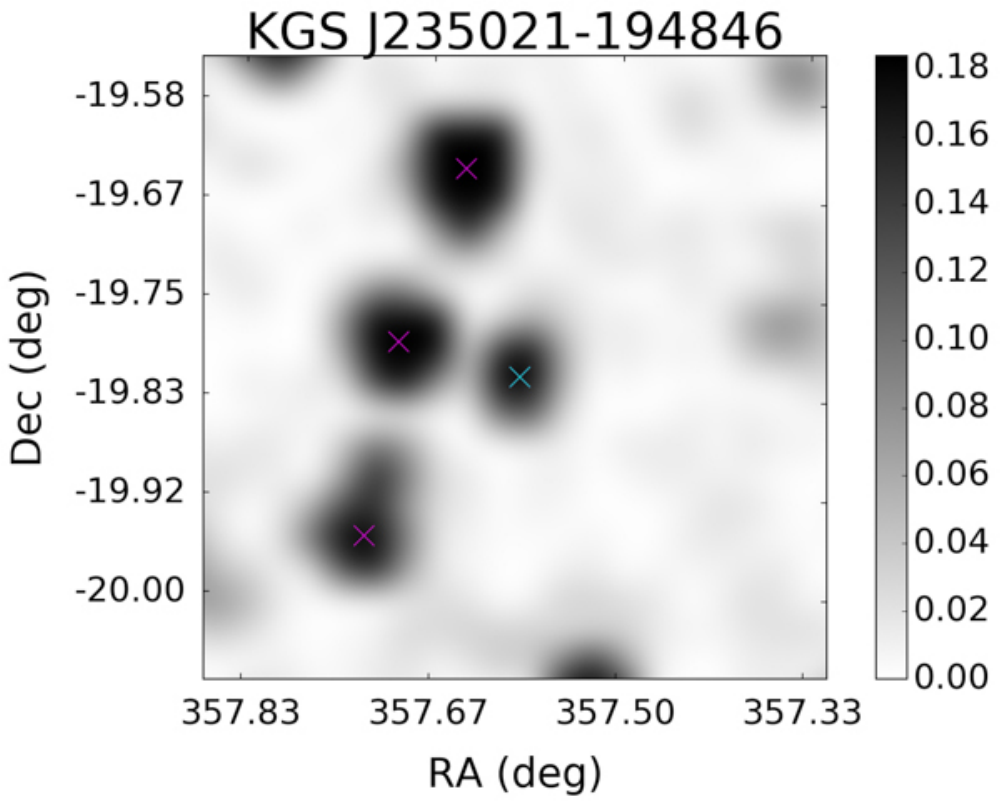}}\\
	
	\subcaptionbox{}{\includegraphics[width=0.3\textwidth]{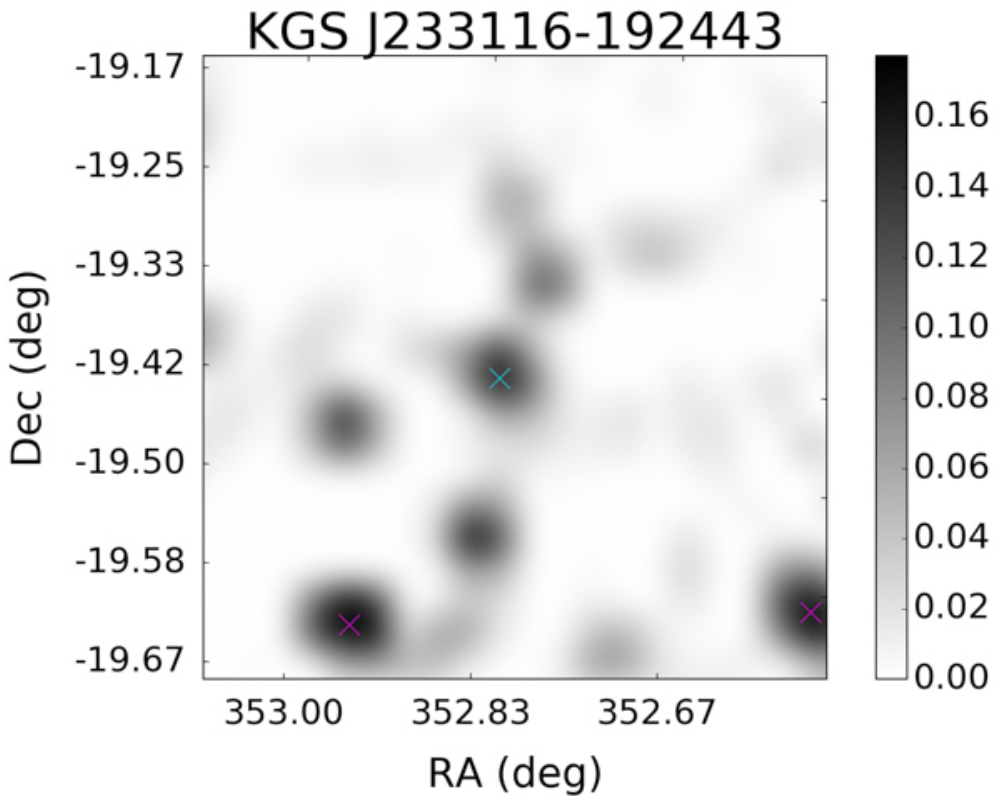}}
	\subcaptionbox{}{\includegraphics[width=0.3\textwidth]{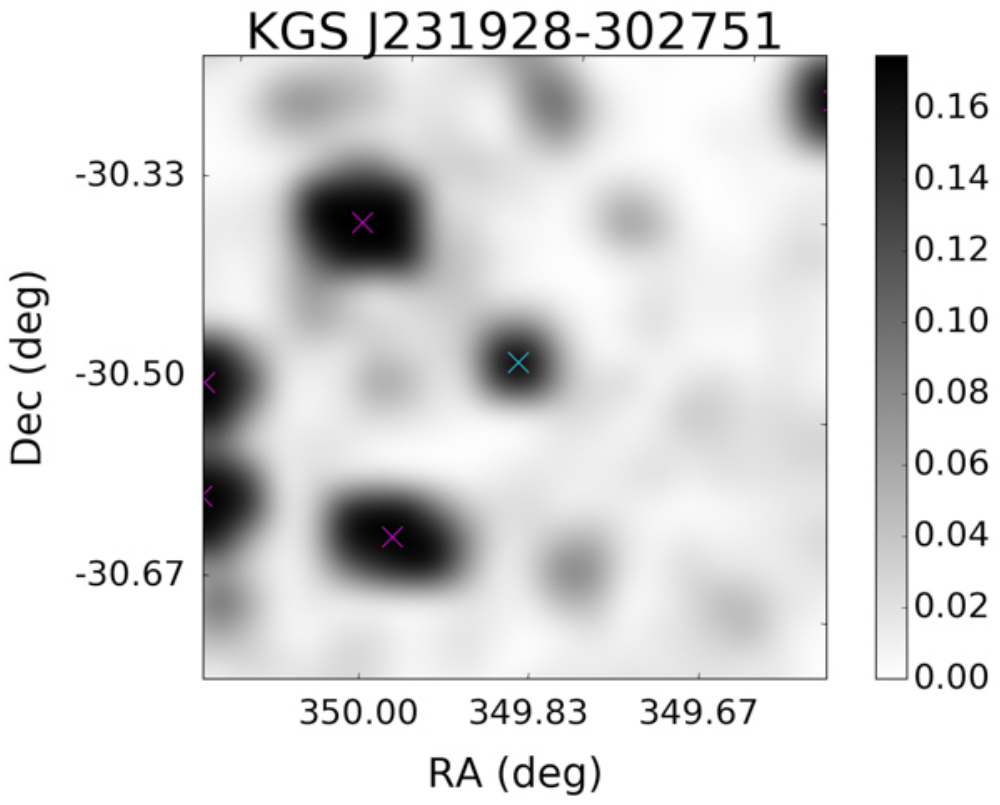}}
	\subcaptionbox{}{\includegraphics[width=0.3\textwidth]{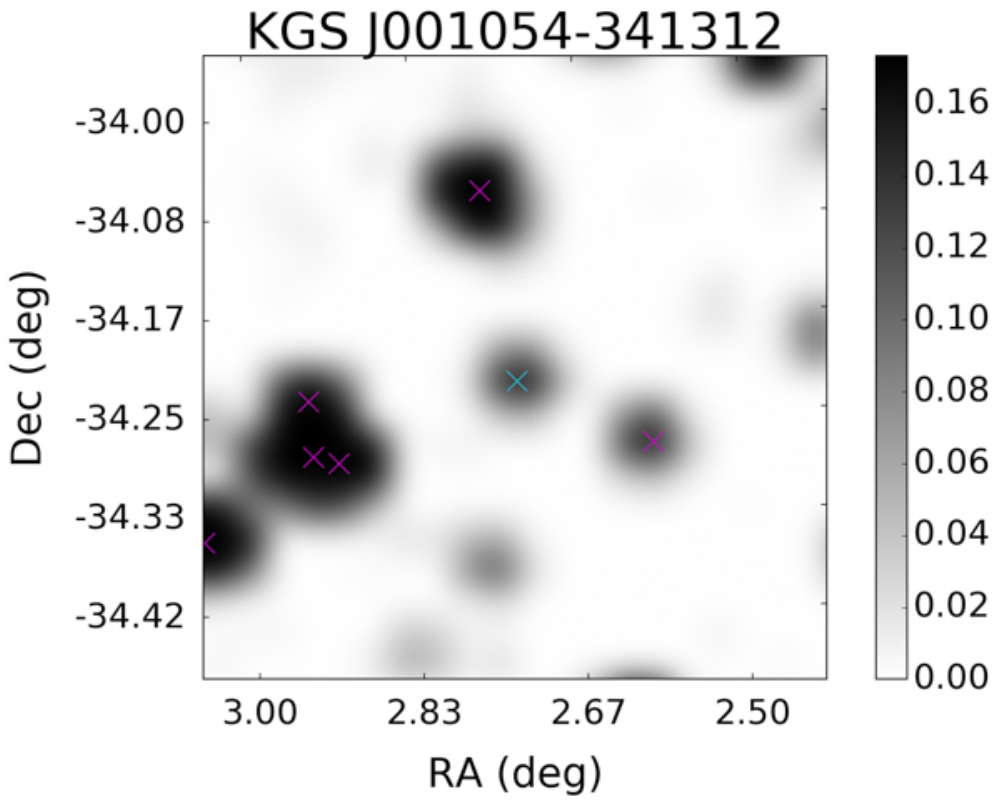}}\\
	
	\subcaptionbox{}{\includegraphics[width=0.3\textwidth]{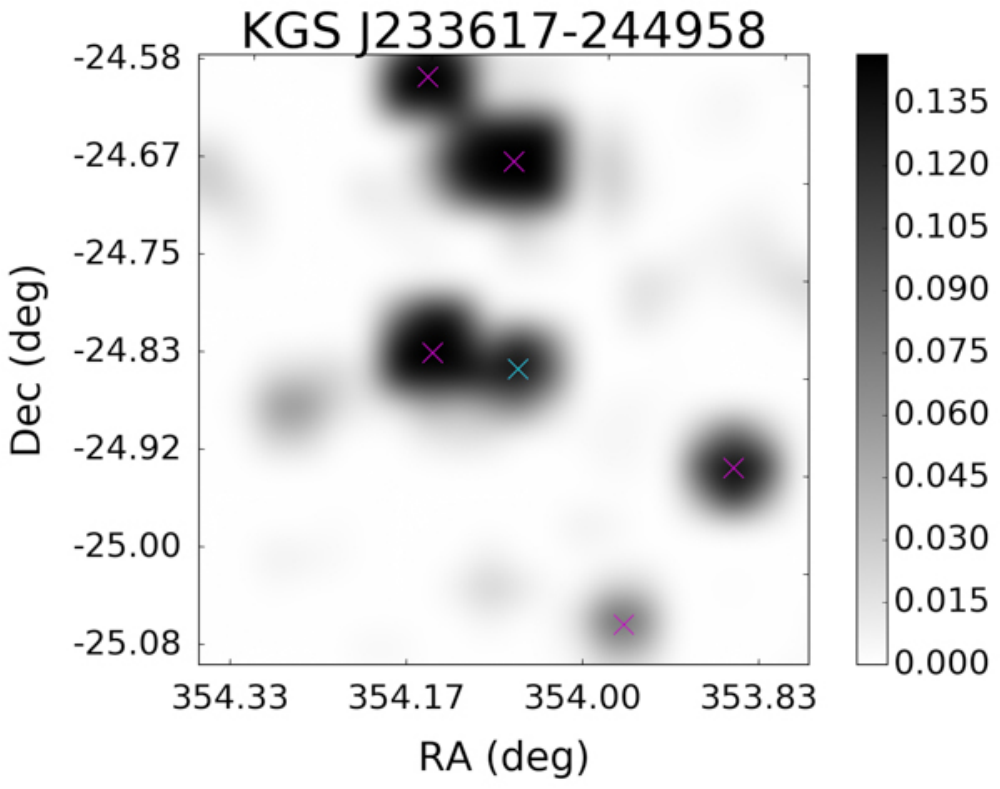}}
	\subcaptionbox{}{\includegraphics[width=0.3\textwidth]{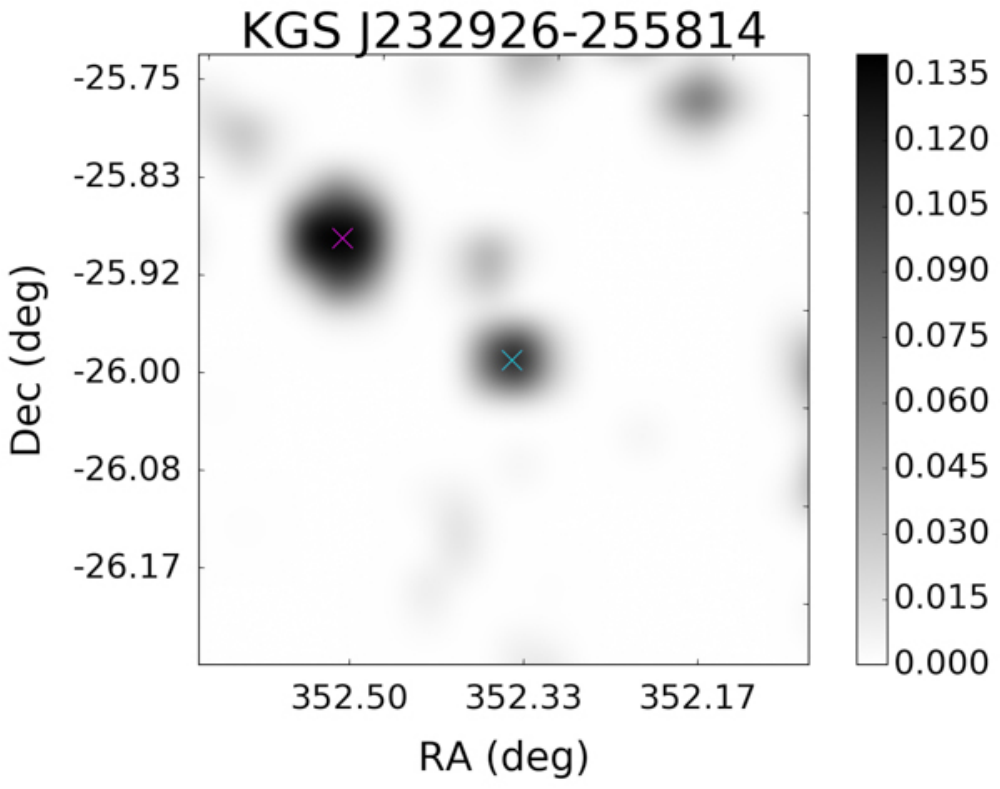}}
	\subcaptionbox{}{\includegraphics[width=0.3\textwidth]{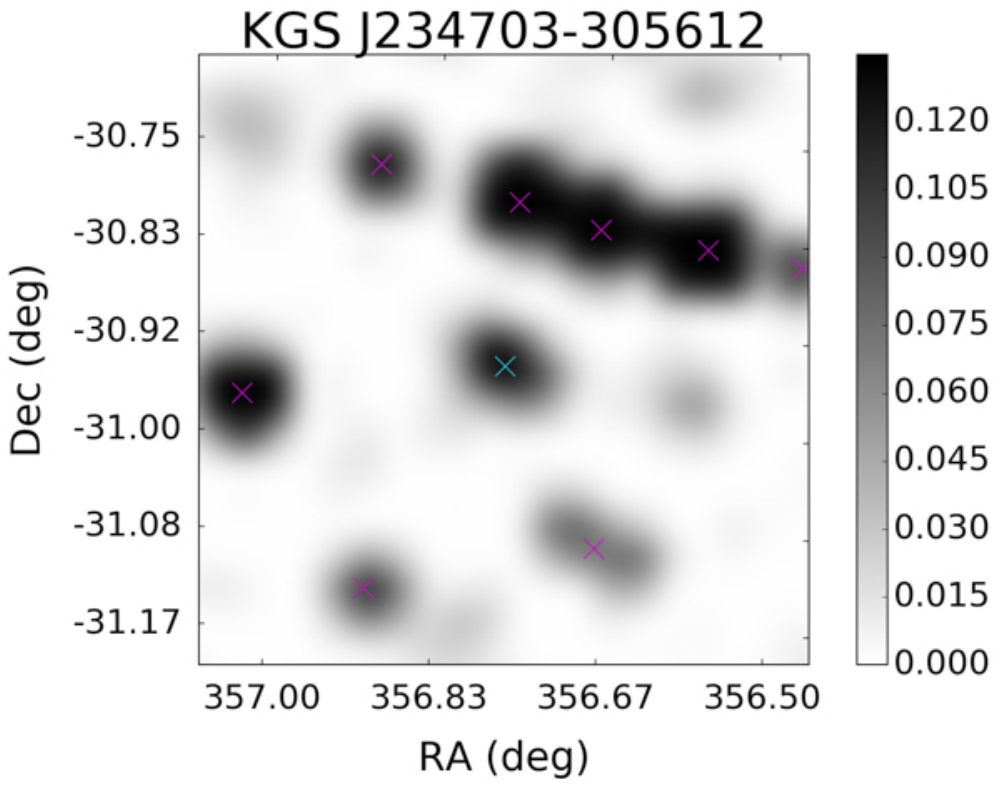}}
\caption{Postage stamp images of sources undetected in any of the comparison surveys. The fist 20 are isolated sources ordered by flux density. Markers indicate the KGS mean positions of the unmatched source (cyan) and other catalogue sources (magenta). Images are 20x20 pixels, smoothed with a cubic interpolation. Units are approximately Jy/beam. This figure continues onto page~\pageref{fig:new-dets-contd}.}
\label{fig:new_detections}
\end{figure*}

There are \NewSrcs sources with no previous radio detection identified. The properties of these sources are listed in Table \ref{tab:nomatches} and postage stamp images are shown in Figure~\ref{fig:new_detections}. A search was made around the positions of these sources in the NASA Extragalactic Database\footnote{The NASA/IPAC Extragalactic Database (NED) is operated by the Jet Propulsion Laboratory, California Institute of Technology, under contract with the National Aeronautics and Space Administration.} to find potential associations within a 2.3' search radius. Eleven sources are unmatched at any wavelength within 30". Eight possible associations with galaxy clusters are identified as well as three with galaxy groups. Others seem to trace galaxy over-densities, or may be high redshift (z) radio galaxy (HzRG) candidates. As these are likely to be some of the most interesting objects in the field, a proposal will be submitted to make follow up observations with the Australia Telescope Compact Array (ATCA). A summary for each source is described below (note the quoted flux densities are corrected for Eddington bias). 

\begin{description}
\item[{\bf KGS J233620-313606:}]
A 424~mJy source most likely associated with the galaxy cluster Abell~S1136. The elongated shape suggests it may be a blended double. The cluster center is located at a distance of 0.92' with a radius $R_c \equiv 1.72/z =$~27.5'. The source is most closely matched to GALEXASC J233618.66-313604.6 at 17" and the x-ray source SW J233617-313626 at 0.7'. Several other cluster members and sources at all wavelengths are also found within the 2.3' search radius. 
\hfill
\item[{\bf KGS J232803-145208:}]
A 249~mJy source most closely matched to the galaxy APMUKS(BJ) B232526.08-150914.0 at 0.6' from the source position. Five APMUKS(BJ) galaxies are found within 2.3' including one extended IR source 2MASX J23280750-1452221 at 1' separation.
\hfill
\item[{\bf KGS J000958-353932:}]
A 164~mJy source 19" from the galaxy cluster member EDCC 408:[CGN95] 000726.8-3556 and extended IR source 2MASX J00095865-3539515. EDCC~408 is cross-identified with Abell~2730, centered 1.8' from the source position with a 14' estimated cluster radius. Many other sources are found within 2.3' including 12 other cluster members. 
\hfill
\item[{\bf KGS J231311-230716:}]
A 147~mJy blended double most closely matched to the UV source GALEXASC J231312.42-230715.1 with 20" separation. The center of the galaxy cluster ABELL S1099 is found at a distance of 1.8' with a cluster radius $R_c=$~15.6'. Two cluster members CMW2004~388 and 339 are cross-identified with 2MASX extended IR sources and located at 1.35' and 1.82' respectively from the source position.
\hfill
\item[{\bf KGS J235156-165850:}]
A 133~mJy source unmatched within 30". The extended IR source 2MASX J23515772-1657374 is 1.3' from the source position and many MRSS galaxy and GALEXASC UV sources are located with the search radius.
\hfill
\item[{\bf KGS J235021-194846:}]
A 123~mJy source unmatched within 30". Eleven MRSS and APMUKS identified galaxies are found within 2.3'. 
\hfill
\item[{\bf KGS J233116-192443:}]
A 113~mJy source matched most closely to the UV source GALEXASC J233115.57-192441.2 at 7". Several other UV sources and three MRSS galaxies are located within 2.3'.
\hfill
\item[{\bf KGS J231928-302751:}]
A 114~mJy source most closely matched to the quasar 2QZ J231927.7-302845 at a 0.9'. Several other UV sources and one MRSS galaxy are found within 2.3'.
\hfill
\item[{\bf KGS J001054-341312:}]
A 112~mJy source most closely matched to 2MASX J00105255-3413132 at 19" and 2dFGRS S495Z294 at 28". The latter is one of 4 members of the 14-member galaxy group 2PIGG SGPGAL~5843 within 1.8'. The galaxy cluster and x-ray source APMCC 014 is 1.8' from the source position. A GALEXASC UV source and 1WGA X-ray source are also found at 20" and 26" respectively.
\hfill
\item[{\bf KGS J233617-244958:}]
A 98~mJy source. A total of 24 galaxies and UV sources are found between 11" and 2.3'.
\hfill
\item[{\bf KGS J232926-255814:}] 
A 92~mJy source most closely matched to the galaxy 2dFGRS S128Z262 at a distance of 40". Five other galaxies and 4 UV sources are also located within the search radius.
\hfill
\item[{\bf KGS J234703-305612:}]
A 88~mJy source unmatched within 30". Seven galaxies are found within the search radius including two un-grouped 2dFGRS identifications at 1' and 2' from the source position. Two additional UV sources are also located within the search radius.
\hfill
\item[{\bf KGS J001640-215455:}]
An 80~mJy source most closely matched to the UV source GALEXASC J001640.19-215516.7 at 21". Two optical galaxies and 26 UV sources are located within the search radius.
\hfill
\item[{\bf KGS J000215-275242:}]
A 79~mJy source unmatched within 30". The galaxy 2dFGRS S198Z035 2' from the source position is part of the 32~member galaxy group 2PIGG~SGP~2684. Five optical galaxies and 4 UV sources are found within the search radius.
\hfill
\item[{\bf KGS J002837-261426:}]
A 66~mJy source unmatched within 30". Four optical galaxies and many UV sources are found within the search radius. 
\hfill
\item[{\bf KGS J234344-263049:}]
A 68~mJy source most closely matched to the galaxy 2dFGRS S194Z277 at 0.84'. There are 5 galaxies and 4 UV sources within the search radius.
\hfill  
\item[{\bf KGS J235556-224242:}]
A 60~mJy source with 4 galaxies and 8 UV sources within the search radius. 
\hfill
\item[{\bf KGS J234709-281746:}]
A 69~mJy source most closely matched to the X-ray source 2XMM J234709.2-281814 at a distance of 27". The radio source ABELL 4038:[SPS89] 06iii is 860~mJy at 1.5~GHz lcoated 1.735' from the source position. Abell~4037 is centered 1.67' from the source position with a radius $R_c =$~59'. Numerous other sources including several cluster members and are also found.
\hfill
\item[{\bf KGS J234851-232934:}]
A 67~mJy source matched to two GALEXASC UV sources within 30". The radio source NVSS J234845-232827 at 1.69' has a low probability of association. Two galaxies and 4 other UV sources are found within 2.3'. 
\hfill
\item[{\bf KGS J002451-204048:}]
A 240~mJy confused source blended with a 2~Jy source just beyond the 2.3' search radius. It is coincident with the center of the galaxy cluster Abell~0027 at only 14" separation. Several cluster members and numerous other sources are found within the search radius.
\item[{\bf KGS J000821-193833:}]
A 217~mJy confused source possibly associated with the galaxy cluster Abell~0002. It is most closely matched to GALEXASC and 2MASX identified galaxies at 7.4" and 8" respectively from the source position. Abell~0002 has a cluster radius $R_c=$~14' centered 1.2' from the source position. This source may be associated with the cluster member and radio source PKS 0005-199 but is not matched at 1.67' separation. Numerous other sources at all wavelengths are found within the search radius.
\hfill
\item[{\bf KGS J000412-151811:}]
A 211~mJy confused source possibly associated with PMN J0004-1518 (cross identified in the NVSS and VLSS) but unmatched at 1.5'. Many galaxies and UV sources are found within the search radius.
\item[{\bf KGS J001702-312239:}]
A 155~mJy confused source. It is most closely matched to 2dFGRS~S436Z221, part of the 116-member galaxy group 2PIGG~SGP~7135, at 1.2' from the source position. Several other sources are found within the search radius, including the 5 other 2PIGG member galaxies within 1.7'. 
\item[{\bf KGS J001007-282942:}]
A 132~mJy confused source most closely matched to a UV source at 27" separation. The galaxy 2dFGRS S279Z095 is found at 0.79' and is part of the 2-member galaxy group 2PIGG~SGP~8480. Two NVSS radio sources lie within the search radius but are not matched. The x-ray source 1WGA J0009.9-2829 and numerous galaxies and UV sources are also found.
\hfill
\item[{\bf KGS J235139-255937:}]
An 81~mJy confused source closely matched to 2dFGRS S132Z149 at 22" separation. This is cross identified as an X-ray and extended IR source, and is a member of the galaxy cluster Abell~2667. The nearest radio source is NVSS J235145-260038 at 1.8' from the source position. Many other sources at all wavelengths are found within the search radius.  
\hfill
\end{description}

\begin{figure*}
\captionsetup[subfigure]{labelformat=empty}
\label{fig:new-dets-contd}
\begin{flushleft}
{\bf Figure~\ref{fig:new_detections} continued.} The final 5 images are new detections that appear to be associated with a brighter counterpart. The low surface-brightness sensitivity of the MWA allows for the detection of diffuse and extended emission that may be be resolved out of the comparison surveys.
\linebreak
\linebreak
\end{flushleft}
\centering
    	\subcaptionbox{}{\includegraphics[width=0.3\textwidth]{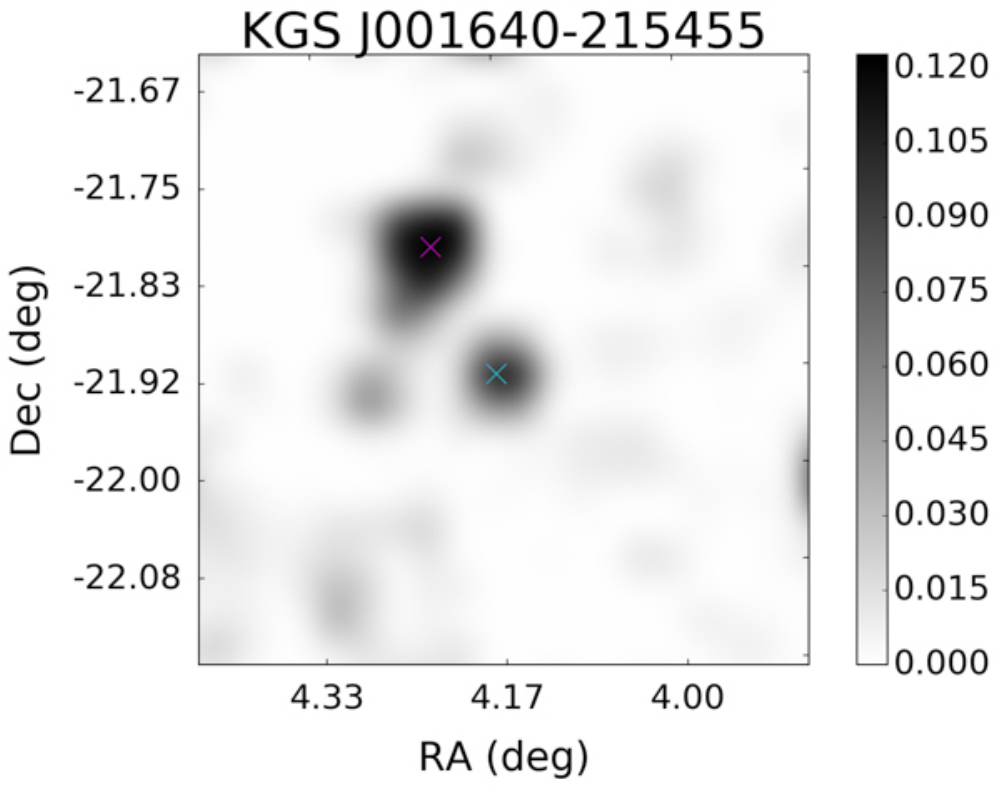}}
    	\subcaptionbox{}{\includegraphics[width=0.3\textwidth]{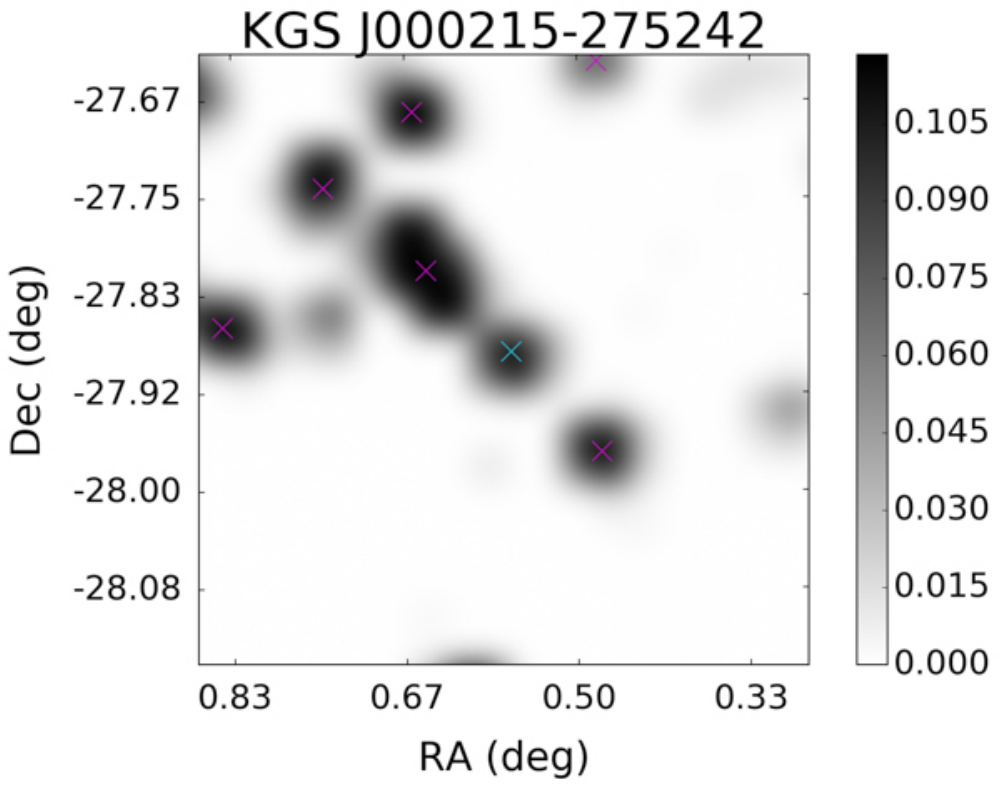}}
	\subcaptionbox{}{\includegraphics[width=0.3\textwidth]{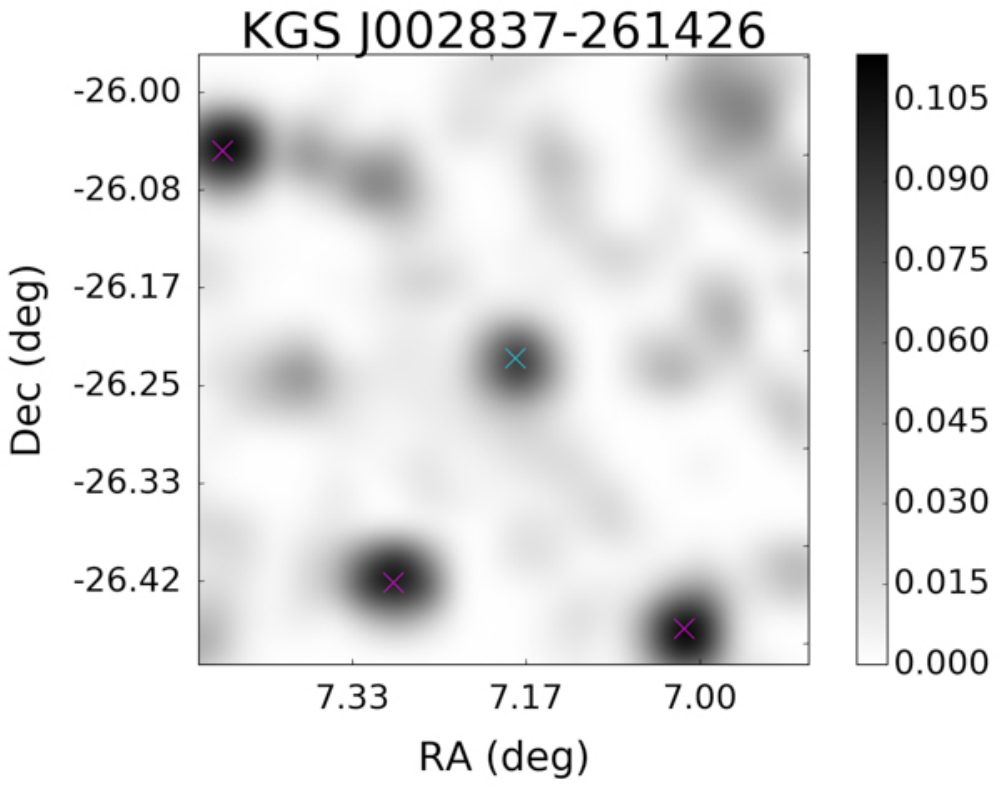}}\\
	\subcaptionbox{}{\includegraphics[width=0.3\textwidth]{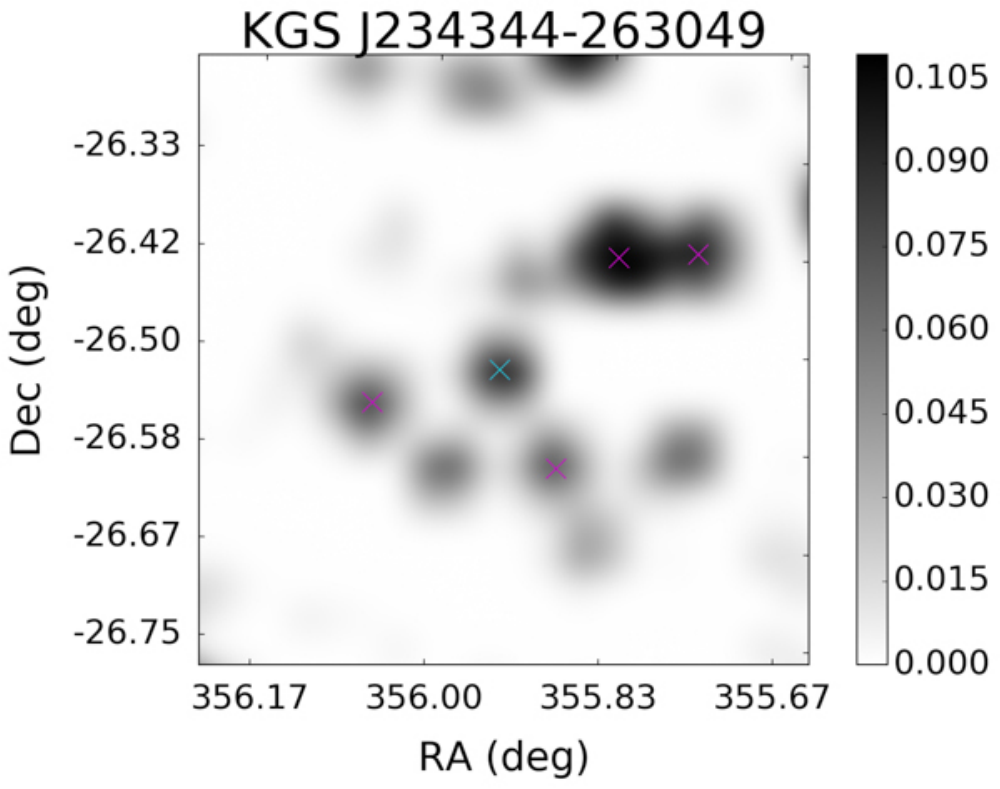}}
	\subcaptionbox{}{\includegraphics[width=0.3\textwidth]{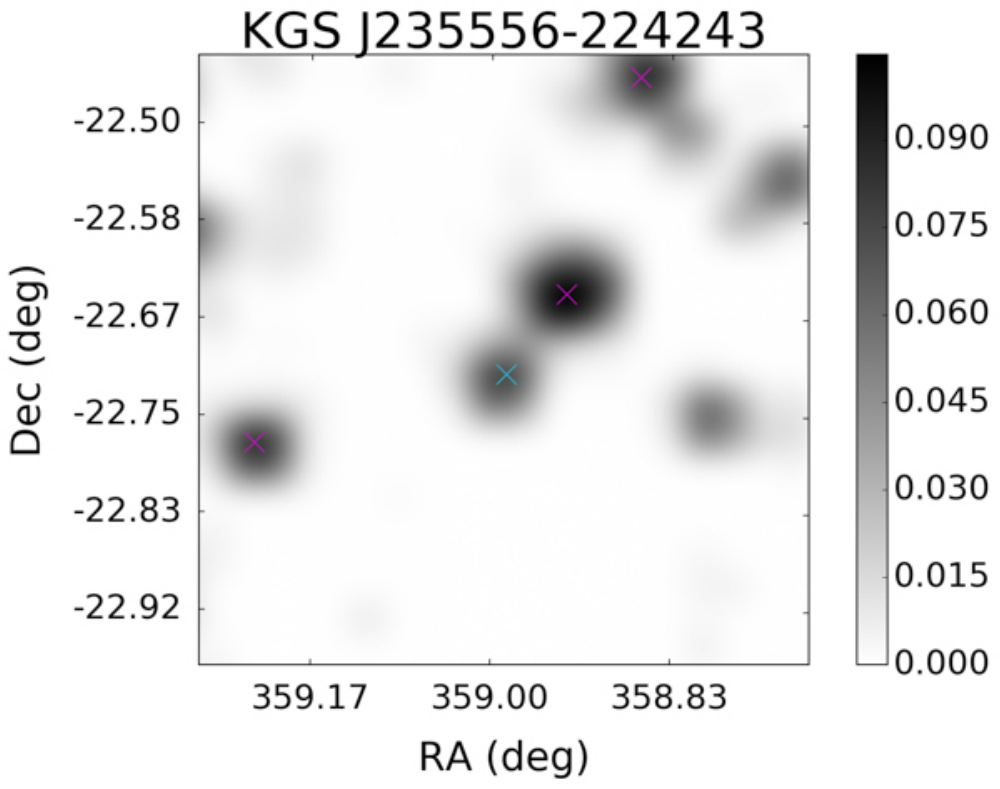}}
	\subcaptionbox{}{\includegraphics[width=0.3\textwidth]{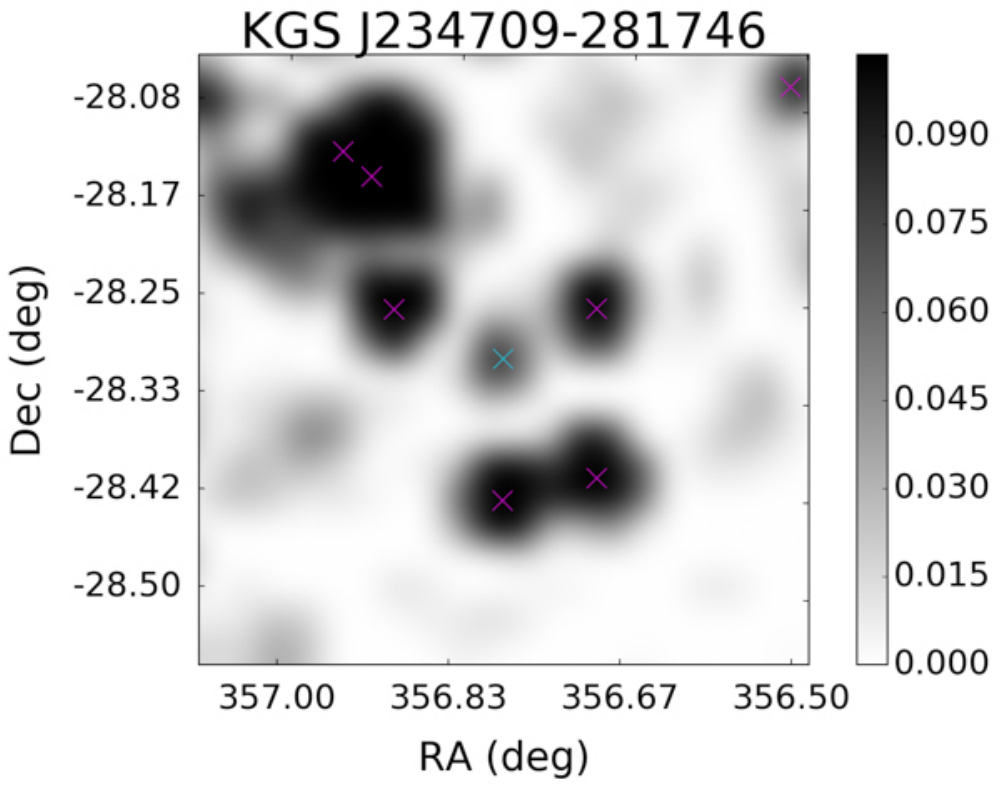}}\\	
	\subcaptionbox{}{\includegraphics[width=0.3\textwidth]{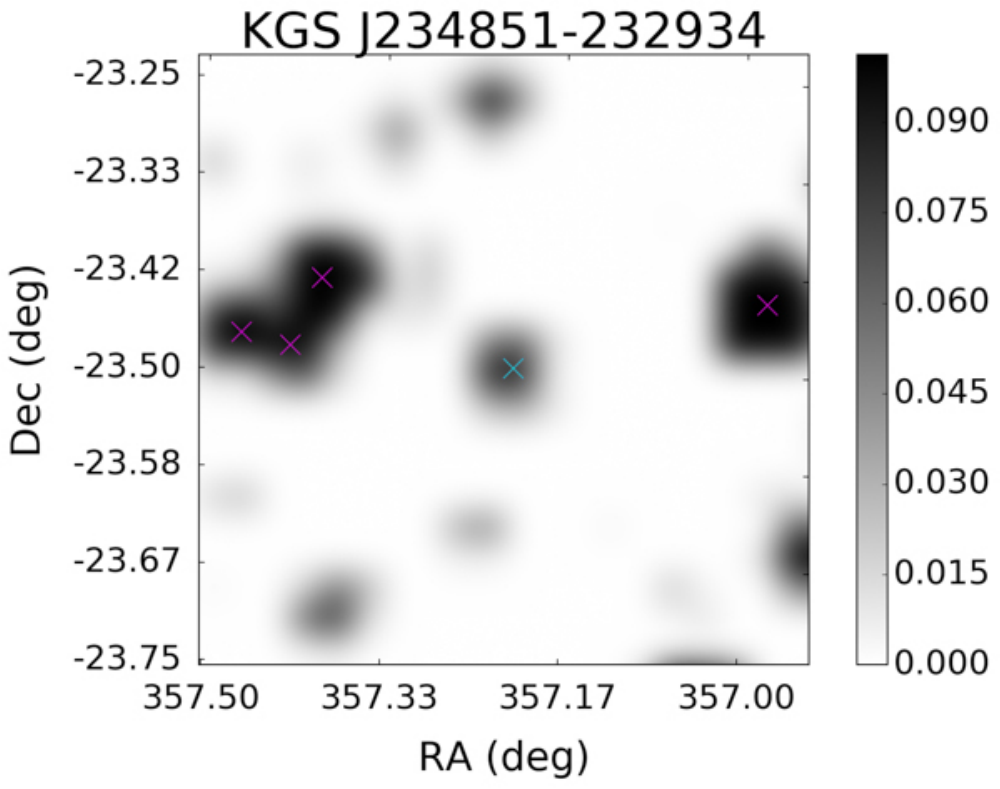}}
	\subcaptionbox{}{\includegraphics[width=0.3\textwidth]{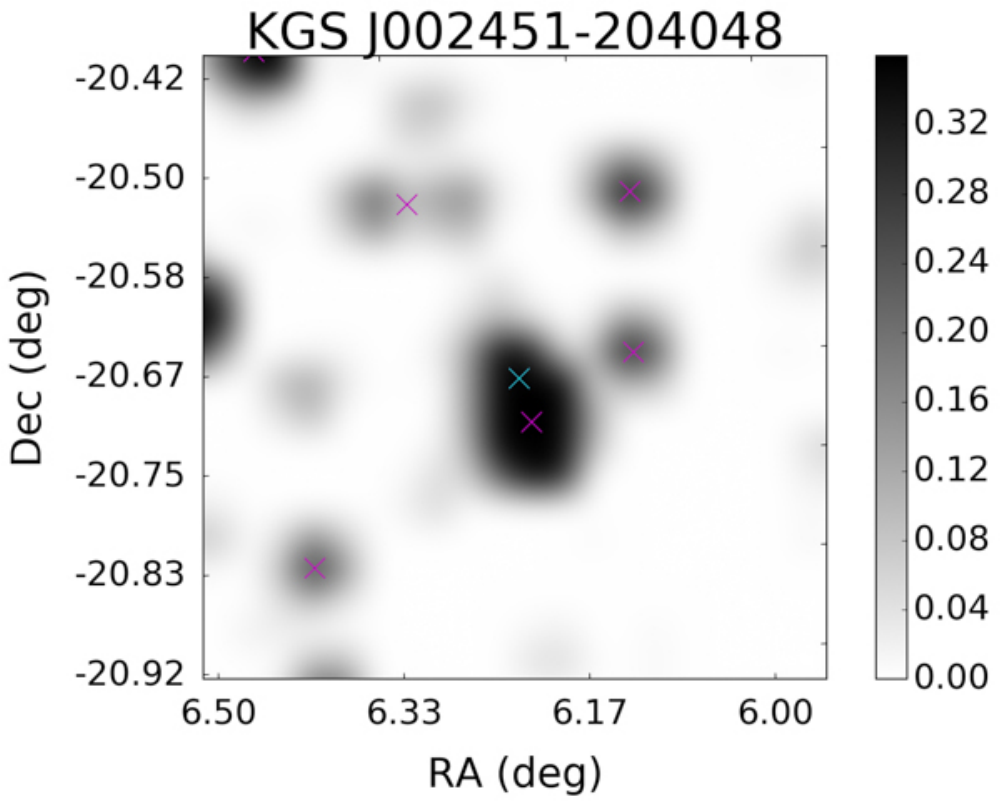}}
	\subcaptionbox{}{\includegraphics[width=0.3\textwidth]{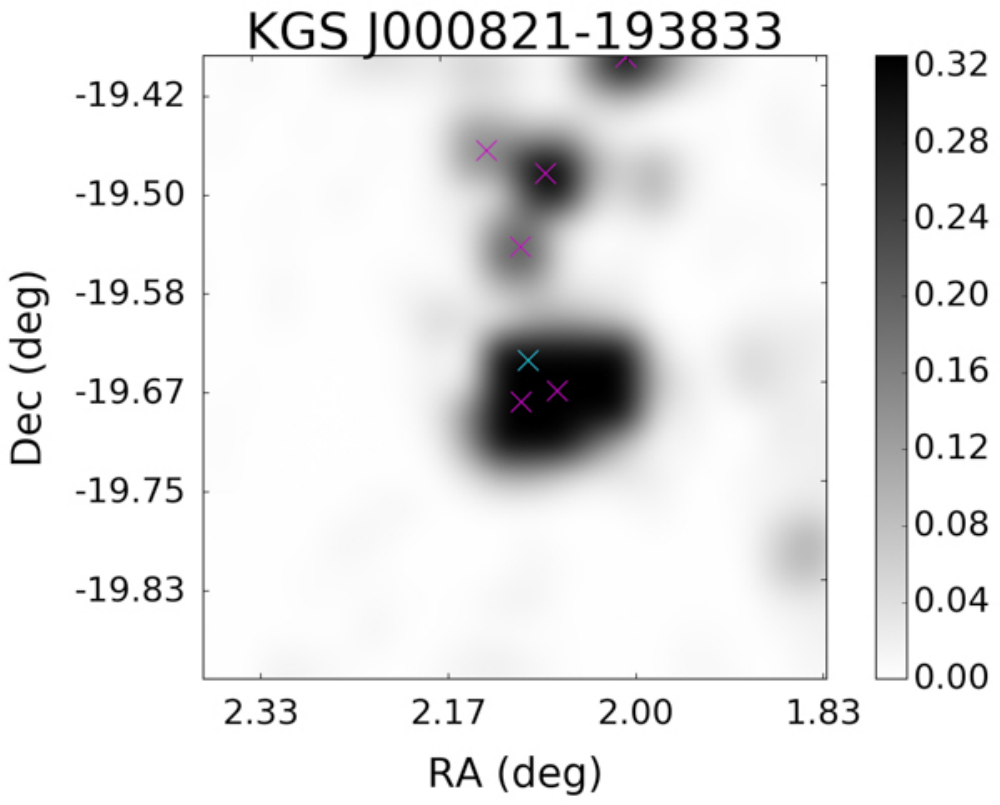}}\\	
	\subcaptionbox{}{\includegraphics[width=0.3\textwidth]{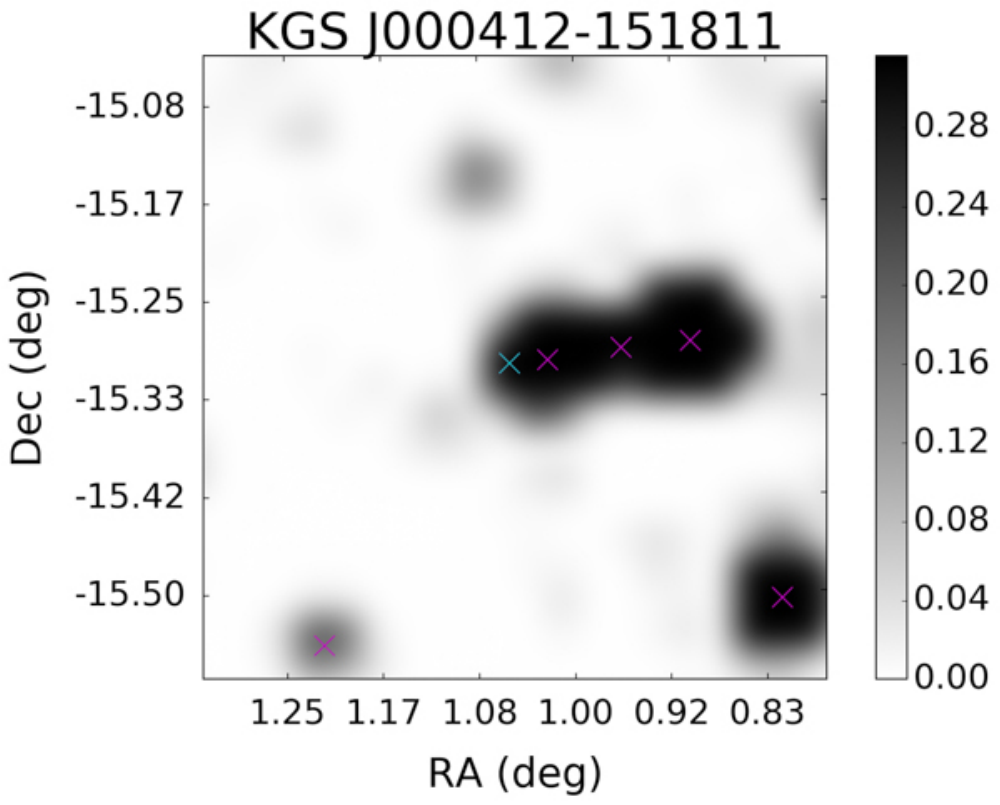}}
	\subcaptionbox{}{\includegraphics[width=0.3\textwidth]{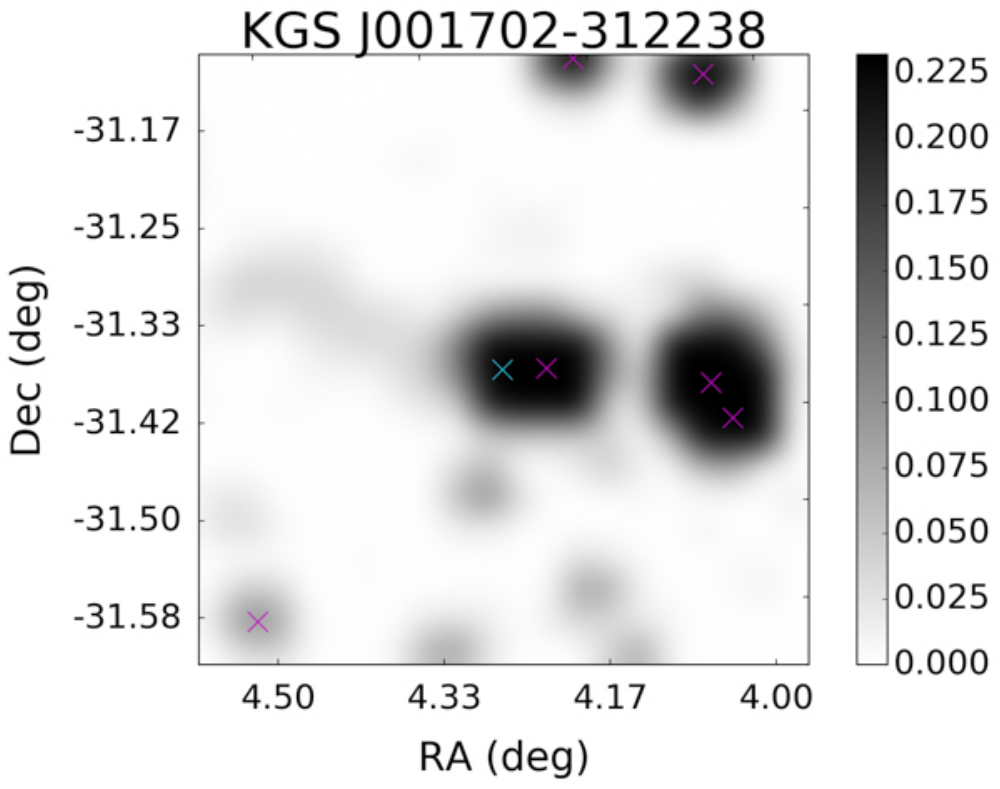}}
	\subcaptionbox{}{\includegraphics[width=0.3\textwidth]{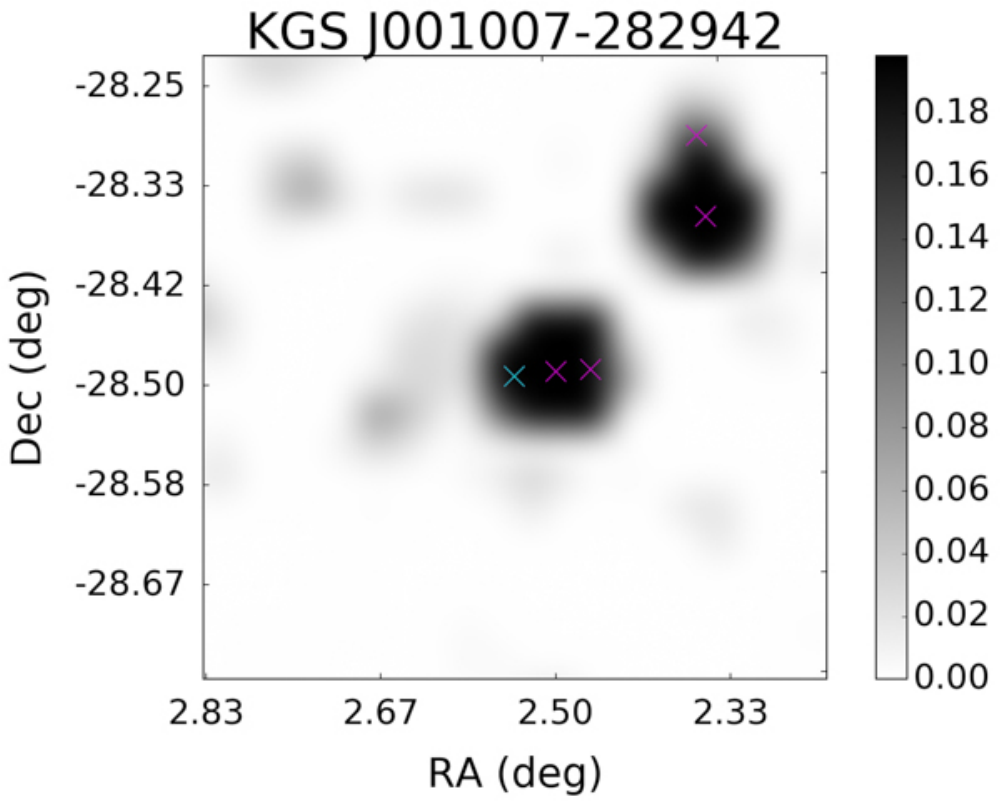}}\\
    \subcaptionbox{}{\includegraphics[width=0.3\textwidth]{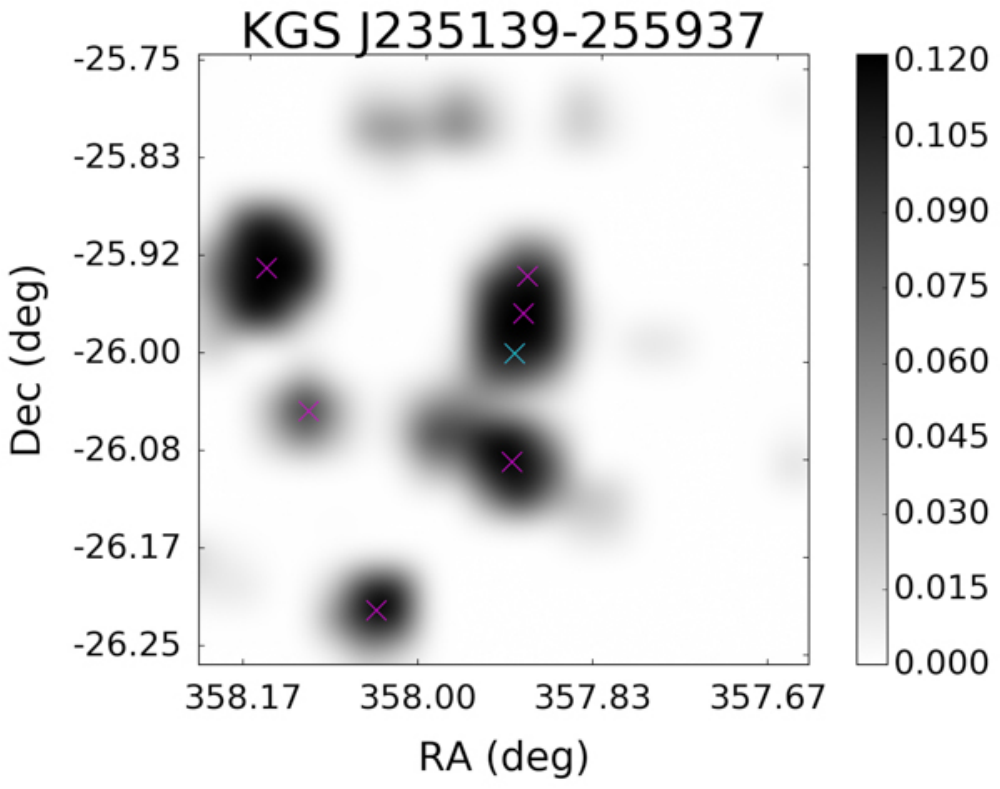}}
\end{figure*}

\begin{table*}
\resizebox{\textwidth}{!}{
\begin{tabular}{lrrrrcccc}
\hline

Name & RA (deg)& Dec (deg) & $S$ (mJy) & $\sigma_S$ (mJy) & $f_{\rm EB}$ &  $N_{\rm det}$ & $\rm Beam_{mean}$ &$R_{\rm class}$\\ \hline \hline

	KGS J233620-313606 &	354.08570 &	$-$31.60165 &	424 &	57 &	1.00 &	71 &	0.67 &	0 \\
	KGS J232803-145208 &	352.01385 &	$-$14.86894 &	270 &	59 &	0.93 &	30 &	0.25 &	4 \\
	KGS J000958-353932 &	2.49438   &	$-$35.65899 &	164 &	27 &	1.00 &	47 &	0.50 &	4 \\
	KGS J231311-230716 &	348.29806 &	$-$23.12123 &	147 &	47 &	1.00 &	22 &	0.54 &	6 \\
	KGS J235156-165850 &	357.98357 &	$-$16.98081 &	136 &	14 &	0.98 &	14 &	0.51 &	6 \\
	KGS J235021-194846 &	357.58806 &	$-$19.81283 &	123 &	19 &	1.00 &	57 &	0.69 &	2 \\
	KGS J233116-192443 &	352.81680 &	$-$19.41199 &	118 &	20 &	0.95 &	11 &	0.59 &	6 \\
	KGS J231928-302751 &	349.86850 &	$-$30.46438 &	117 &	20 &	0.98 &	19 &	0.61 &	6 \\
	KGS J001054-341312 &	2.72537   &	$-$34.22013 &	116 &	18 &	0.97 &	14 &	0.60 &	6 \\
	KGS J233617-244958 &	354.07316 &	$-$24.83280 &	98  &	16 &	1.00 &	47 &	0.82 &	4 \\
	KGS J232926-255814 &	352.36171 &	$-$25.97067 &	93  &	 8 &	0.99 &	17 &	0.80 &	6 \\
	KGS J234703-305612 &	356.76343 &	$-$30.93677 &	90  &	16 &	0.98 &	32 &	0.81 &	4 \\
	KGS J001640-215455 &	4.16737   &	$-$21.91550 &	82  &	12 &	0.97 &	7  &	0.84 &	8 \\
	KGS J000215-275242 &	0.56359   &	$-$27.87857 &	79  &	14 &	0.99 &	44 &	0.94 &	4 \\
	KGS J002837-261426 &	7.15760   &	$-$26.24063 &	76  &	13 &	0.87 &	5  &	0.85 &	8 \\
	KGS J234344-263049 &	355.93651 &	$-$26.51384 &	73  &	17 &	0.94 &	14 &	0.87 &	6 \\
	KGS J235556-224242 &	358.98621 &	$-$22.71193 &	70  &	 9 &	0.85 &	5  &	0.88 &	8 \\
	KGS J234709-281746 &	356.78764 &	$-$28.29634 &	69  &	 7 &	0.84 &	6  &	0.94 &	8 \\
	KGS J234851-232934 &	357.21345 &	$-$23.49292 &	68  &	 6 &	0.66 &	6  &	0.91 &	8 \\ \hline
	KGS J002451-204048 &	6.21621   &	$-$20.68013 &	240 &	23 &	1.00 &	49 &	0.61 &	4 \\
	KGS J000821-193833 &	2.09041   &	$-$19.64273 &	217 &	65 &	0.99 &	34 &	0.68 &	4 \\
	KGS J000412-151811 &	1.05055   &	$-$15.30305 &	211 &	40 &	1.00 &	13 &	0.39 &	6 \\
	KGS J001702-312239 &	4.25956   &	$-$31.37749 &	155 &	31 &	0.99 &	63 &	0.74 &	3 \\
	KGS J001007-282942 &	2.53171   &	$-$28.49514 &	132 &	28 &	1.00 &	37 &	0.91 &	4 \\
	KGS J235139-255937 &	357.91289 &	$-$25.99377 &	81  &	20 &	0.98 &	11 &	0.92 &	6 \\ \hline

\end{tabular}
}
\caption{Properties of the new radio detections. There are 19 isolated sources (top) and 6 sources of apparently extended or diffuse emission (bottom). Postage stamp images are shown in Figure~\ref{fig:new_detections}.}
\label{tab:nomatches}
\end{table*}

\section{Conclusions}
\label{sec:Conclusions}

We have presented a catalogue of \Total extragalactic radio sources in the MWA EoR RA=0 field at 182~MHz. This survey was motivated by the EoR analysis and the need for an accurate foreground model. The foreground catalogue is used for the purposes of calibration and subtraction, and is the predominant systematic hurdle to making an EoR detection. A catalogue of high precision, reliability, and completeness at low frequencies in the southern sky is required. To this aim, new methods were tested for deconvolution and source finding. An in-depth analysis confirmed source reliability and excluded contamination from noise and side lobe sources.

Seventy-five consecutive snapshot observations were processed, covering 2.5 hours approximately centered on zenith. These were independently deconvolved using FHD, resulting in an array of centroided positive components. Source finding was done by spatially clustering the deconvolved components into source candidates for each observation. This approach was chosen to reduce errors inherent to the process of producing a restored image and fitting simplified morphological shapes to sources in the image plane. By source finding independently for each snapshot we retain information on the detection frequency, a valuable diagnostic of source reliability. We identified 9490 unique source candidates detected in at least two snapshots. 

Radio surveys are afflicted by contamination from both noise and side lobes. This is especially troublesome outside of the typical half-power cutoff of the primary beam response. We wished to push this boundary to 5\% beam response while maximizing both completeness and reliability. Given the basic knowledge of how true sources, noise, and side lobes are expected to behave in the data, we used machine learning methods to categorize the source candidates into 10 "reliability" classes based on their observed properties. This gave us a more informative indication of confidence than SNR or detection frequency alone, while remaining entirely self-consistent. 

For the purposes of this work, we selected only \Cand robustly detected above $5\sigma$ or reliability class $R_{\rm class}<7$. These were probabilistically cross-matched to overlapping radio surveys from 74~MHz to 1400~MHz using both positional and broad-band spectral information. Outliers, complex sources, and unmatched sources were flagged and individually investigated. The reliability classification and postage stamp images further aided the decision to include, modify, or exclude a source. Individually inspecting outliers for consistency resulted in the serendipitous identification of variable, peaked or steep spectrum, as well as morphologically complex sources.

Among sources unmatched to another catalogue, \NewSrcs are included in the final catalogue. Some of these detections can be attributed to the low surface-brightness sensitivity of the MWA. Others may be ultra steep spectrum HzRG candidates. We identify several possible associations with galaxy clusters or groups. 

The final catalogue contains \Total sources. Compared to NVSS or SUMSS the median absolute offset is only 7" compared to a 2.3' beam FWHM. No significant evidence is found for flux scale bias in the catalogue. The median broad-band spectral index is found to be $-$0.85 but is dependent on the catalogues matched for each source. A trend in median spectral index is observed as a function of frequency, possibly indicating spectral flattening among a sub-population of sources. The median spectral index at 182~MHz is estimated at $-$0.71 using a second order polynomial fit to sources detected in 3 or more surveys. 

This catalogue provides the most reliable discrete source sky model available to date in the MWA EoR0 field for foreground subtraction. The impact of the improved foreground model on the EoR power spectrum will be presented in Carroll et al. ({\it in prep.}). Based on the lessons learned through the creation of this catalogue, particularly the value of repeated snapshots in assessing source reliability, we are currently taking new observations to extend the EoR foreground survey across the southern sky.

\section{Acknowledgements}

This scientific work makes use of the Murchison Radio-astronomy Observatory, operated by CSIRO. We acknowledge the Wajarri Yamatji people as the traditional owners of the Observatory site.  Support for the operation of the MWA is provided by the Australian Government Department of Industry and Science and Department of Education (National Collaborative Research Infrastructure Strategy: NCRIS), under a contract to Curtin University administered by Astronomy Australia Limited. We acknowledge the iVEC Petabyte Data Store and the Initiative in Innovative Computing and the CUDA Center for Excellence sponsored by NVIDIA at Harvard University.

P.C. would like to acknowledge the support of the American Australian Association Sir Keith Murdoch fellowship and the University of Washington Graduate Opportunities and Minority Achievement Program dissertation fellowship. This work was supported by National Science Foundation grants AST-0847753, AST-1410484, and AST-1506024.

This research has made use of the NASA/IPAC Extragalactic Database (NED), which is operated by the Jet Propulsion Laboratory, California Institute of Technology, under contract with the National Aeronautics and Space Administration. We also acknowledge the use of NASA's SkyView facility (\url{http://skyview.gsfc.nasa.gov}) located at NASA Goddard Space Flight Center.

\appendix

\section{Machine learning}
\label{apx:machine-learning}

Here we expand on the description of the reliability classification summarized in \S\ref{sec:Classification} and present intermediate results. The classification steps can be broken down as follows; feature Selection and standardization, dimensionality reduction, initial cluster finding (unsupervised classification), and building an ensemble classifier. This process makes ample use of the Python software package Scikit-learn \citep{Pedregosa2011}. 

\subsection{Feature Standardization and Dimensionality Reduction}

\begin{figure*}
\centering
\includegraphics[width=\textwidth]{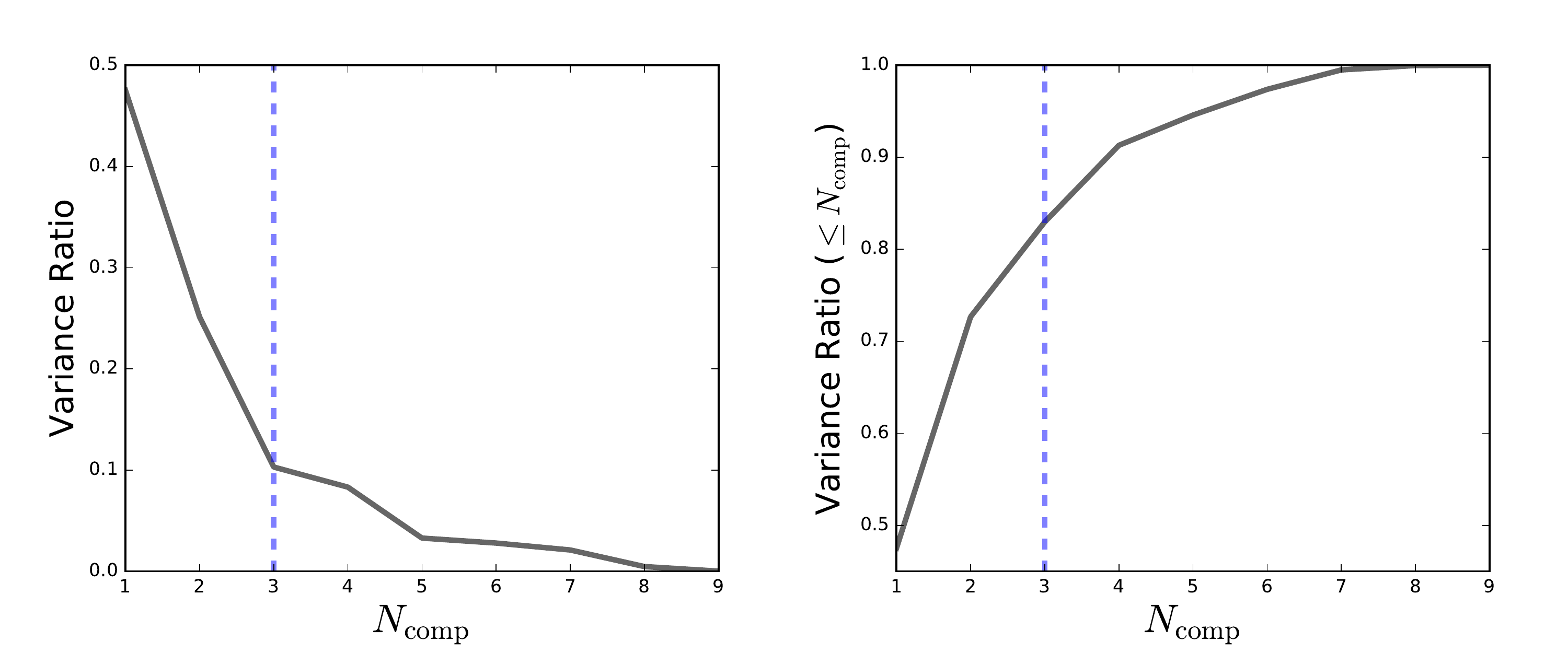}
\caption{The fraction (left) and cumulative fraction (right) of total feature variance explained by each PCA component. The first component explains nearly half of the variance, while the first three explain 83\%.}
\label{fig:pca-var-ratio}
\end{figure*}

Feature selection is described in \S\ref{sec:Classification}. The features were standardized by subtracting the mean and dividing by the standard deviation to put them on to approximately the same scale. Principle component analysis (PCA) was used to reduce the parameter space prior to fitting a model to the distribution. Figure~\ref{fig:pca-var-ratio} shows the fraction of the total variance explained by each PCA component. The first component accounts for nearly 50\% of the total variance, while the first thee account for 83\%. We select the first three components, beyond which the variance ratio begins to level out. Reducing the parameter space to three dimensions allows for much simpler model fitting and visualization compared to higher dimensions and the 17\% information loss is recovered at a later stage. 

Each principle component (denoted $C_0$, $C_1$, and $C_2$) is a linear combination of the input features weighted by the set coefficients given in table \ref{tab:pca}. The density distribution of the principal components is shown in the top row of Figure~\ref{fig:pca}.

\begin{table*}
\resizebox{\textwidth}{!}{
\begin{tabular}{|l|l|l|l|l|l|l|l|l|l|}
\hline
      & $S$ (Jy) & $S/\sigma_S$      & $S/\sigma_{\overline{S}}$ & $r_{\rm det}$      & $N_{\rm det}$ & $N_{\rm exp}$ & $\rho_N$ & $d_{\rm bright}$ & $S/S_{\rm bright}$ \\ 
Scale & $\log_{10}$ & $\log_{10}$ & $\log_{10}$       & linear & linear       & linear         & linear & linear     & $\log_{10}$     \\ \hline \hline

$C_0$	&	$-$0.391	&	0.034	&	$-$0.462	&	$-$0.458	&	$-$0.081	&	$-$0.245	&	$-$0.479	&	0.01	&	$-$0.355	 \\ 

$C_1$	&	0.341	&	$-$0.544	&	$-$0.181	&	$-$0.202	&	$-$0.482	&	0.051	&	$-$0.15	&	0.36	&	0.356	 \\ 

$C_2$	&	0.058	&	0.111	&	$-$0.235	&	$-$0.232	&	$-$0.049	&	0.835	&	$-$0.064	&	$-$0.408	&	0.062	 \\ \hline

\end{tabular}
}
\caption{The principle component coefficients of all input features described in \S\ref{sec:Classification}. }
\label{tab:pca}
\end{table*}

\subsection{Initial Unsupervised Classification}

\begin{figure}
\centering
 \includegraphics[width=\columnwidth]{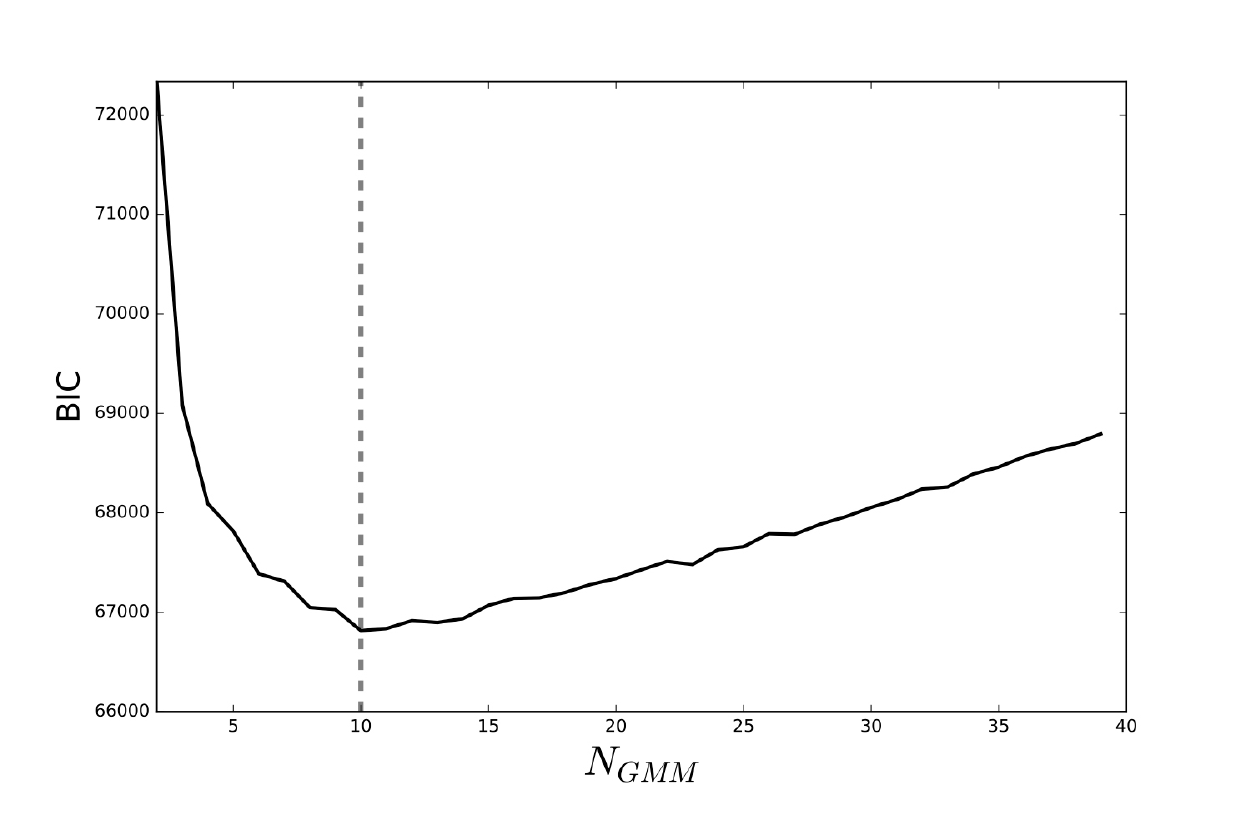}
\caption{The optimum number of Gaussian components was chosen to minimize the Bayes Information Criterion. A strong minimum is found at 10 components. }
\label{fig:bic}
\end{figure}

We fit a Gaussian mixture model (GMM) to the data in the reduced parameter space. Various other methods may be used to make this initial classification (e.g. k-means, nearest neighbors, quadratic discriminant analysis) however we found that the distribution was sufficiently represented by a ten component GMM. At ten components a sharp minimum is observed in the Bayes information Criterion (BIC; Figure \ref{fig:bic}), an indicator of goodness of fit that is sensitive to over-fitting. Every source candidate was labelled according to the GMM class it most probably belonged to. The distribution of classifications in the reduced space is shown in the middle row of Figure~\ref{fig:pca}. 

While the BIC suggests the GMM is a sufficient approximation of the data distribution, it is naive to assume Gaussianity with full knowledge that input feature distributions are non-Gaussian. The boundaries between classes appear to be forced by the Gaussian assumption rather than true to the underlying distribution. We therefore use the GMM classes only as input to train a Decision Tree ensemble classifier. 

\subsection{Decision Tree Ensemble Classifier}

Both Random Forest (RF) and AdaBoost (adaptive boosting) classifiers were tested. RF averages over many decision trees created on sub-samples of the data to build a more accurate classifier \citep{Breiman2001}. RF is robust against label noise (mis-classifications on the training set). AdaBoost builds a strong classifier from a decision tree by iteratively adjusting feature weights to focus on the outliers \citep{Freund1995,Zhu2009}. AdaBoost is therefore sensitive to mis-classifications. Because the decision tree favours the most distinguishing features, we used all nine original inputs to minimize information loss. 

The source candidates were split randomly 9:1 into training and testing sets. The classifier learns on the training set and is used to predict the test set. This was repeated for 5000 iterations so that each source was independently classified an average of 500 times and assigned to the cluster with the highest mean probability. 		

The variable input parameters to the RF and AdaBoost classifiers were not fully optimized, but variations on the number of trees or iterations were explored. The RF classifier tended to change the GMM classifications minimally or result in odd boundaries that were not easily interpretable. The AdaBoost classifier was subject to over-fitting if too many iterations were allowed. An AdaBoost classifier with 50 iterations reduced the artificial footprint of the Gaussian model assumption while maintaining reasonable agreement and an interpretable structure. 

\begin{figure*}
\captionsetup[subfigure]{labelformat=empty}
\centering
\subcaptionbox{}{\includegraphics[width=\textwidth]{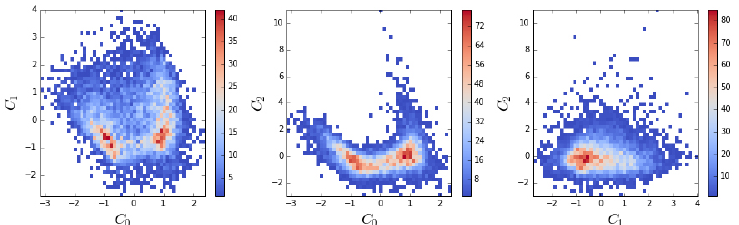}}
\subcaptionbox{}{\includegraphics[width=\textwidth]{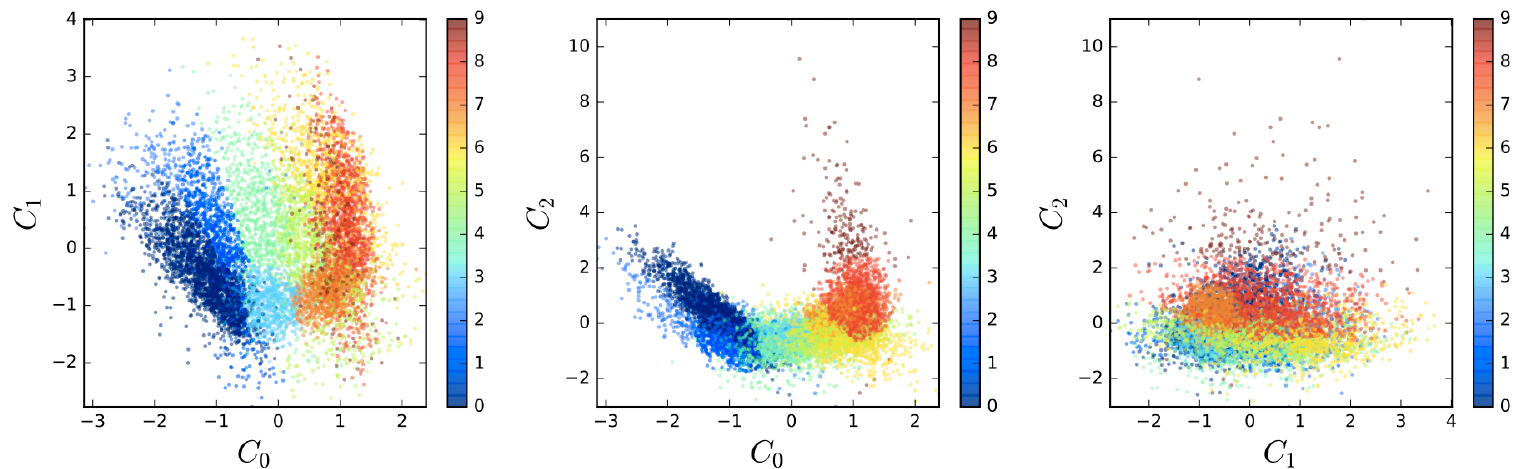}}
\subcaptionbox{}{\includegraphics[width=\textwidth]{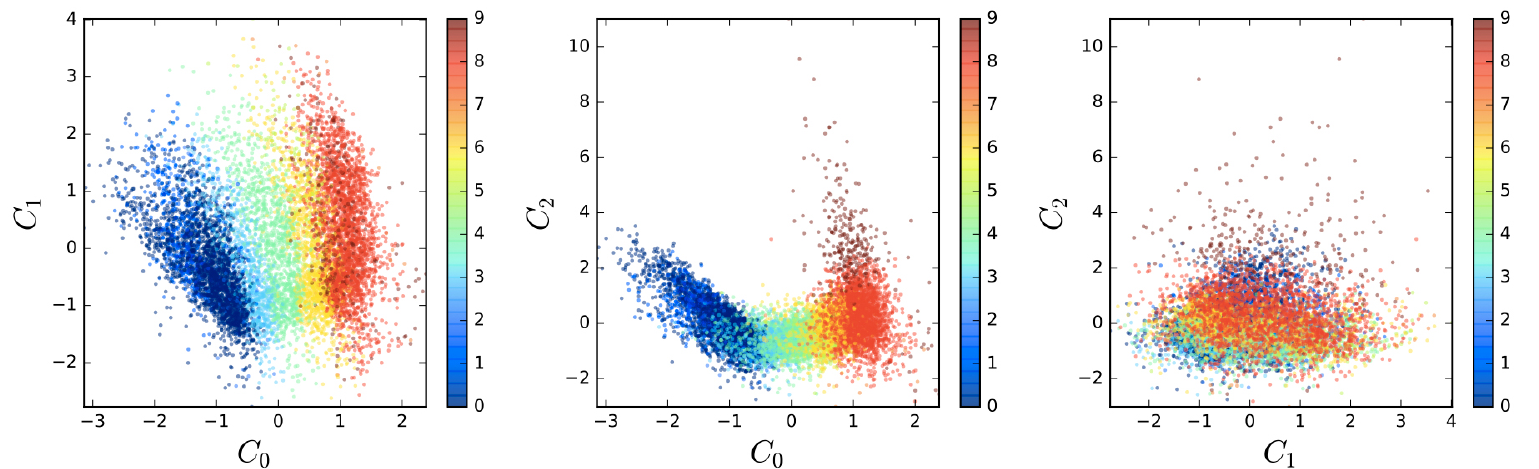}}
\caption{The distributions of the principle components for all source candidates. Top row: the density distribution. Middle row: Scatter plot colored by GMM classification. Bottom Row: Scatter plot colored by the AdaBoost classification. }
\label{fig:pca}
\end{figure*}

\begin{figure*}
\centering
\includegraphics[width=\textwidth]{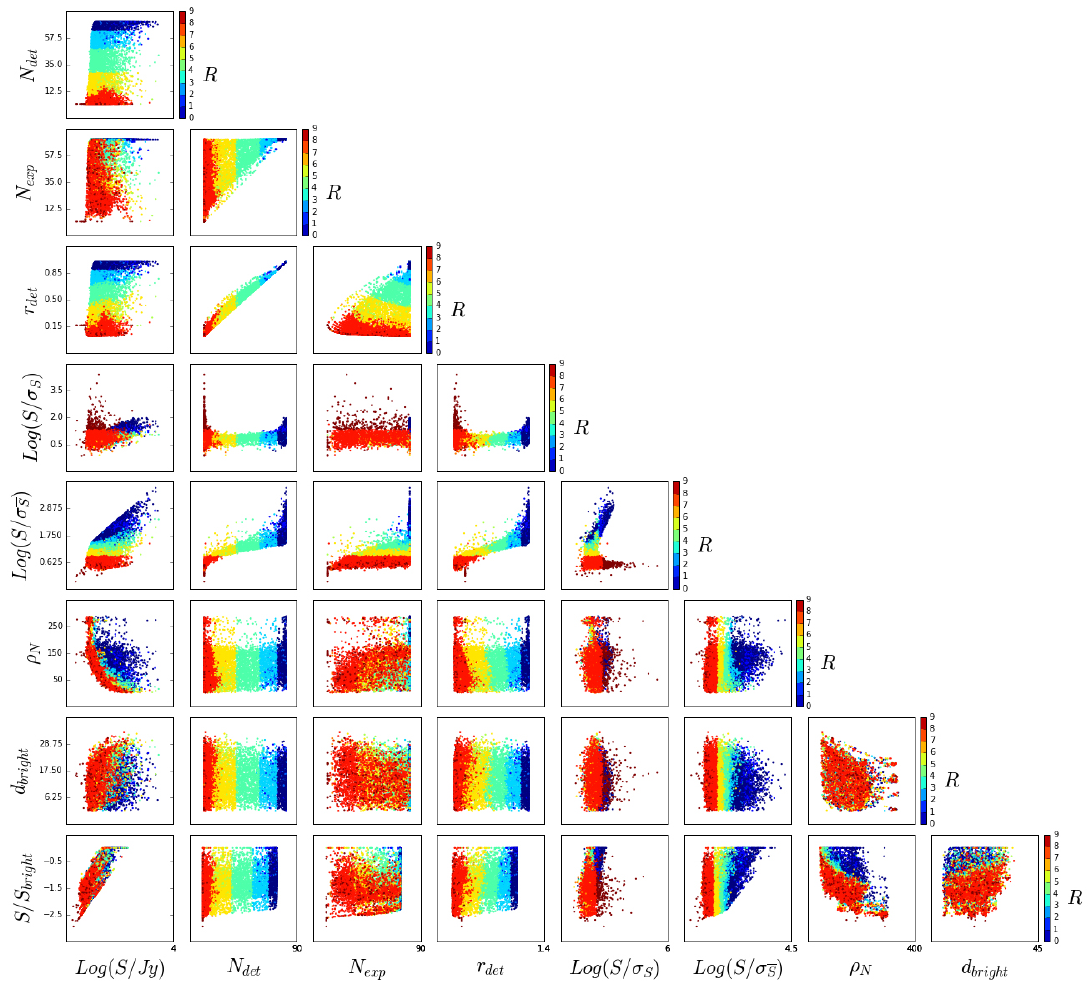}
\caption{The input features and their final classifications ($R0-R9$). Blue are the most reliable and red the least.}
\label{fig:triclass}
\end{figure*}

The final reliability classifications are shown in the bottom row of Figure~\ref{fig:pca}. To interpret these we look at their average properties, feature distribution, and spatial distribution. A selection of the input features are shown in Figure~\ref{fig:triclass} and the median values for each class are listed in table \ref{tab:classes}. Lower numbered classes tend to be the most reliable overall in terms of detection rate and SNR. Classes $R0-R2$ are highly reliable sources observed and detected in nearly all snapshots. Classes $R3-R6$ capture fainter sources with reliability decreasing further afield. Classes $R5$ and $R6$ appear to identify real sources in high density regions, whereas false side lobe sources are mostly restricted to classes $R8$ and $R9$. Class $R9$ sources stand out, having artificially high SNR due to a very small number of detections. Sources with $R_{class}>6$ may be real but observed few times near the detection threshold. 

The reliability classification is informative but not infallible, as discussed in \S\ref{sec:unmatched}. Only in the case of a source unmatched to another radio catalogue is the reliability class used to include or exclude it from the final catalogue. Future work will explore the usefulness of this classification in determining an optimal EoR foreground model. 

\section{Bayesian matching}
\label{app_bayesmatch}

In \S\ref{sec_XMatch} we cross match to multiple other radio catalogs using the Positional Update and Matching Algorithm (Line et al., {\it in prep.}). PUMA uses positional and spectral information to statistically test whether sources from multiple surveys in close proximity to one-another are true matches.

When matching $N_{\rm cat}$ catalogues, it can be shown \citep[see][for further details]{Budavari2008} that the Bayes Factor is given by

\begin{align}
B = 2^{N_{\rm cat}-1} \frac{\prod w_i}{\sum w_i} \exp \left( -\frac{\sum_{i<j} w_iw_j\psi^2_{ij}}{2\sum w_i} \right),
\label{bayes_fact}
\end{align}

where $\psi_{ij}$ is the angular separation between sources in the $i^{th}$ and $j^{th}$ catalogues, and $w_i$ is the weighting for the $i^{th}$ catalogue. This is given by $w = 1/\sigma^2$, where $\sigma^2$ is the astrometric error (taken to be $\sigma^2 = \sigma_{\rm RA}^2 + \sigma_{\rm Dec}^2$). $B$ can be used to estimated to posterior probability that the catalogues are describing the same source through 
\[P(H|D) = \dfrac{BP(H)}{1+BP(H)}. \]

The prior, $P(H)$, is given by

\begin{align}
P(H) = \frac{n_0}{\prod_{i=1}^{N_{\rm cat - 1}} n_i} ,
\label{prior}
\end{align}

where the scaled full sky number of sources in each catalogue is $n_i = 4\pi N_i/ \Omega_i$, with $N_i$ equal to the number of sources in the catalogue, and $\Omega_{i}$ the catalogue survey area. $n_0$ is the scaled full sky number of sources in the base catalogue.

If $P<0.95$, the SED is fit with a power law model of the form $\log S \propto \alpha \log \nu$ using linear least squares. The fit is considered good if the reduced Chi-square statistic $\chi_{\rm red}^2$ is less than 10. Due to uncertainty on $\chi_{\rm red}^2$ given the small number of data points and uncertainty on the errors, the fit is additionally considered good if the residuals $\epsilon$ are less than 0.1 (\ref{eqn:residual-metric}).

\begin{equation}
\label{eqn:residual-metric}
\epsilon = \frac{1}{N_{\rm cat}} \sum_{i=1}^{N_{\rm cat}}\left(\frac{|f_i- \left < f_i \right >|}{f_i} \right )
\end{equation}

\bibliography{kgs_ms}
\bibliographystyle{mnras}

\label{lastpage}
\end{document}